\documentclass[a4paper,11pt,final]{article}

\usepackage{epsfig}
\usepackage[cmex10]{amsmath}
\usepackage{amssymb,amsthm}
\usepackage{mathtools}
\usepackage{nccmath}
\usepackage{soul}
\usepackage[normalem]{ulem}
\usepackage{cite}
\usepackage{todonotes}
\usepackage{tikz}
\usepackage{hyperref}
\usetikzlibrary{hobby,decorations.markings,arrows,intersections,shapes}
\usepackage{pgfplots}
\usepgfplotslibrary{fillbetween}
\pgfplotsset{compat=newest}
\usepgfplotslibrary{fillbetween}
\usepackage{verbatim}

\begin{document}

% paper title: Must keep \ \\ \LARGE\bf in it to leave enough margin.
\title{\ \\ \LARGE\bf Constructing transient amplifiers for death-Birth updating: A case study of cubic and quartic regular graphs}

\author{Hendrik~Richter \\
HTWK Leipzig University of Applied Sciences \\ Faculty of
Electrical Engineering and Information Technology\\
        Postfach 301166, D--04251 Leipzig, Germany. \\ Email: 
hendrik.richter@htwk-leipzig.de. }

\maketitle

\begin{abstract}

A central question of evolutionary dynamics on graphs is whether or not a mutation introduced in a population of residents survives and eventually even spreads to the whole population, or gets extinct.  The outcome naturally depends on the fitness of the mutant and the rules by which mutants and residents may propagate on the network, but arguably the most determining factor is the network structure.  Some structured networks are transient amplifiers. They increase for a certain fitness range  the fixation probability of beneficial mutations as compared to a well-mixed population. 
We study a perturbation methods for identifying transient amplifiers for death-Birth updating. The method includes calculating the coalescence times of random walks on  graphs and finding the vertex with the largest remeeting time. If the graph is perturbed by removing an edge from this vertex, there is a certain likelihood that the resulting perturbed graph is a transient amplifier. We test all pairwise nonisomorphic cubic and quartic regular graphs up to a certain size and thus cover the whole structural range expressible by these graphs. We carry out a spectral analysis and show that the graphs from which the transient amplifiers can be constructed  share certain structural properties. The graphs are path-like, have low conductance and are rather easy to divide into subgraphs by removing edges and/or vertices. This  is connected with the subgraphs being identical (or almost identical) building blocks and the frequent occurrence of cut and/or hinge vertices. Identifying spectral and structural properties may promote finding and designing such networks.

\end{abstract}

\section*{Author summary}
Until recently it was assumed that amplifiers of natural selection for the death-Birth Moran process are either very rare or do not exist.  Newer results, however, have modified this assumption in two respects. The first result is that if amplifiers for death-Birth updating exist, they must be transient, which means the amplification only applies to a limited range of the mutant's fitness. The second result is a rather simple numerical test to decide whether or not a graph is an amplifier of weak selection, which means for a small  effect of the mutant's fitness. This test involves to perturb the network describing the interactions between the mutant and the  population of residents.    
We study this perturbation methods
for identifying transient amplifiers for death-Birth updating and study their network structure. Thus we identify structural properties of transient amplifiers which may promote finding and designing such networks.

\section*{Introduction}
  
An important measure for the success of an initially rare mutant among a resident population on an evolutionary graph is the fixation probability of the mutation.  The evolutionary dynamics associated with the mutant's spread most likely depends on its fitness, with a  beneficial mutant possessing a higher fitness than the resident individuals, while a deleterious mutant has a lower fitness.
Previous works have shown that compared to the complete graph representing a well-mixed population, some graphs produce a higher fixation probability for a beneficial mutant, thus amplifying the effect of selection~\cite{adlam15,hinder15,jam15,pav17}. For some other graphs we find the opposite with an increased fixation probability for a deleterious mutation, and thus suppressing selection. Apart from these two groups of graphs, a third category has been identified which is transient amplifiers characterized by an increased fixation probability for some range of the mutant's fitness.     

Fixation probabilities are determined by the structure of the evolutionary graph, but also by the rules by which mutants and residents propagate on the graph.
Thus, whether an evolutionary graphs is a transient  amplifier, or an amplifier, or a suppressor, does not only depend on the graph but also on the updating rule describing the dynamics of  mutants and residents. Two updating rules frequently studied are Birth-death (Bd) updating and death-Birth (dB) updating.   
As transient amplifiers provide a mechanism for shifting the balance between natural selection and genetic drift, and thus have considerable significance in evolutionary dynamics, they have been studied intensively~\cite{alcalde17,allen20,hinder15,pav18,tka20,adlam15,pav17,monk18,jam15}.  In these works there are two main approaches for identifying transient amplifiers. One approach is to check a large number of random graphs, for instance Erd{\"o}s-Rényi or Barabási-Albert graph~\cite{alcalde18,hinder15,moell19,tka19}. Another approach is to design candidate graphs out of structural aspects and considerations~\cite{adlam15,jam15,pav17,pav18}. An example for the first approach is a comprehensive
 numerical study checking a larger number of random graphs with $N\leq 14$ vertices, which found a multitude of amplifiers of selection for Bd, but none for dB updating~\cite{hinder15}. From these results it was assumed that either there are no amplifiers for dB, or they are very rare, at least for graphs with small size.   Examples for the second approach involve, for instance, constructing arbitrarily strong amplifiers for star and comet graphs, particularly by additionally designing weights~\cite{pav17,pav18}.
 
 There are two recent works on amplifiers for dB updating, upon which this work is particularly based. In Tkadlec et al.~\cite{tka20} it is shown that for dB updating no universal amplification is possible and at most evolutionary graphs can be transient amplifiers. In Allen at al.~\cite{allen20} 
 a method is devised to check whether or not a given graph is at least an amplifier for weak selection. This method involves calculating the coalescence times of random walks on the graph~\cite{allen17} and finding the vertex with the largest remeeting time. If subsequently the graph is perturbed by removing an edge from this vertex, there is a certain likelihood that the resulting perturbed graph is a transient amplifier.
 
 In this paper it is proposed to take as an input to this perturbation method all pairwise nonisomorphic cubic and quartic regular graphs up to a certain size.
 For a small size $N$ the exact number of connected (pairwise nonisomorphic) cubic and quartic graphs is known and all graphs can be generated algorithmically~\cite{mer99}. With the available numerical resources results have been obtained for $N\leq 22$ for cubic and $N\leq 16$ for quartic graphs.  Thus,  the study is conducted for the whole set of these regular graphs and ensures
that the whole structural range expressible by these graphs is covered. 
The results show that a substantial number of graphs obtained by perturbing cubic and quartic regular graphs are transient amplifiers. Thus, the method discussed in the paper yields a sufficiently large set of amplifier graphs, which suggests to consider these graphs as an ensemble to be treated statistically.
Moreover, we study structural graph properties  by the spectrum of the normalized Laplacian, thus adding to the applications of spectral analysis of evolutionary graphs~\cite{rich17,rich19a,rich19b,rich20,allen19}.  In the analysis we use a  smoothed spectral density which convolves the eigenvalues with a Gaussian kernel and show that cubic or quartic regular graphs which allow to construct transient amplifiers have a characteristic spectral density curve.   Thus, it becomes feasible to deduce from the spectrum of the graph if a transient amplifier can be constructed.

\section*{Methods}
\subsection*{Constructing transient amplifiers}
Our topic is evolutionary dynamics on graphs with a population of $N$ individuals on an undirected graph  $\mathcal{G}=(V,E)$ with  vertices $v_i \in V$ and edges $e_{ij}\in E$. Each individual is represented by a vertex $v_i$ and an edge $e_{ij}=e_{ji}=1$ shows that the individuals associated with $v_i$ and $v_j$ are  mutually interacting  neighbors~\cite{allen17,lieb05,ohts07,patt15,rich17}. 
We consider
the graph $\mathcal{G}$ to be simple and  connected with each vertex $v_i$ having degree $k_i$, which implies that there is no self-play and the individual on $v_i$ has $k_i$ neighbors. The test set used as an input to the perturbation method for constructing transient amplifiers mostly consists of cubic and quartic regular graphs, for which consequently there is $k_i=k=\{3,4\}$ for all vertices $v_i$. 

Individuals can be of two types, mutants and residents. Residents have a constant fitness normalized to unity, while  mutants have a fitness $r>0$. An individual can change from mutant to resident (and back) in a fitness-dependent selection process. We consider a death--birth (dB) process, e.g.~\cite{allen14,patt15}. The type of a vertex becomes vacant as the  occupying
individual is assumed to die, which happens uniformly at random. One of the neighbors is chosen to give birth with a probability depending on its fitness. It hands over its type,  thus effectively replacing the death individual. 

We are interested in the fixation probability $\varrho_{\mathcal{G}}$ defined as the expected probability that starting with a single mutant appearing at a vertex uniformly at random all vertices of the graph $\mathcal{G}$ eventually become the mutant type. In particular, we want to know how this fixation probability $\varrho_{\mathcal{G}}(r)$ compares to the fixation probability $\varrho_{\mathcal{N}}(r)$ for the complete graph with $N$ vertices for varying fitness $r$. More specifically, a graph $\mathcal{G}$ is called an amplifier of selection if $\varrho_{\mathcal{G}}(r)<\varrho_{\mathcal{N}}(r)$ for $0<r<1$ and $\varrho_{\mathcal{G}}(r)>\varrho_{\mathcal{N}}(r)$ for $r>1$. Correspondingly, a graph $\mathcal{G}$ is a suppressor of selection if $\varrho_{\mathcal{G}}(r)>\varrho_{\mathcal{N}}(r)$ for $0<r<1$ and $\varrho_{\mathcal{G}}(r)<\varrho_{\mathcal{N}}(r)$ for $r>1$. 
Finally, a transient amplifier is characterized by
$\varrho_{\mathcal{G}}(r)<\varrho_{\mathcal{N}}(r)$ for $r_{min}<r<1$ and $r>r_{max}$, while there is also $\varrho_{\mathcal{G}}(r)>\varrho_{\mathcal{N}}(r)$ for $1<r<r_{max}$ for some $0<r_{min}<1<r_{max}$~\cite{allen20,hinder15,adlam15,pav17}.

There are two recent works on amplifiers for dB updating, which inspired and motivated this work. In Tkadlec et al.~\cite{tka20} it is shown that for dB updating no universal amplification is possible and at most evolutionary graphs can be transient amplifiers. In Allen at al.~\cite{allen20} it is demonstrated that for weak selection, that is $r=1+\delta$ for $\delta \rightarrow 0$, the question of whether or not a graph $\mathcal{G}$ is an amplifier can be answered by a numerical test executable with polynomial time complexity. The test relies upon coalescing random walks~\cite{allen17} and involves 
calculating 
the effective population size $N_{eff}$ from the relative degree $\pi_i=k_i/\sum_{j \in \mathcal{G}} k_j$ and the remeeting time $\tau_i$ of vertex $v_i$ by \begin{equation} N_{eff}= \sum_{i \in \mathcal{G}} \pi_i \tau_i. \label{eq:neff} \end{equation}
The remeeting time $\tau_i$ can be obtained via
\begin{equation} \tau_i= 1+ \sum_{j \in \mathcal{G}} p_{ij}\tau_{ij}  \end{equation}
from the coalescence times $\tau_{ij}$ and the step probabilities 
$p_{ij}=e_{ij}/k_i$ (implying $p_{ij}=1/k_i$ if $e_{ij}=1$ and $p_{ij}=0$ else). The remeeting times observe the condition $\sum_{i \in \mathcal{G}}  \pi_{i} ^2\tau_{i}^{}=1$.
Finally, the coalescence times $\tau_{ij}$ can be computed by solving the system of $\left(\begin{smallmatrix}N\\2\end{smallmatrix}\right)$ linear equations
\begin{equation} \tau_{ij}= \left\{ \begin{array}{cc} 0 & i=j\\ 1+ \frac{1}{2} \sum_{k \in \mathcal{G}} (p_{ik}\tau_{jk}+p_{jk}\tau_{ik}) & i\neq j  \end{array} \right.  .\end{equation}
In~\cite{allen20} it is shown that a graph $\mathcal{G}$ is an amplifier of weak selection if \begin{equation}N_{eff}>N. \label{eq:amplifiercondition} \end{equation}
Furthermore, it is argued that an amplifier of weak selection can be constructed by the following perturbation method. If the graph 
$\mathcal{G}$  is $k$-regular, then $k_i=k$ and $\pi_i=1/N$ for all $i=1,2,\ldots,N$. Thus, with the identity condition $\sum_{i \in \mathcal{G}}  \pi_{i} ^2\tau_{i}^{}=1$  we have   $N_{eff}= \sum_{i \in \mathcal{G}} \tau_i/N=N\sum_{i \in \mathcal{G}} \tau_i/N^2=N \sum_{i \in \mathcal{G}}  \pi_{i} ^2\tau_{i}^{}=N$, which according to Eq. \eqref{eq:amplifiercondition} means that $k$--regular graphs  cannot be amplifiers of weak selection. But disturbing the regularity can change the equality $N_{eff}=N$.
Moreover, it has been analyzed that most promising for such a perturbation is to remove a single (or even more than one) edge from a vertex $v_i$ with the largest remeeting time $\tau_i$, that is $\max_i \tau_i$, on the $k$--regular graph to be tested. The argument is that if we take a regular graph and induce a small perturbation by removing  an edge, 
the relative degree $\pi_i$ and the remeeting time $\tau_i$ experience  small deviations $\Delta \pi_i$ and $\Delta \tau_i$. For the identity condition  $\sum_{i \in \mathcal{G}}  \pi_{i} ^2\tau_{i}^{}=1$ it follows \begin{equation} 
 \sum_{i \in \mathcal{G}} \Delta (\pi_{i} ^2\tau_{i}^{})= \sum_{i \in \mathcal{G}} (2 \pi_i \Delta \pi_{i} \tau_{i}^{}+ \pi_{i} ^2 \Delta \tau_{i}^{}) \approx 0.
\end{equation}
As there is $\pi_i=1/N$ for the unperturbed regular graph,  we obtain \begin{equation}  1/N  \sum_{i \in \mathcal{G}} \Delta \tau_{i}^{} \approx   -2 \sum_{i \in \mathcal{G}} \Delta \pi_{i} \tau_{i}^{}. \label{eq:perturb2} \end{equation}
For the calculation of the effective population size, Eq. \eqref{eq:neff}, the perturbation yields
\begin{equation} \Delta N_{eff}= \sum_{i \in \mathcal{G}} \Delta(\pi_i \tau_i) \approx \sum_{i \in \mathcal{G}} (\Delta \pi_i \tau_i+\pi_i \Delta \tau_i). \end{equation}
By observing $\pi_i=1/N$ for the unperturbed regular graph and inserting Eq. \eqref{eq:perturb2}, we get  
\begin{equation} \Delta N_{eff} \approx -\sum_{i \in \mathcal{G}} \Delta\pi_i \tau_i.\end{equation}
This relationship implies that a positive perturbation of the effective population size (and thus the possibility to obtain $N_{eff}>N$) is obtained if for a large $\tau_i$ the perturbation induces a decrease of the relative degree $\pi_i$, that is a negative $\Delta \pi_i$. 

\subsection*{Spectral analysis}
For the  spectral analysis of evolutionary graphs we consider the spectrum of the  normalized Laplacian $L_{\mathcal{G}}$ of the graph $\mathcal{G}$, which is $L_{\mathcal{G}}=I-D^{-1/2}AD^{-1/2}$;  $A$ is the adjacency matrix of the graph $\mathcal{G}$ and $D$ is the degree matrix.  The spectrum is denoted by $\lambda(\mathcal{G})$ and consists of $N$ eigenvalues $0=\lambda_1\leq \lambda_2 \leq \ldots \lambda_N\leq 2$. The spectrum of the  normalized Laplacian  is used
rather than the spectrum of the adjacency matrix as it has been shown that  
it captures well some geometric properties~\cite{ban08,ban09,ban12,gu16}. In addition, the spectrum is contained in $[0,2]$ for any graph size and degree, which makes is convenient to compare for varying graphs. From the spectrum  a spectral distance $d$ can be defined, which can be used to compare two  families (or classes) of graphs $(\mathcal{G})$ and $(\mathcal{G}')$. For each family only containing a single member, we in fact can also compare two graphs $\mathcal{G}$ and $\mathcal{G}'$.  The comparison can be done directly by viewing the histograms of the eigenvalue distribution $f_{(\mathcal{G})}(x)$ on a discrete variable $X$ as spectral plots~\cite{ban07}. A more refined comparison can be achieved by considering a smoothed spectral density which convolves the eigenvalues $\lambda_i$ with a Gaussian kernel with standard deviation $\sigma$~\cite{ban09,ban12,gu16} 
\begin{equation}
    \rho_{\mathcal{G}}(x)= \sum_{i=1}^{n} \frac{1}{\sqrt{2 \pi \sigma^2}} \exp{\left(\frac{(x-\lambda_i)^2}{2 \sigma^2} \right)}. \label{eq:density}
\end{equation} We set $\sigma=1/(3N)$.
From this continuous spectral density we can define a pseudometric on graphs by the distance~\cite{gu16} 
\begin{equation}
d(\mathcal{G},\mathcal{G}')= \int_0^2|\rho_{\mathcal{G}}(x)-\rho_{\mathcal{G}'}(x)|dx. \label{eq:distance}
 \end{equation}

\section*{Results}
\subsection*{Regular graphs as amplifier constructors}
Previous approaches to finding amplifiers of selection for dB updating focused on checking numerically generated random graphs, for instance Erd{\"o}s-Rényi or Barabási-Albert graphs with prescribed expected degree or linking number~\cite{alcalde18,hinder15,moell19,tka19}. As we are interested in how population structure relates to evolutionary dynamics, it would be most desirable to study differences in the graph structure among the realizations of random graphs; ideally, these  structural differences would encompass all what is structurally possible. There are, however, some problems with such an approach. Suppose we generate random graphs and two of them are structurally the same.  In an experimental setting, however, this might be hard to detect  as  the computational problem of determining whether two finite graphs are isomorphic is not solvable in polynomial time~\cite{arv05,bab18}.  On the other hand, isomorphic graphs have the same fixation properties. Related to the problem addresses in this paper, the numerical procedure would yield for two random but isomorphic graphs the same $N_{eff}$. Moreover, 
even for a relatively small number of vertices, the number of nominally different random graphs is huge, see for instance the results for the class of labelled regular graphs~\cite{worm99}.   In addition, 
algorithms producing random graphs, regular or otherwise, may have a bias towards certain graph structures~\cite{bay10,klein2012}. Thus, even if a large number of random graphs is produced and checked, 
there might be isomorphic graphs that have the same structural properties or
there might be
``blind spots'' for certain structural types of graphs. Thus, as long as not all graphs of a certain class are enumerated, it is far from certain that by checking a finite number of random graphs from this class the relevant search space of graph structures has been adequately covered. 

The numerical procedure suggested here aims for 
a more systematically conducted search for a certain class of graphs. 
For a small number  of vertices $N$ the exact number of connected (pairwise nonisomorphic) cubic and quartic graphs is  known, see e.g.~\cite{mer99} and also Tab. \ref{tab:graphs}. With the numerical resources available  for this study cubic graphs could be checked for $N \leq 22$ and quartic graphs for $N \leq 16$. The complete set is tested. Thus, the study ensures that the whole structural range of these regular graphs is covered. We take these graphs as the input for the perturbation method described in the previous section to find transient amplifiers of weak selection.  

 \begin{table}[htb]
\centering
\caption{\small{The  number $\mathcal{A}_k(N)$ of simple, connected, pairwise nonisomorphic $k$-regular graphs on  $N$ vertices that construct transient amplifiers, also called amplifier constructors. Compare to the total number of  nonisomorphic $k$-regular graphs on  $N$ vertices in Tab. \ref{tab:graphs} of the Supporting Information.  The number of parenthesis are the graphs for which 2 edges can be removed to produce an amplifier.   }}
\label{tab:graphs_amp}
\center
\begin{tabular}{ccccc}

\hline
 \boldmath{$\: {}_N \: \backslash \: {}^k$} & \textbf{3} & \textbf{4} & \textbf{5} &\textbf{6-12} \\ \hline 11 &
 & 1 &  & 0 \\
 12 & 1 & 4 & 0 & 0  \\
 13 &  & 23  &  & 0   \\
 14 &7 & 108 / (3) & 30 / (5) & 0 \\
 15 &  & 562 / (16) \\
 16 & 42& 3.129 / (36)  \\
 18 & 265 \\
 20 & 1.822 \\
 22 & 13.889\\

\hline

\end{tabular}
\end{table}

\begin{figure}[htb]
\centering
\includegraphics[trim = 55mm 110mm 50mm 110mm,clip, width=6.5cm, height=5cm]{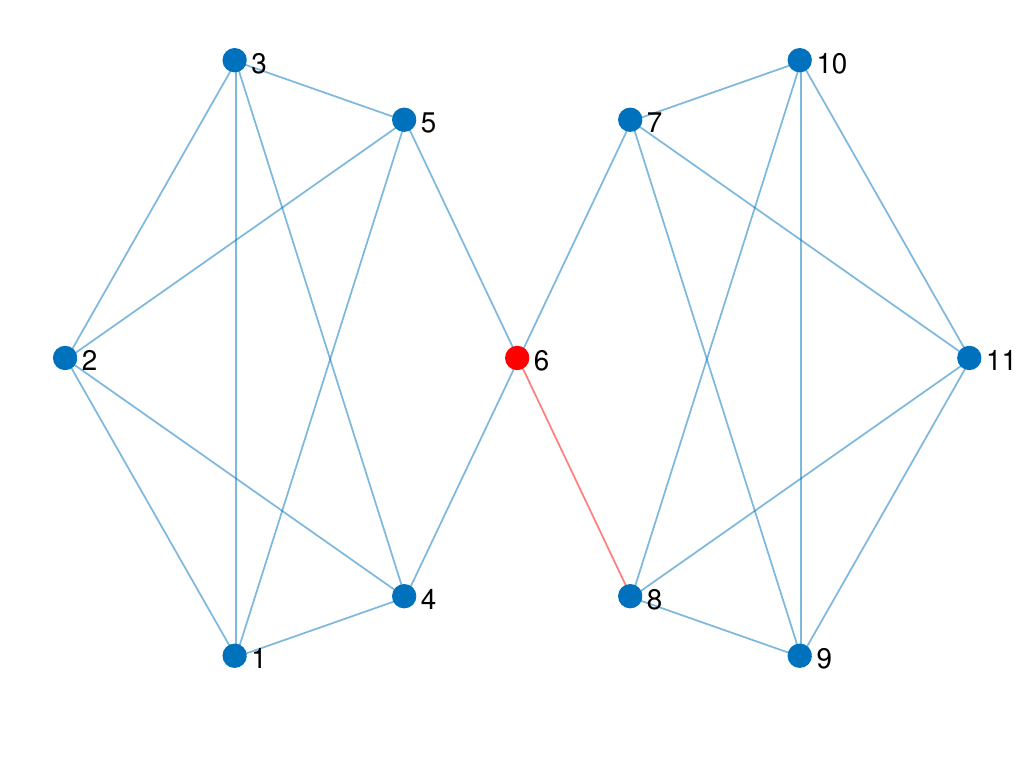}
\includegraphics[trim = 25mm 90mm 20mm 80mm,clip, width=6.5cm, height=5.5cm]{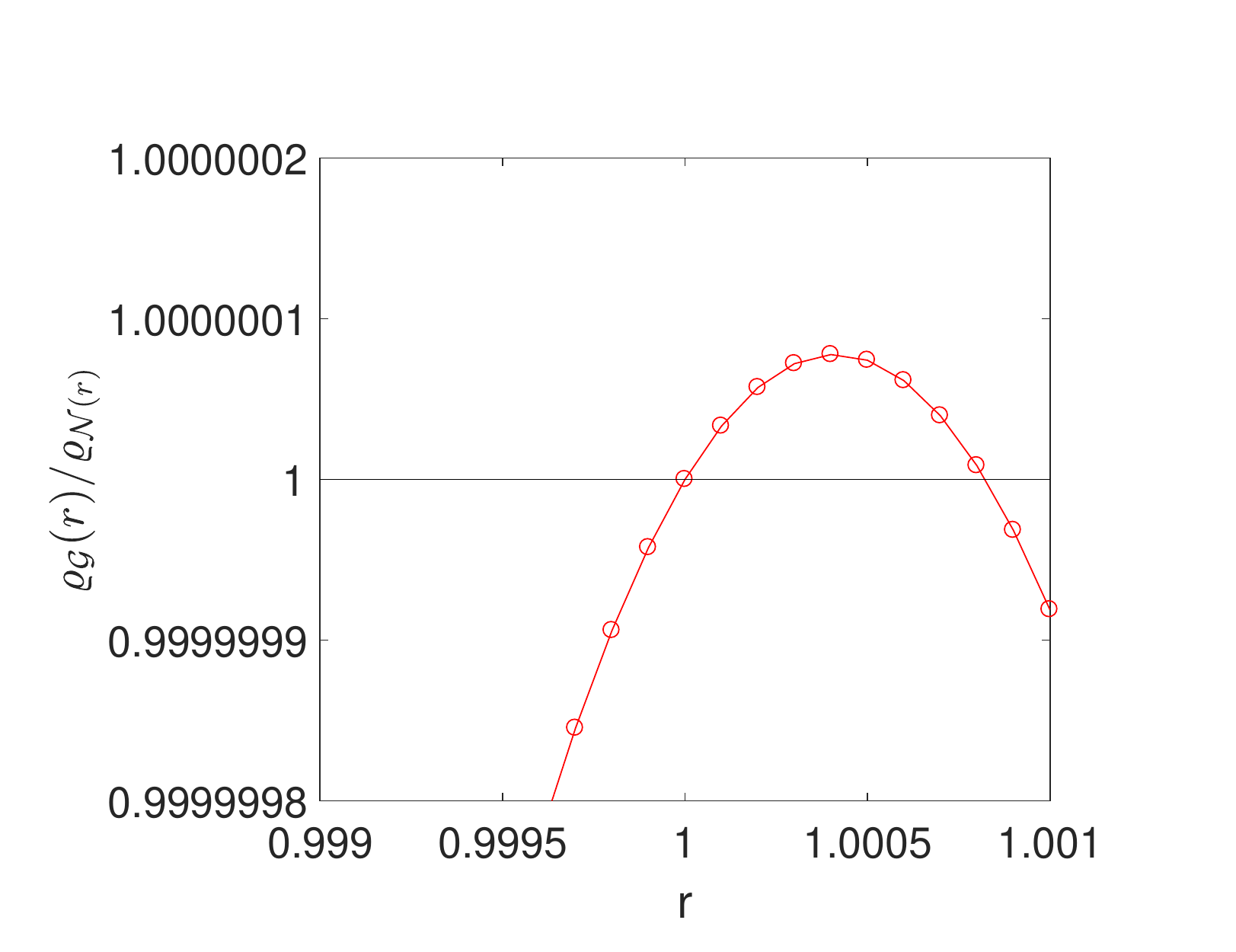}

 (\textbf{a}) \hspace{4.7cm} (\textbf{b})

\caption{\small{The $4$-regular graph of size $11$ used to construct a transient amplifier for $1<r<r_{max}\approx 1.00075$. (a) The vertex $v_6$ has the maximal remeeting time $\tau_6=17.6997$ and a transient amplifier arises if an edge from this vertex is removed. Due to the symmetry removing any of the 4 edges yields a graph with $N_{eff}=11.0008>11$. (b) The ratio between the fixation probability of the graph $\mathcal{G}$ and the complete graph which indicates a transient amplifier for $\varrho_{\mathcal{G}}(r)/\varrho_{\mathcal{N}}(r)>1$.  }  }
\label{fig:graph_11_4}
\end{figure}

Tab. \ref{tab:graphs_amp} gives the numbers $\mathcal{A}_k(N)$ of regular graphs that produce $N_{eff}>N$ if the perturbation method is applied. We call these graphs amplifier constructors. Apart from the cubic and quartic graphs, also the graphs with degree  $4<k<N-1$ have been tested for $N\leq 14$. Viewing these results, several observations can be made. Within the bounds of the experimental setting, only for $k=\{3,4,5\}$ regular graphs produce transient amplifiers. For the tested graphs with $k\geq 6$ no such graphs have been found for $N\leq 14$. It is, however, quite possible that this is due to the number of vertices not large enough and that a certain difference $(N-d)$ is needed for the property to be an amplifier constructor. Another possibility is that for graphs with a higher degree $k$ the perturbation method only gives transient amplifiers if a larger number of edges is removed. This should to be clarified by future work.
The smallest graph for which $N_{eff}>N$ was found is the quartic graph with size $N=11$ given in Fig. \ref{fig:graph_11_4}a. It should be noted that although the results given in this paper apply to unlabeled pairwise nonisomorphic graphs, the vertices of the graph in Fig. \ref{fig:graph_11_4}a, as well as the vertices of graphs in other figures, are labeled with consecutive integers. This is solely done to ease addressing and communicating certain vertices or edges, for instance for indicating which vertex has the largest remeeting time, or which edge is removed from a given graph.

A second observation is that 
the absolute number $\mathcal{A}_4(N)$ is larger than $\mathcal{A}_3(N)$. But as  $\mathcal{L}_4(N)$ grows much faster than $\mathcal{L}_3(N)$, compare Tab. \ref{tab:graphs}, the relative number is not. For cubic graphs, we have  a ratio $\mathcal{A}_3(N)/\mathcal{L}_3(N)= \{1/85,7/509,42/4.060, 265/41.301, 1.822/510.489, 13.889/7.319.447 \}$ for $N=\{12,14,\ldots,22 \}$, which shows that the percentage of regular graphs with amplifier construction properties starts from more than $1\%$ for $N=12$ to go to $0.35\%$ and $0.19\%$ for $N=\{20,22\}$. For quartic graphs with $N=\{11,12,\ldots,16 \}$ the ratio $\mathcal{A}_4(N)/\mathcal{L}_4(N)$ sets out with $0.37\%$ for $N=11$ to end with $0.07\%$ and $0.04\%$ for $N=\{15,16\}$. In other words, regular graphs that construct transient amplifiers are rare but not unprecedented.

 Tab. \ref{tab:graphs_amp} gives the numbers of graphs from which at least one transient amplifier graph can be constructed. As cubic graphs can be perturbed by removing $3$ edges from the vertex with the largest remeeting time, frequently from the same graph more than one amplifier is obtained. Sometimes, the resulting amplifiers are the same due to symmetry (as for instance the graph in Fig. \ref{fig:graph_11_4}a), but there are also cases where transient amplifiers with different $N_{eff}$ arise. The same applies for quartic graphs, which additionally have the property that a small number of these graphs allow to remove 2 egdes to produce an amplifier.   This has been observed for $14 \leq N \leq 16 $ and also for the quintic graphs with $N=14$. Tab.  \ref{tab:graphs_amp} gives the number of these graphs with 2 removable edges as the quantities in parenthesis.  
 
\begin{figure}[htb]
\centering
\includegraphics[trim = 25mm 90mm 40mm 20mm,clip, width=6.5cm, height=9.5cm]{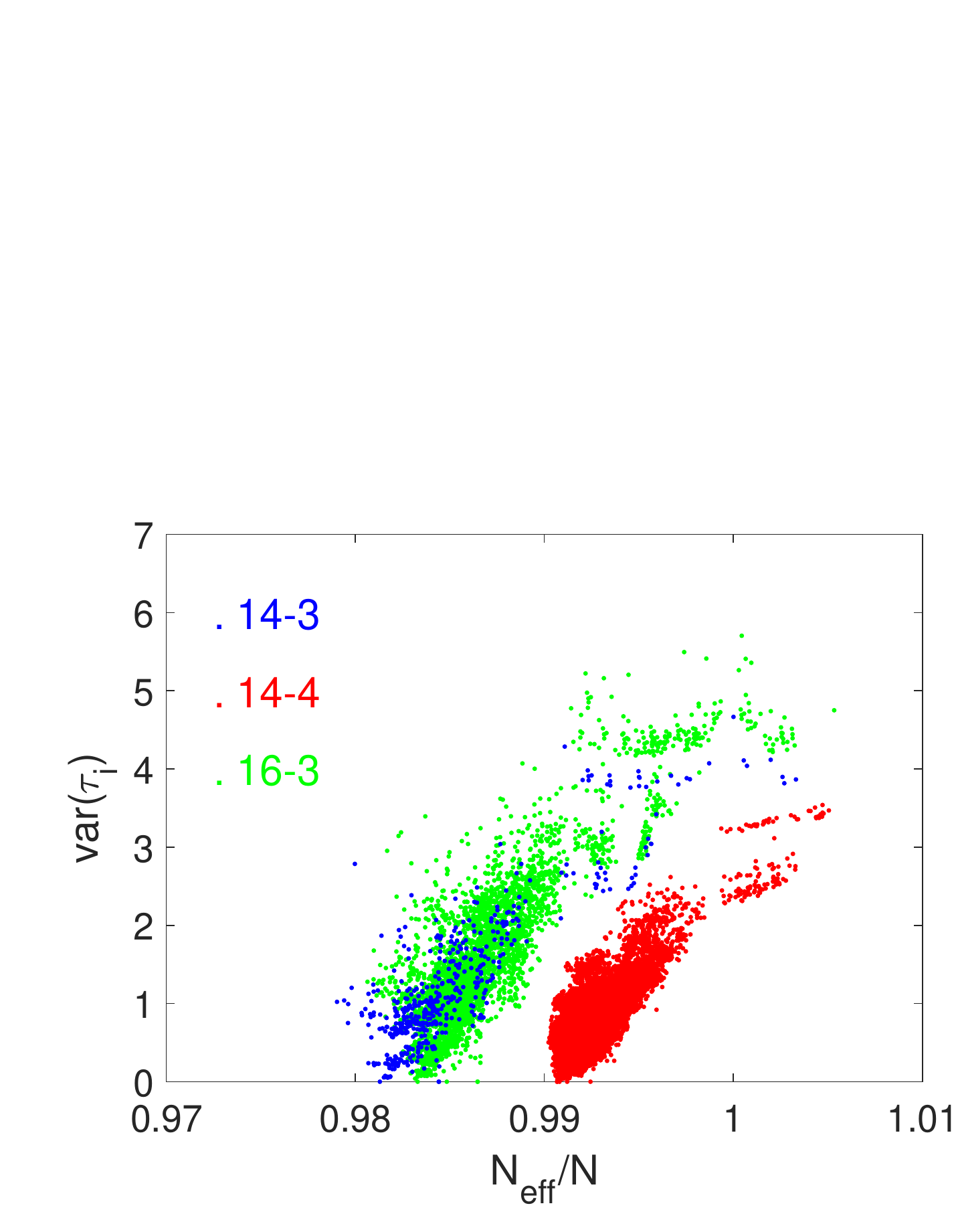} 
\includegraphics[trim = 25mm 90mm 40mm 20mm,clip, width=6.5cm, height=9.5cm]{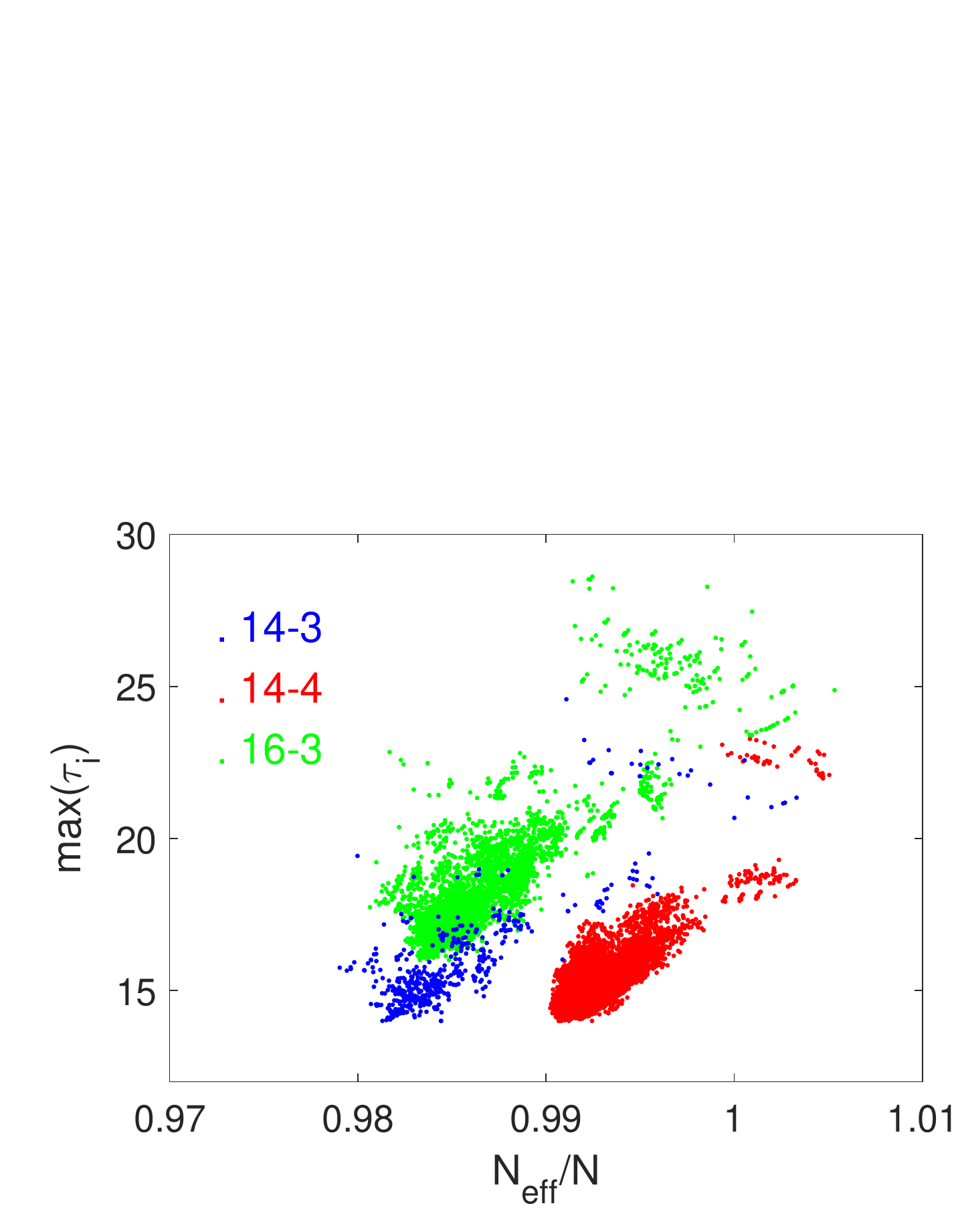} 

  (\textbf{a}) \hspace{4.7cm} (\textbf{b})

\caption{\small{Statistics of remeeting times $\tau_i$ for cubic graphs with $N=\{14,16\}$ and the quartic graphs with $N=14$. Each data point represents a graph. The variance $var(\tau_i)$ (a) and the maximum $max(\tau_i)$ (b) are calculated over the vertices of each graph and given over $N_{eff}/N$.     }}
\label{fig:tau}
\end{figure}

 We next look at how the remeeting times $\tau_i$ are distributed over graphs, see Fig. \ref{fig:tau} for  results on the cubic graphs with $N=\{14,16\}$ and the quartic graphs with $N=14$. For other $N$ the results are similar. The figures show the variance $var(\tau_i)$ and maximum $max(\tau_i)$ over $N_{eff}/N$. We see that amplifier constructors $N_{eff}/N>1$ have generally large $var(\tau_i)$ and $max(\tau_i)$, but these relation is not very strict. In other words, the remeeting times alone allow no clear conclusions about the graph's ability to become an amplifier constructor. Another result is that the remeeting times of most of these cubic and quartic graphs have non-negligible variance  $var(\tau_i)$. This means a mean-field approximation~\cite{foto19}, which assumes that the remeeting times have rather equal values, does mostly not apply.    
 
The results so far are for weak selection, that is $r=1+\delta$ with $\delta \rightarrow 0$. In other words, the graphs that have $N_{eff}>N$ are tangential amplifiers. 
In the following, we check for selected graphs with $ N_{eff}>N$ the range $1<r<r_{max}$ for which they are transient amplifiers. This is in line with the recent finding that for dB updating there are no universal amplifiers~\cite{tka20} and thus there is a finite $r<r_{max}$ for which $\varrho_{\mathcal{G}}(r)>\varrho_{\mathcal{N}}(r)$.  The study requires to calculate the fixation probabilities $\varrho_{\mathcal{G}}(r)$ and compare them to the fixation probability of the complete graph with $N$ vertices, which is~\cite{kav15,hinder15,allen20} \begin{equation}
\varrho_{\mathcal{N}}(r)=\frac{N-1}{N} \frac{1-r^{-1}}{1-r^{-(N-1)}}.  \label{eq:complete}  
\end{equation}
The fixation probabilities $\varrho_{\mathcal{G}}(r)$ are calculated by evaluating a Markov  state transition matrix~\cite{hinder15,hinder16,hinder19}. This methods is an exact computation and does not involve Monte Carlo simulations. Such  Monte Carlo simulations rely upon repeating a numerical experiment~\cite{broom09,hinder19}. A mutant with given fitness $r$ is placed randomly on a vertex of the graph. The evolutionary dynamics specified by the dB updating process leads for a sufficiently large number of iterations to either the mutant taking over the entire graph, or the mutant getting extinct. The simulation is repeated many times, which yields a relative frequency of the mutation getting fixated, which in turn is interpreted as the fixation probability. Such a numerical procedure involves a large number of repetitions until an acceptable precision is obtained, and convergence to the fixation probability is sometimes difficult to ascertain~\cite{hinder19}. 
For the exact numerical calculation method used here there are no repetitions and no convergence issues. Apart from the problem of numerical round--off, the method yields an exact value.  The limitation of the method is that it requires to solve a system of $2^N-2$ linear equations, which  becomes infeasible for $N$ getting large. For $N\leq 22$, however, the calculations were still possible. Fig. \ref{fig:graph_11_4}b shows the result for the quartic graph with size $N=11$, which is the smallest graph that is an amplifier constructor, see Tab. \ref{tab:graphs_amp}. For $r=1$  we have $\varrho_{\mathcal{G}}(r)=\varrho_{\mathcal{N}}(r)=1/N$.  

 \begin{figure}[htb]
\centering
\includegraphics[trim = 55mm 110mm 60mm 110mm,clip, width=6.5cm, height=5cm]{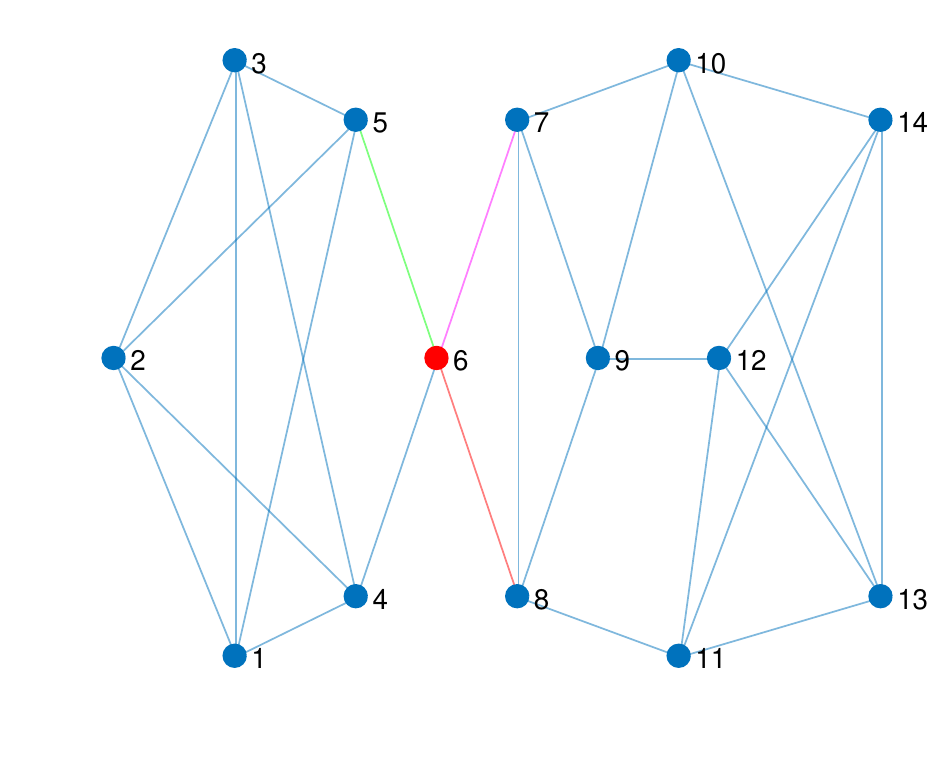} 
\includegraphics[trim = 25mm 90mm 40mm 90mm,clip, width=6.5cm, height=5.5cm]{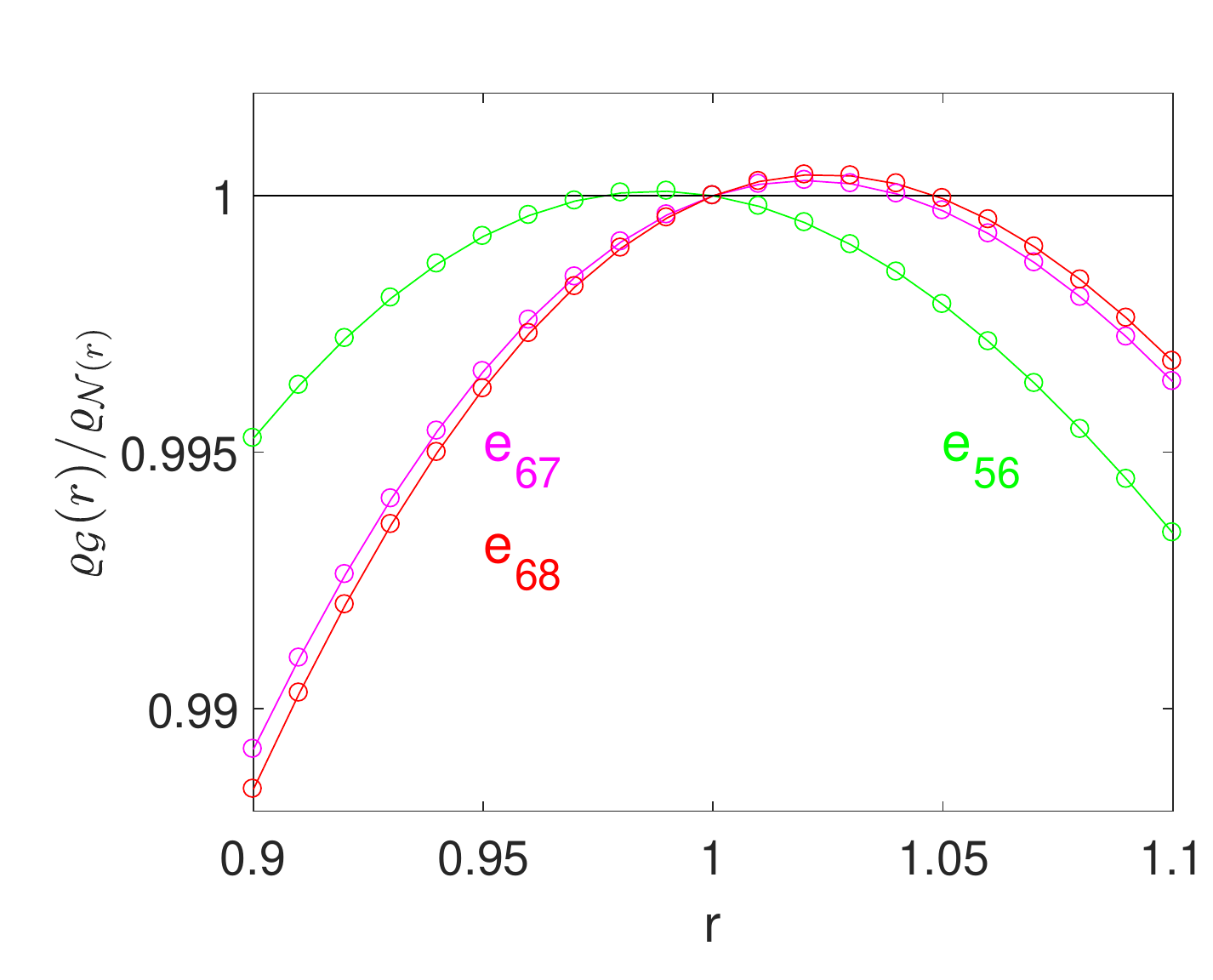} 

 (\textbf{a}) \hspace{4.7cm} (\textbf{b})  
 
\caption{\small{Constructing transient amplifiers from a quartic graph of size $N=14$. {\bf(a)} The graph with the largest remeeting time   $\tau_6=22.0892$ for vertex $v_6$, marked with red. The effective population size $N_{eff}=14.0708$ is obtained by removing the edge $e_{68}$, marked with red, and the  removal of the edge $e_{67}$, marked with magenta, also yields a $N_{eff}=14.0594>N$. Taking away the edges  $e_{56}$ or $e_{46}$, however, gives the same value $N_{eff}=13.9705<N$. {\bf (b)} The quantity $\varrho_{\mathcal{G}}(r)/\varrho_{\mathcal{N}}(r)$  over $r$. The graphs obtained by removing the edges $e_{68}$ and $e_{67}$ are transient amplifiers (characterized by values $\varrho_{\mathcal{G}}(r)/\varrho_{\mathcal{N}}(r)>1$ for $1\leq r \leq r_{max}$), taking away $e_{56}$ or $e_{46}$ yields a behavior, which can be seen as a transient suppressor. }  }
\label{fig:graph_14_4}
\end{figure}

For the interval 
$1<r<r_{max}\approx 1.00075$, the fixation probability of the graph is larger than the fixation probability of the complete graph with $N=11$. As the interval is small with respect to the mutant fitness $r$ as well as with respect to the ratio $\varrho_{\mathcal{G}}(r)/\varrho_{\mathcal{N}}(r)$, it could be suspected that the result might be a numerical artefact. Surely, it could be lengthy and arduous to obtain the results by Monte Carlo simulation as the number of repetitions needed to achieve convergence to the level of required precision would be rather high. However, the results given in Fig. \ref{fig:graph_11_4}b are obtained by exact calculation of the fixation probabilities. An important argument in favor of the validity of the results is that apart from round--of errors the method does not suffer from convergence issues.
In order to get an estimation of the round--off error, the calculation of the fixation probabilities is compared to the two cases of regular graphs for which analytical results exist. The comparisons are to the complete graph, which is $N-1$-regular, and to the cycle graph, which is $2$-regular. For the complete graph the fixation probability is given by Eq. \eqref{eq:complete}. For the cycle graph it is~\cite{kav15,hinder15,allen20} 
\begin{equation}
\varrho_{\mathcal{C}}(r)= \frac{2(r-1)}{3r-1+r^{-(N-3)}-3r^{-(N-2)}}.  \label{eq:ring}  
\end{equation}
Fig. \ref{fig:error} of the Supporting Information gives the relative error between the calculated fixation probabilities using the Markov state transition matrix calculation used in this paper and the analytical results for a complete  and a cycle  graph of size $N=12$ and $N=14$. We see that the round-off error is in the magnitude of $10^{-14}$, while the differences between the fixation probabilities are in the magnitude of  $10^{-7}$. Therefore, the results given in Fig. \ref{fig:graph_11_4}b should be regarded as valid. 

For the number of vertices $N$ increasing, a larger number of regular graphs are amplifier constructors, and also $r_{max}$ and the maximal ratio $\varrho_{\mathcal{G}}(r)/\varrho_{\mathcal{N}}(r)$ increases, see as another example the quartic graph of size $N=14$ in Fig. \ref{fig:graph_14_4}a, which is the graph with the largest $N_{eff}$ among the $\mathcal{A}_4(14)=109$ quartic graph of size $N=14$ with $N_{eff}>14$.  A main difference to the example of the graph in  Fig. \ref{fig:graph_11_4}a is a lower level of symmetry in the graph. Thus, it matters which of the edges connecting the vertex with the largest remeeting time ($v_6$) with other vertices is removed. For 2 of the 4 edges we obtain a transient amplifier, but with different $r_{max}$. For the other two edges, we obtain $\varrho_{\mathcal{G}}(r)/\varrho_{\mathcal{N}}(r)<1$ for $r>1$, but $\varrho_{\mathcal{G}}(r)/\varrho_{\mathcal{N}}(r)>1$ for some $r_{min}<r<1$. Thus, such a behavior could be seen as a transient suppressor. The Supporting Information contain further examples of transient amplifiers of cubic graphs with $N=\{18,20,22\}$ and quartic graphs with $N=\{15,16\}$, which each produce the largest $N_{eff}$, see Fig. \ref{fig:18_20_22} and Fig.    \ref{fig:15_16}. 

 \begin{figure}[htb]
\centering
\includegraphics[trim = 25mm 90mm 30mm 90mm,clip, width=6.5cm, height=5.5cm]{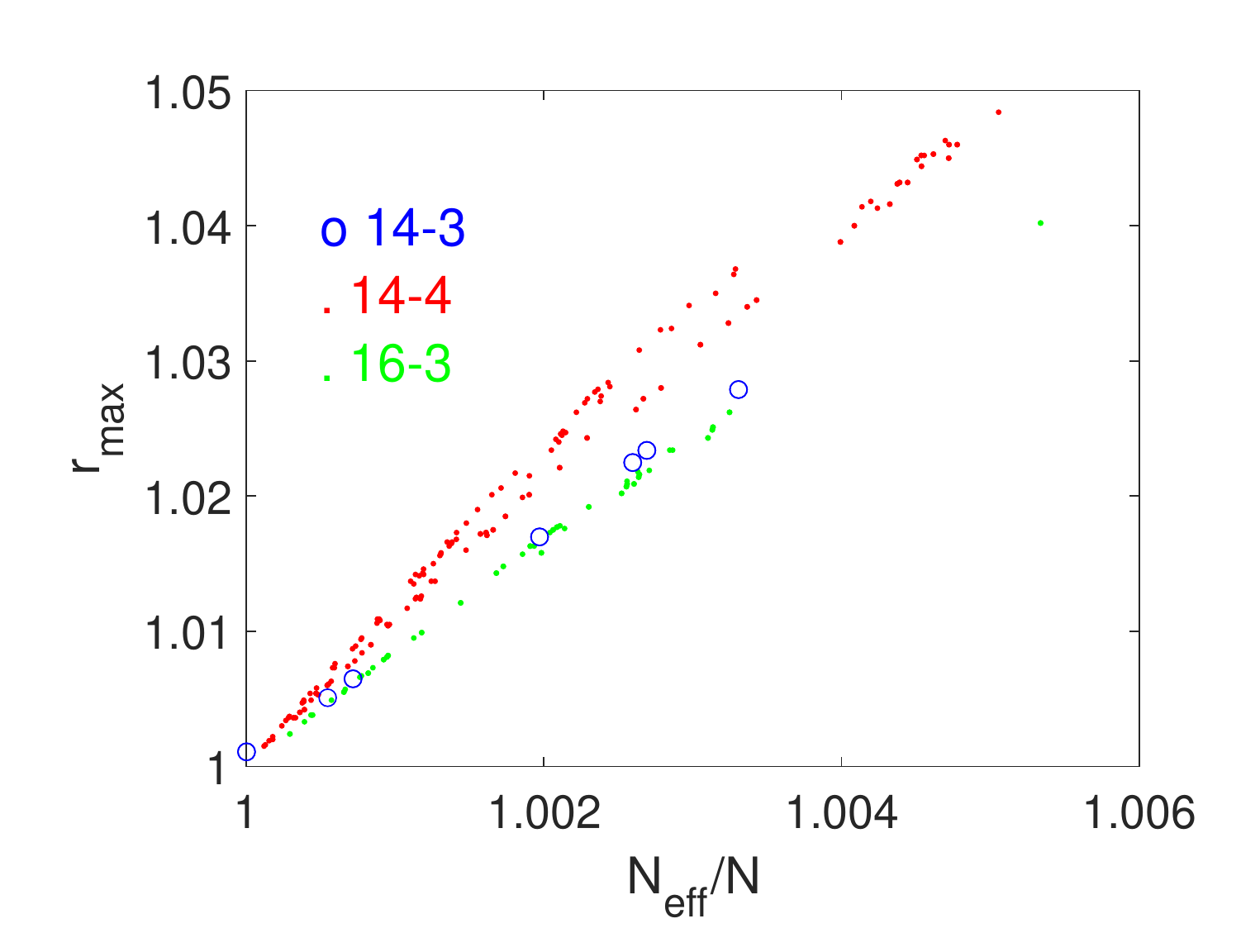}

\caption{\small{Relationship between $N_{eff}$ and $r_{max}$ for cubic graphs with $N=\{14,16\}$ and quartic graph with $N=14$ with $N_{eff}>N$. The critical mutant fitness $r_{max}$ is calculated from   $\varrho_{\mathcal{G}}(r)$, which is equidistantly distributed on $r$, by interpolation using a Lagrange polynomial.}}
\label{fig:neff_rmax}
\end{figure}

We now study the relationship between $N_{eff}$ and $r_{max}$, see Fig. \ref{fig:neff_rmax} showing results for all cubic graphs with $N=\{14,16\}$ and all quartic graph with $N=14$ with $N_{eff}>N$. Again, the results for other $N$ are similar. The values of $r_{max}$ are calculated from  $\varrho_{\mathcal{G}}(r)$ by interpolation using a Lagrange polynomial.   There is an almost linear relationship between $N_{eff}/N$ and $r_{max}$. As $N_{eff}$ is the tangential fixation probability at $r=1$, this means that the curve of $\varrho_{\mathcal{G}}(r)/\varrho_{\mathcal{N}}(r)$ is approximately a 
parabola, whose 
intersection with $\varrho_{\mathcal{G}}(r)/\varrho_{\mathcal{N}}(r)=1$ at $r=r_{max}$ is determined by $N_{eff}$. 
The results additionally suggest that instead of searching a large $r_{max}$, we may use $N_{eff}$ as a proxy. As calculating  $N_{eff}$ has polynomial time complexity, while calculating $r_{max}$ via $\varrho_{\mathcal{G}}(r)$ is exponential, this relationship may save numerical resources. However, it needs to be checked by additional work if the almost linear relationship between $N_{eff}$ and $r_{max}$ is still valid for transient amplifiers that are more strongly perturbed with respect to their regularity as the examples considered here with just a single edge removed.

 \begin{figure}[htb]
\centering
\includegraphics[trim = 53mm 110mm 45mm 90mm,clip, width=7cm, height=4.5cm]{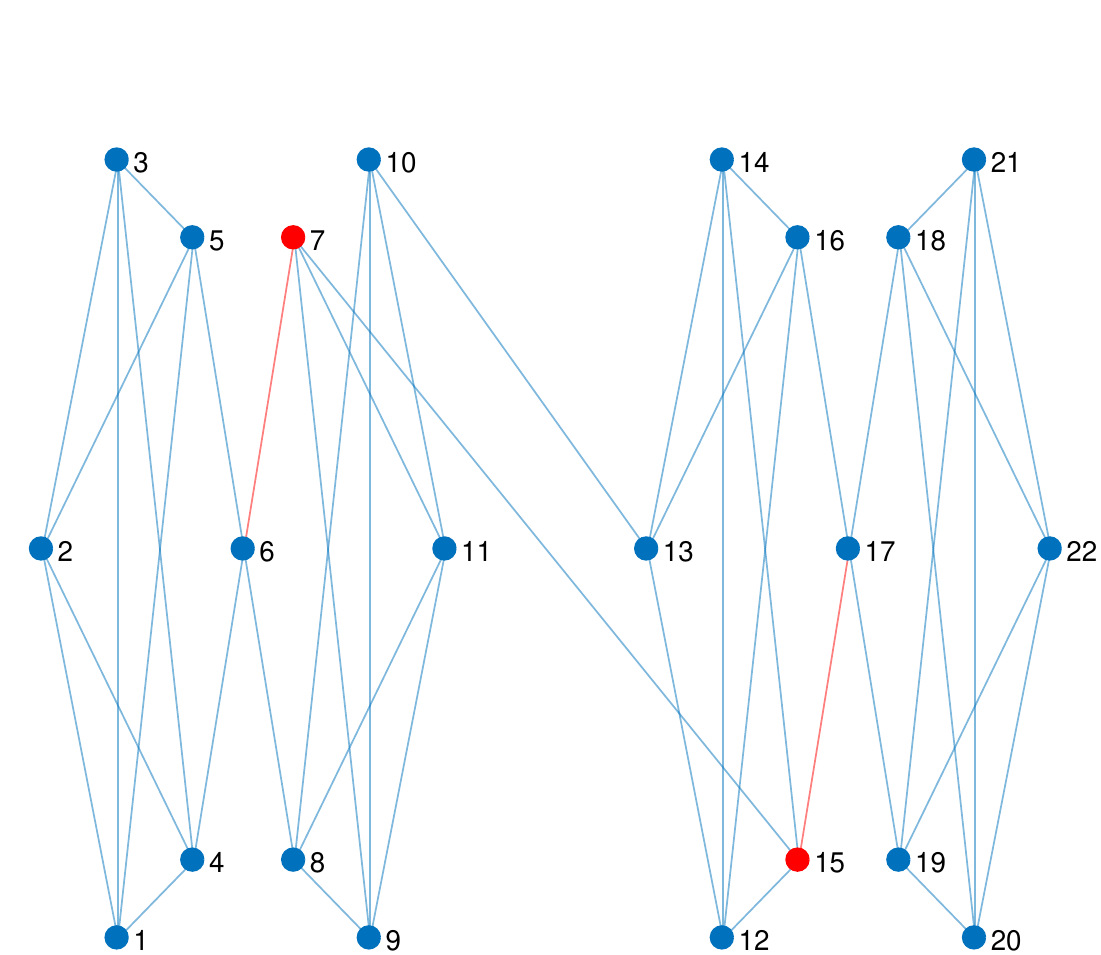}
\includegraphics[trim = 38mm 110mm 45mm 90mm,clip, width=6cm, height=4.45cm]{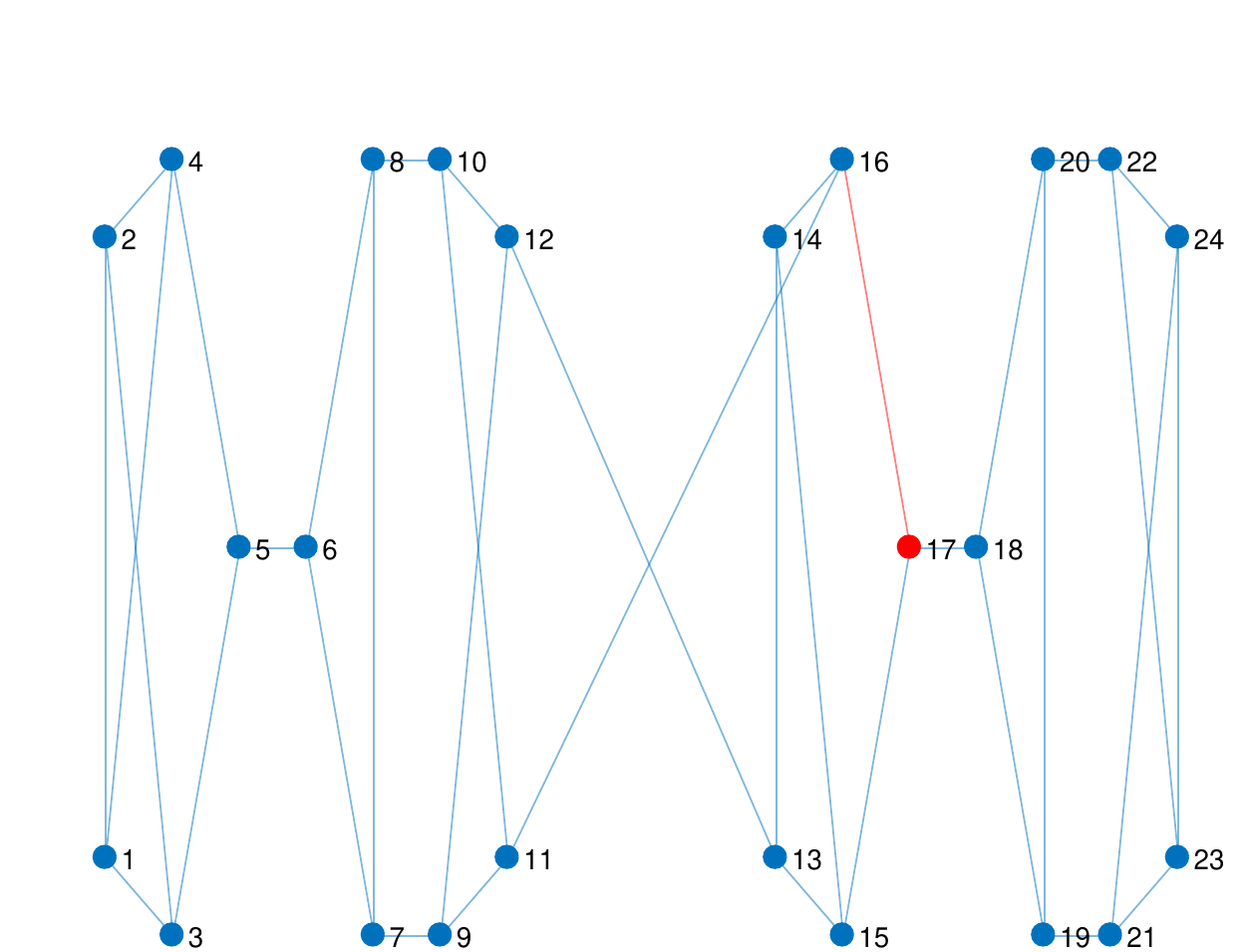}

 (\textbf{a}) \hspace{4.7cm} (\textbf{c})  
 
\includegraphics[trim = 60mm 110mm 50mm 110mm,clip, width=6.5cm, height=5cm]{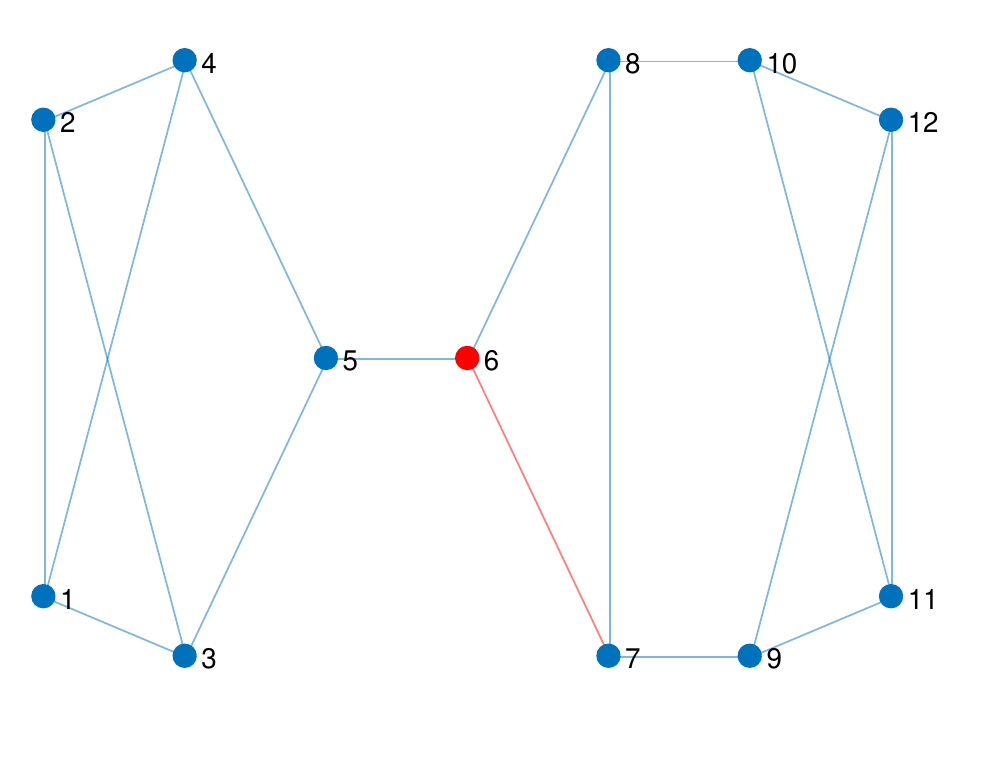} 
\includegraphics[trim = 25mm 90mm 40mm 90mm,clip, width=6.5cm, height=5.5cm]{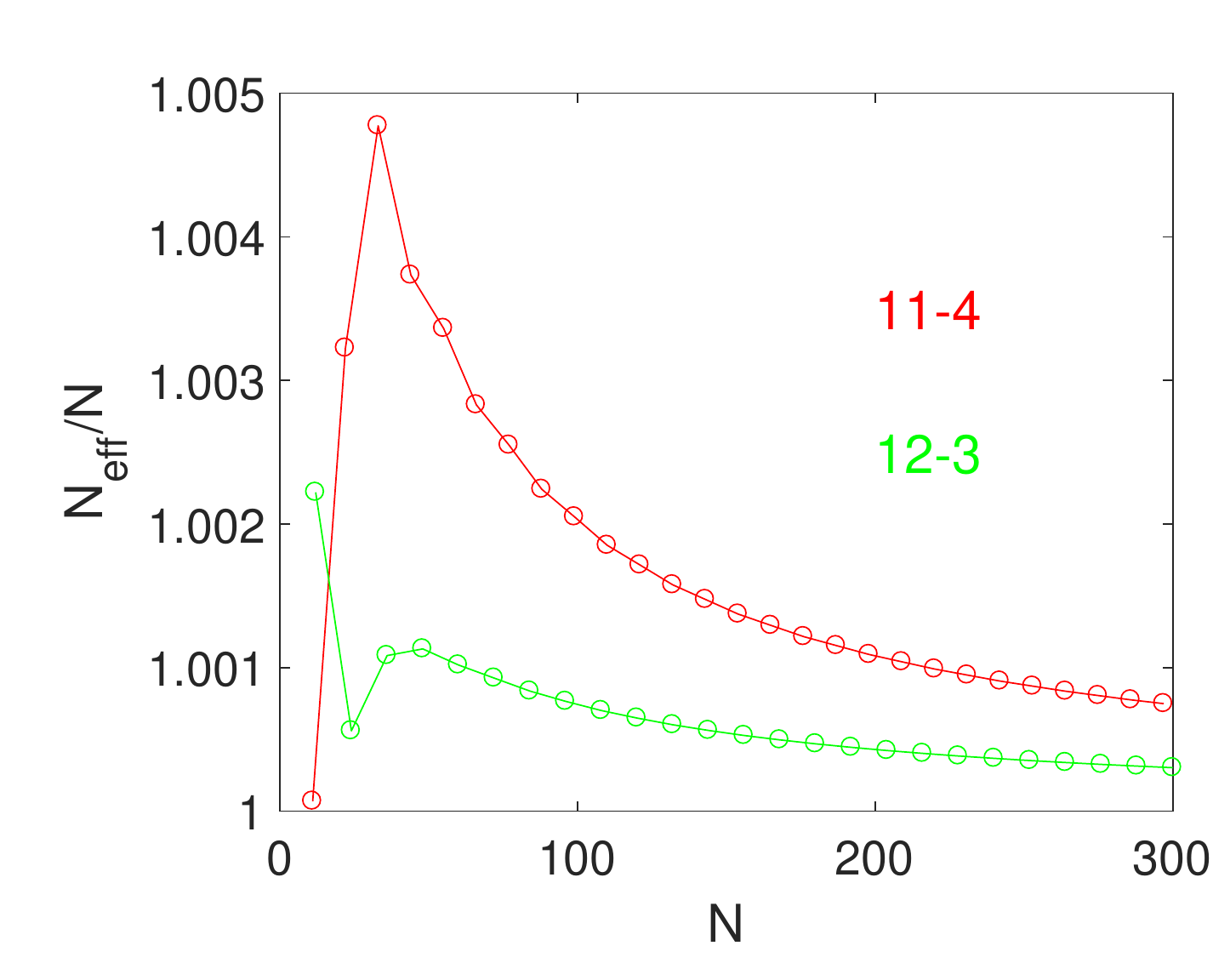} 

 (\textbf{b}) \hspace{4.7cm} (\textbf{d})

\caption{\small{Constructing transient amplifiers by building chains containing a head, a tail and possibly mid-sections. (a) 
A chain containing a head connected with a tail using the quartic graph of size $N=11$ in Fig. \ref{fig:graph_11_4}a. The resulting chain has largest remeeting times $\tau_7=\tau_{15}=33.9780$, and removing the  edge $e_{67}$ or the edge $e_{15\: 17}$ leads to a transient amplifier with  $N_{eff}=22.0709>22$. (b) The only cubic graph of size $N=12$ that is an amplifiers constructor with the largest remeeting time
$\tau_6=18.2911$ and removing the edge $e_{67}$ or  the edge $e_{68}$ produces $N_{eff}=12.0266>12$, compare\cite{allen20}.  (c)  A chain using as building blocks the cubic graph of size $N=12$ shown in (b). The largest remeeting time is $\tau_{17}=38.1158$  and removing the edge $e_{16\: 17}$  gives $N_{eff}=24.0134>24$, while removing the edge $e_{15\: 16}$  yields $N_{eff}=23.9161<24$.
(d) The quantity $N_{eff}/N$ over $N$ for chains with mid-sections. A rising number of mid-section (and thus vertices $N$) gives a series of amplifier constructors. The ratio $N_{eff}/N$ has a peak for a low number of mid-sections and converges to $N_{eff}/N \rightarrow 1$ from above for $N$ getting large.
}  }
\label{fig:chain}
\end{figure}
 
The results in Tab. \ref{tab:graphs_amp} suggest that amplifier constructors are rare compared to the total number of non-isomorphic regular graphs, but their number might not be limited. 
There is another argument for assuming that there are arbitrary many regular graphs that produce transient amplifiers by the perturbation method. Some regular amplifier constructors can be used as building blocks to obtain more amplifier constructors. The simplest example is the quartic graph on $N=11$ vertices shown in Fig \ref{fig:graph_11_4}a. From this graph another quartic graph on $N=22$ vertices can be obtained by the following procedure. We take two copies of graph and call them a head and a tail, respectively. From the head we remove the edge $e_{7 \: 10}$ and from the tail, we remove the edge $e_{24}$. We build a single graph from the head and the tail by  retaining the indices of the vertices $(v_1,v_2,\ldots,v_{11})$ from the head and renaming the vertices of the tail by $(v_{12},v_{13},\ldots,v_{22}):=(v_1,v_2,\ldots,v_{11})$. All edges apart from those removed remain unchanged. We finally connect head and tail by additional edges $e_{7\:15}$ and $e_{10\:13}$, see Fig. \ref{fig:chain}a. From the graph a transient amplifier can be constructed by removing the edge  $e_{67}$ (or $e_{15\: 17}$) to obtain  $N_{eff}=22.0709>22$. 

Another example is the cubic graph in Fig. \ref{fig:chain}b, which is the only cubic graph of size $12$ that is an amplifier constructor, compare also to Fig. 2a in Allen et al.~\cite{allen20}. Here we remove the edge $e_{11\:12}$ from the head and $e_{14}$ from the tail, and connect head and tail by the edges $e_{12 \: 13}$ and $e_{11\:16}$, see Fig. \ref{fig:chain}c. The use of building blocks can be extended by placing mid-sections between the head and the tail. From these mid-sections we need to remove the edges as for the head and the tail, for instance  $e_{7 \: 10}$ and $e_{24}$ for the quartic graph on $N=11$ vertices in Fig \ref{fig:graph_11_4}a. We then connect the mid-section to the head on the one side and to the tail on the other. Fig.  \ref{fig:graph_11_4}d shows the ratio $N_{eff}/N$ over $N$ for such chains using the building blocks of the quartic graph on $N=11$ vertices, Fig \ref{fig:graph_11_4}a, and the  cubic graph on $N=12$ vertices, Fig. \ref{fig:chain}b, for $N=\{11,22,\ldots,297 \}$ and $N=\{12,24,\ldots,300 \}$. For the number of midsections (and thus the number of vertices) increasing we continue to obtain graphs that are transient amplifiers. However, for $N$ getting larger the ratio $N_{eff}/N$ slowly converges to one from above.

 \begin{figure}[htb]
\centering

 \includegraphics[trim = 25mm 90mm 40mm 100mm,clip, width=6.5cm, height=5cm]{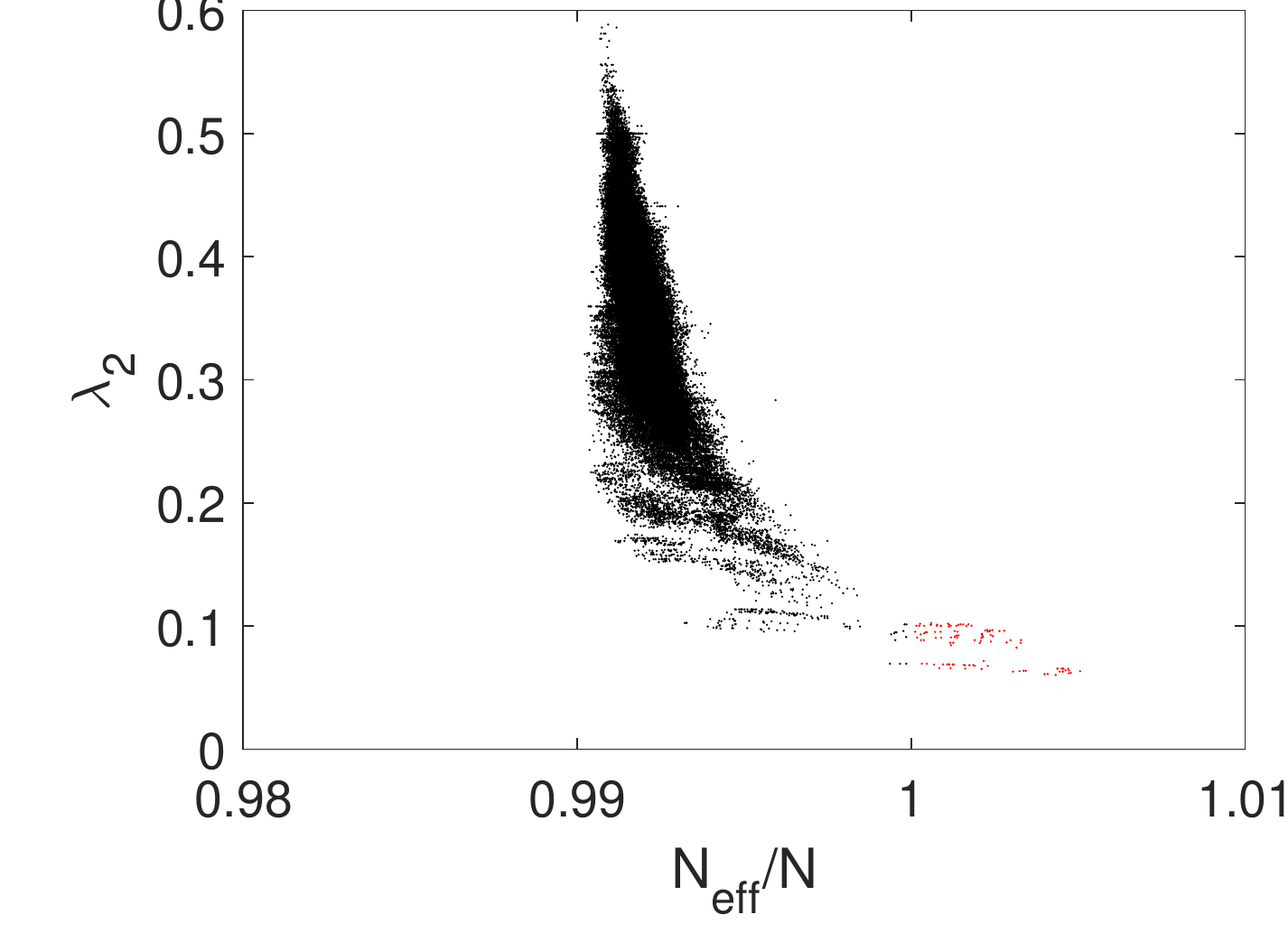} 
\includegraphics[trim = 25mm 90mm 40mm 100mm,clip, width=6.5cm, height=5cm]{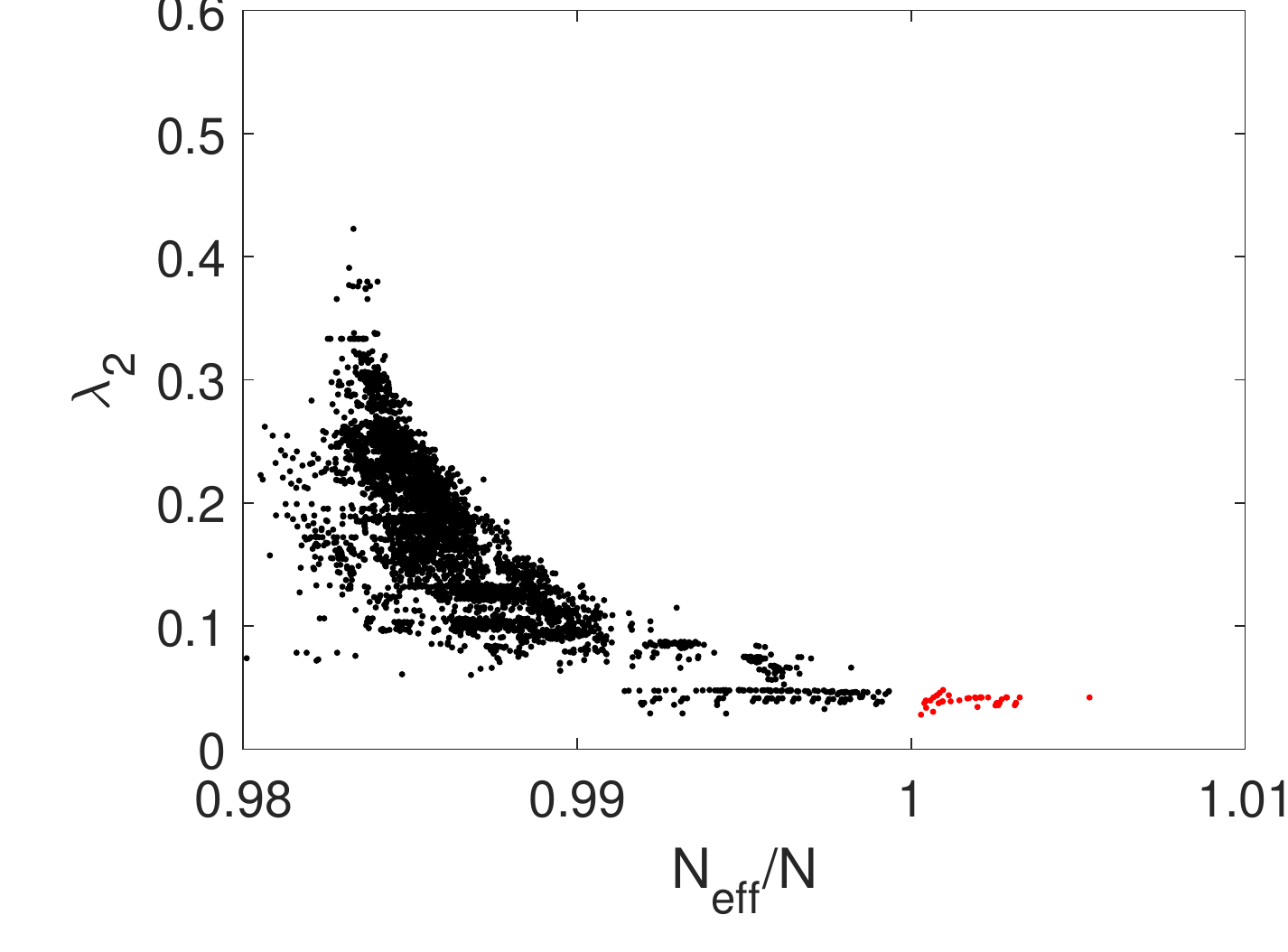} 

(\textbf{a}) 14 4  \hspace{4.7cm} (\textbf{b})  16 3

\caption{\small{The spectral gap $\lambda_2$ versus the ratio $N_{eff}/N$  as a scatter plot for: (a) the quartic graphs of size $N=14$ and (b) the cubic graphs of size $N=16$. The $N_{eff}$ is the maximal value obtained by the perturbation method removing an edge from the vertex $v_i$ with the largest remeeting time $\tau_i$. Transient amplifiers are marked by red dots. See also Fig. \ref{fig:lambda_2} of the Supporting Information for results of the other $N$. }}
\label{fig:lambda_2_1}
\end{figure}

\clearpage
 \begin{figure}[h]
\centering

 \includegraphics[trim = 25mm 90mm 40mm 95mm,clip, width=6.5cm, height=5cm]{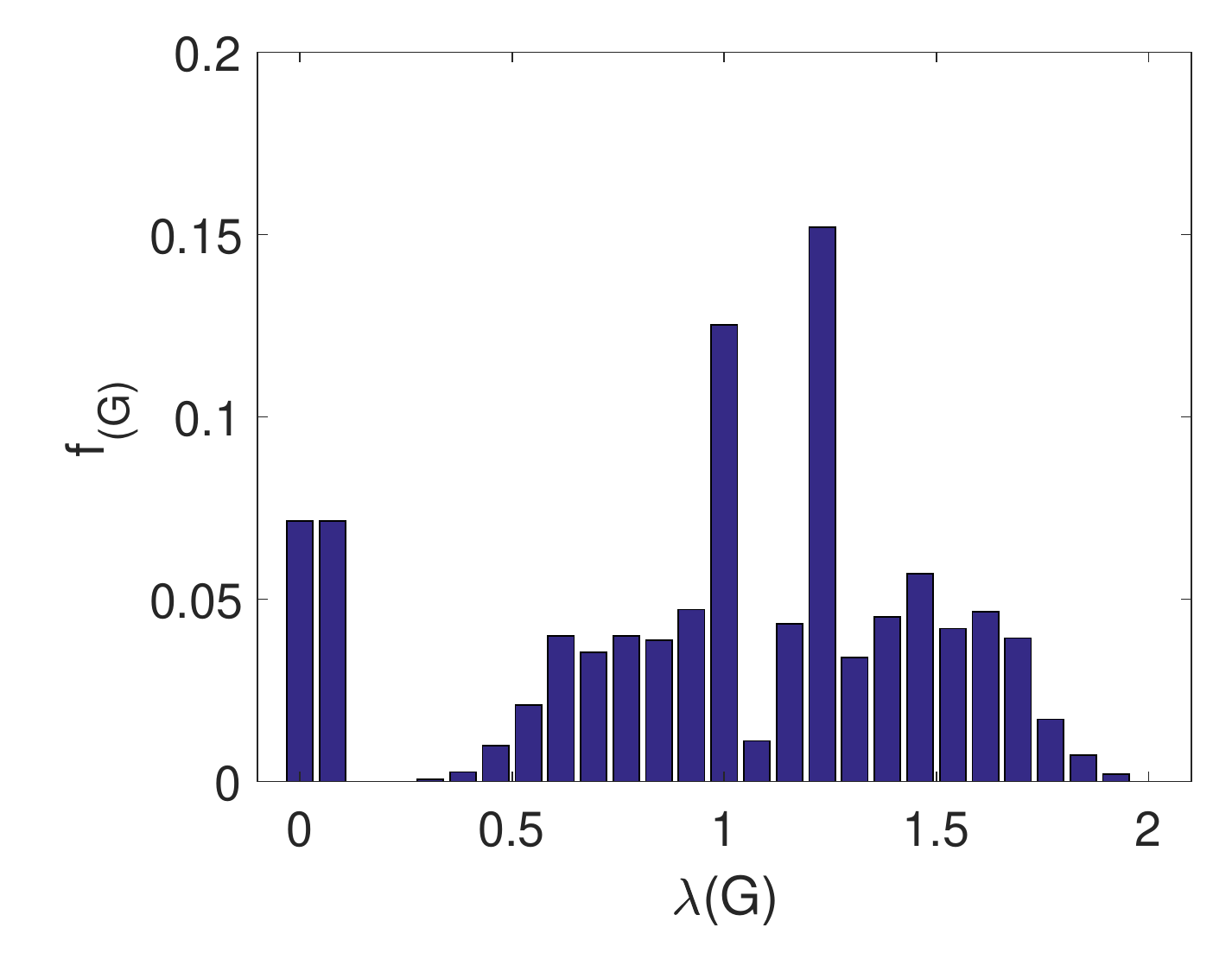} 
\includegraphics[trim = 25mm 90mm 40mm 100mm,clip, width=6.5cm, height=5cm]{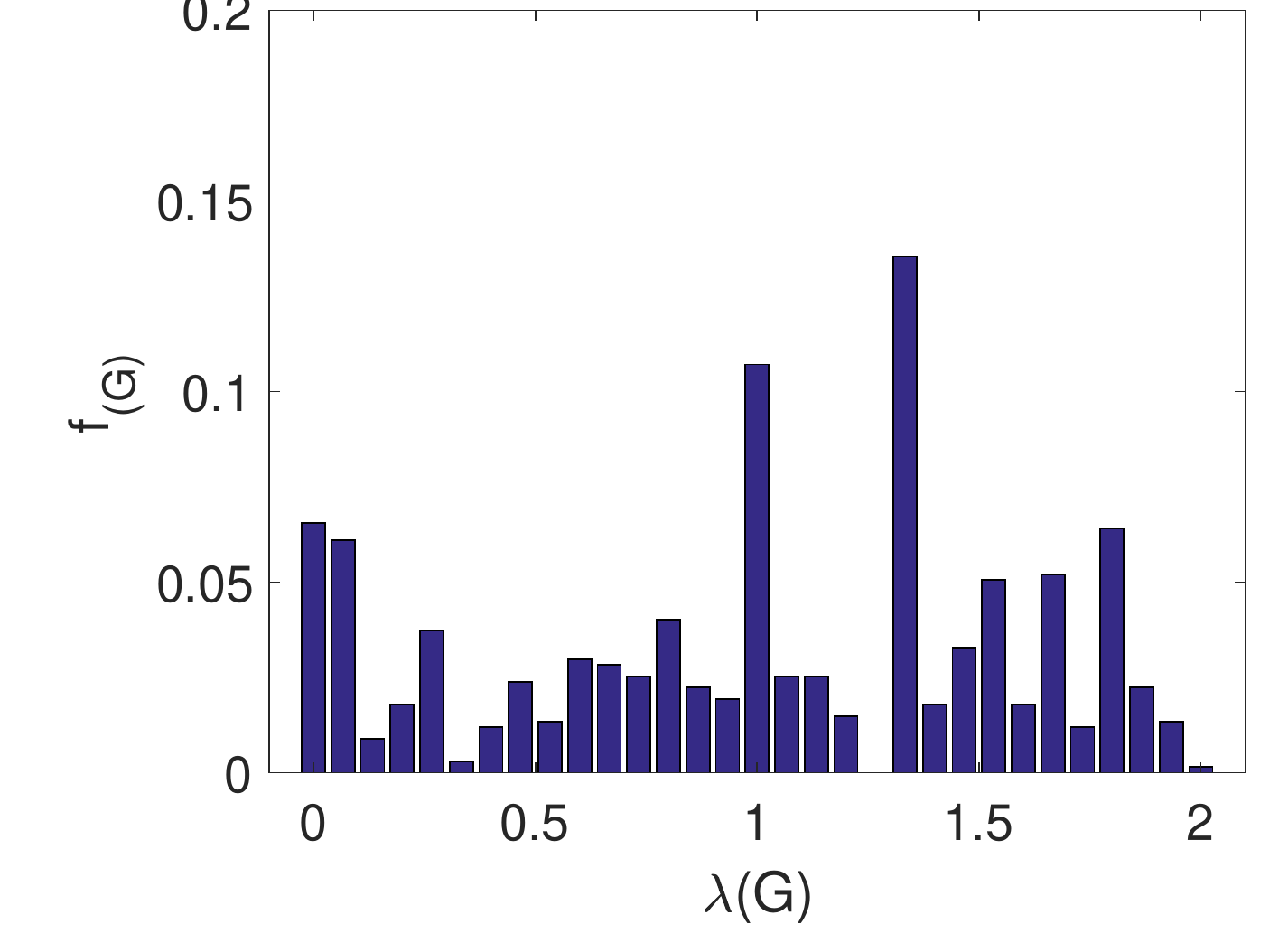} 

(\textbf{a}) 14 4 \hspace{4.7cm} (\textbf{d}) 16 3  \hspace{2.7cm} 
 
 \includegraphics[trim = 25mm 90mm 40mm 95mm,clip, width=6.5cm, height=5cm]{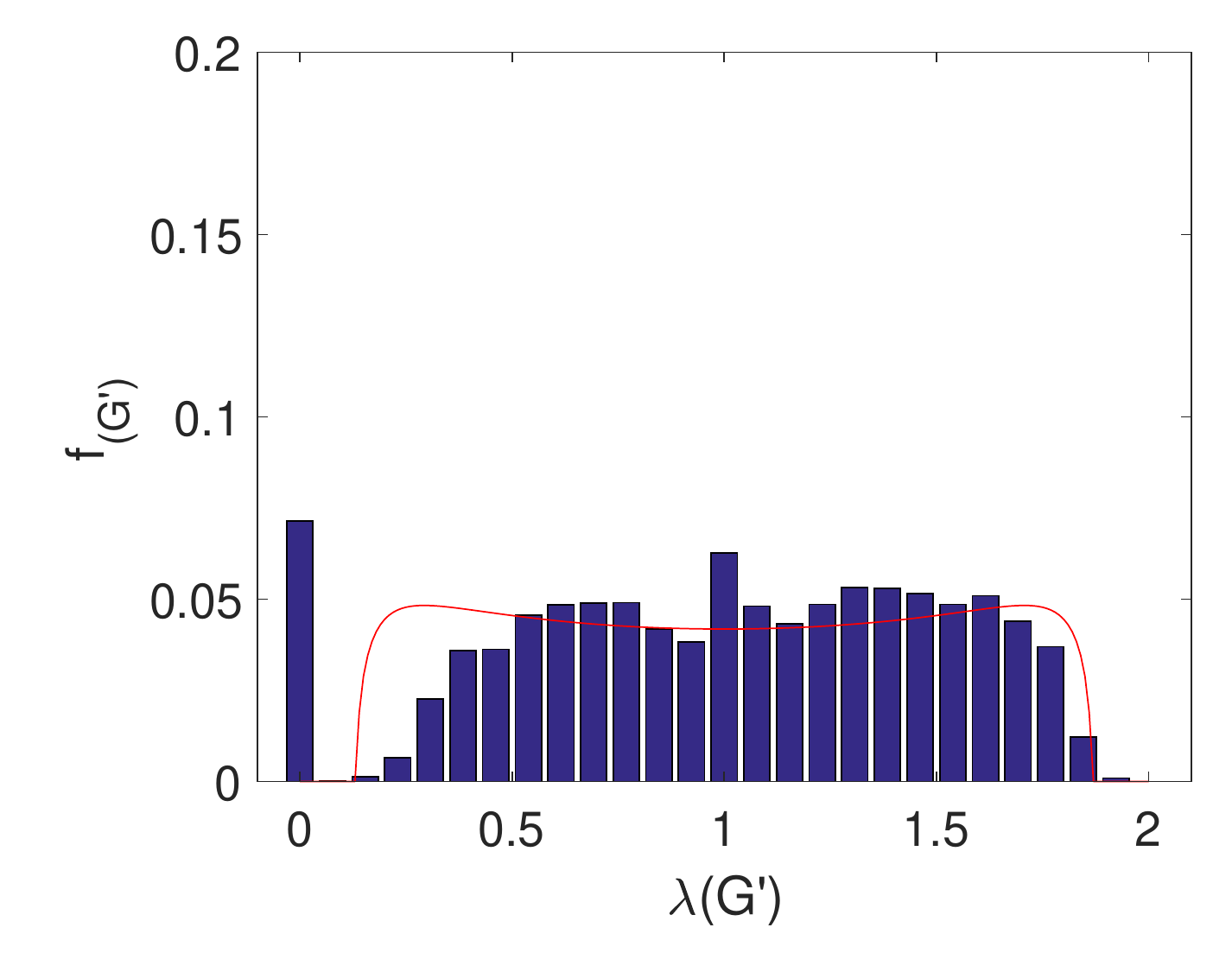} 
\includegraphics[trim = 25mm 90mm 40mm 95mm,clip, width=6.5cm, height=5cm]{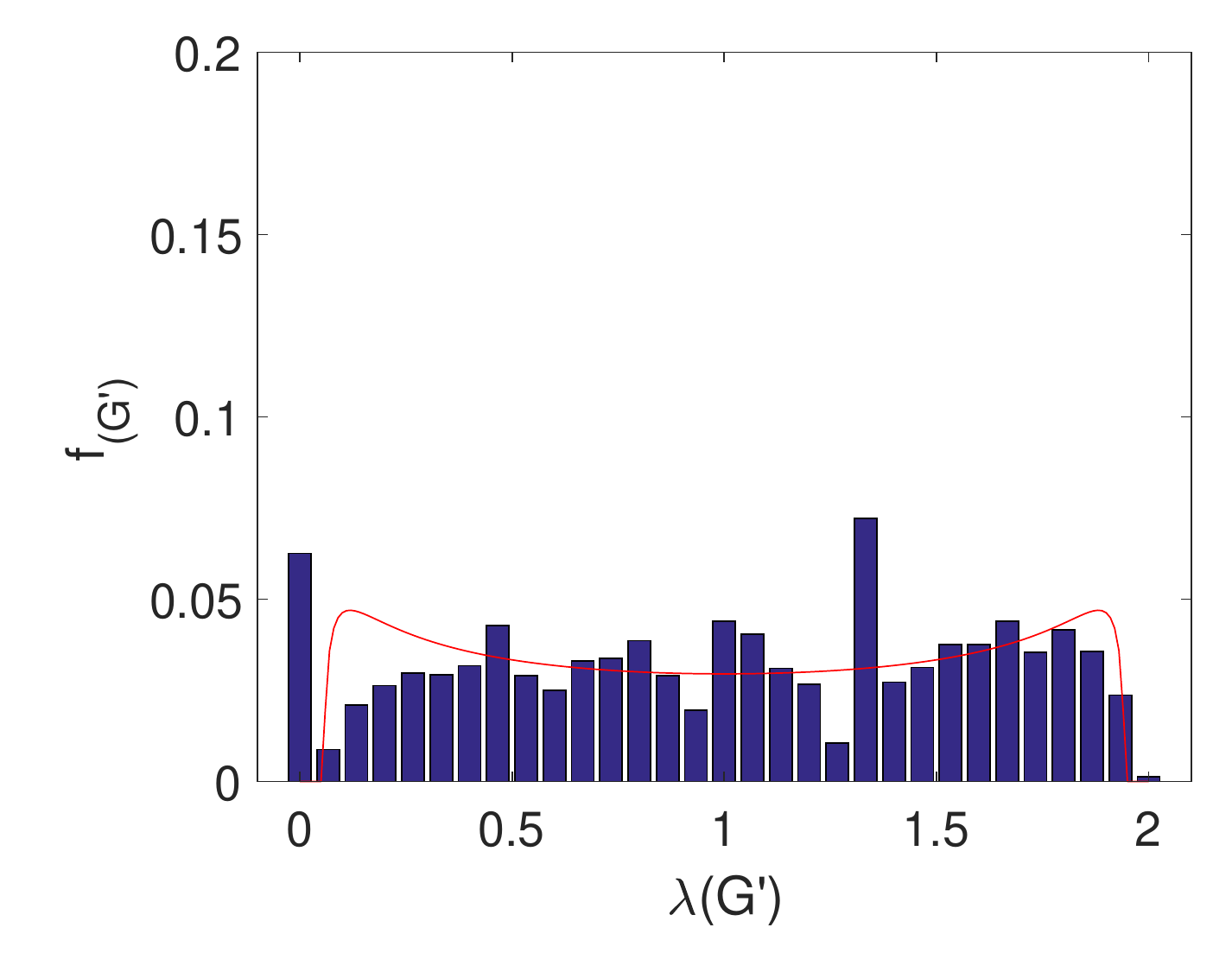} 

(\textbf{b}) 14 4  \hspace{4.7cm} (\textbf{e}) 16 3 \hspace{2.7cm} 

 \includegraphics[trim = 25mm 90mm 40mm 95mm,clip, width=6.5cm, height=5cm]{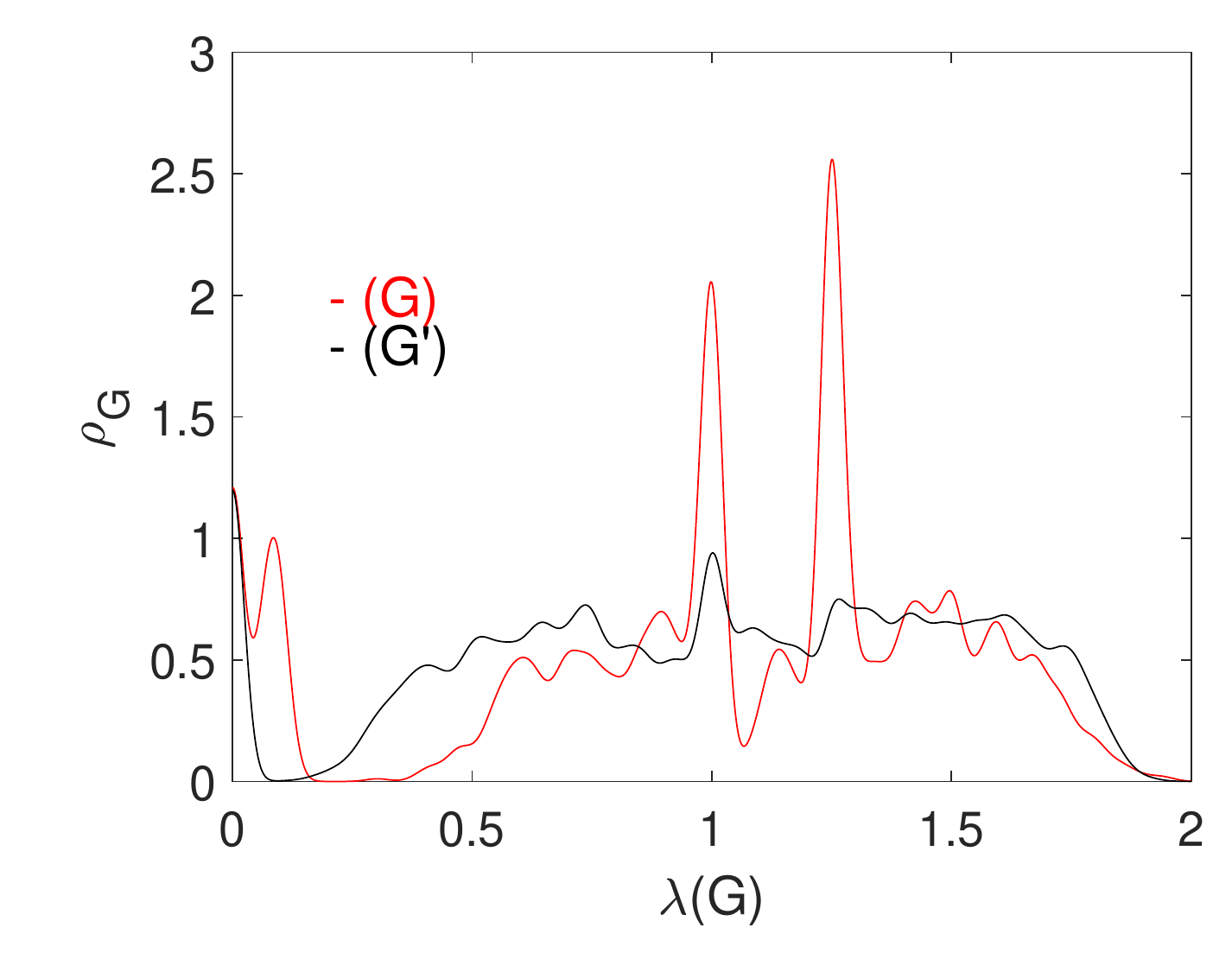} 
\includegraphics[trim = 25mm 90mm 40mm 95mm,clip, width=6.5cm, height=5cm]{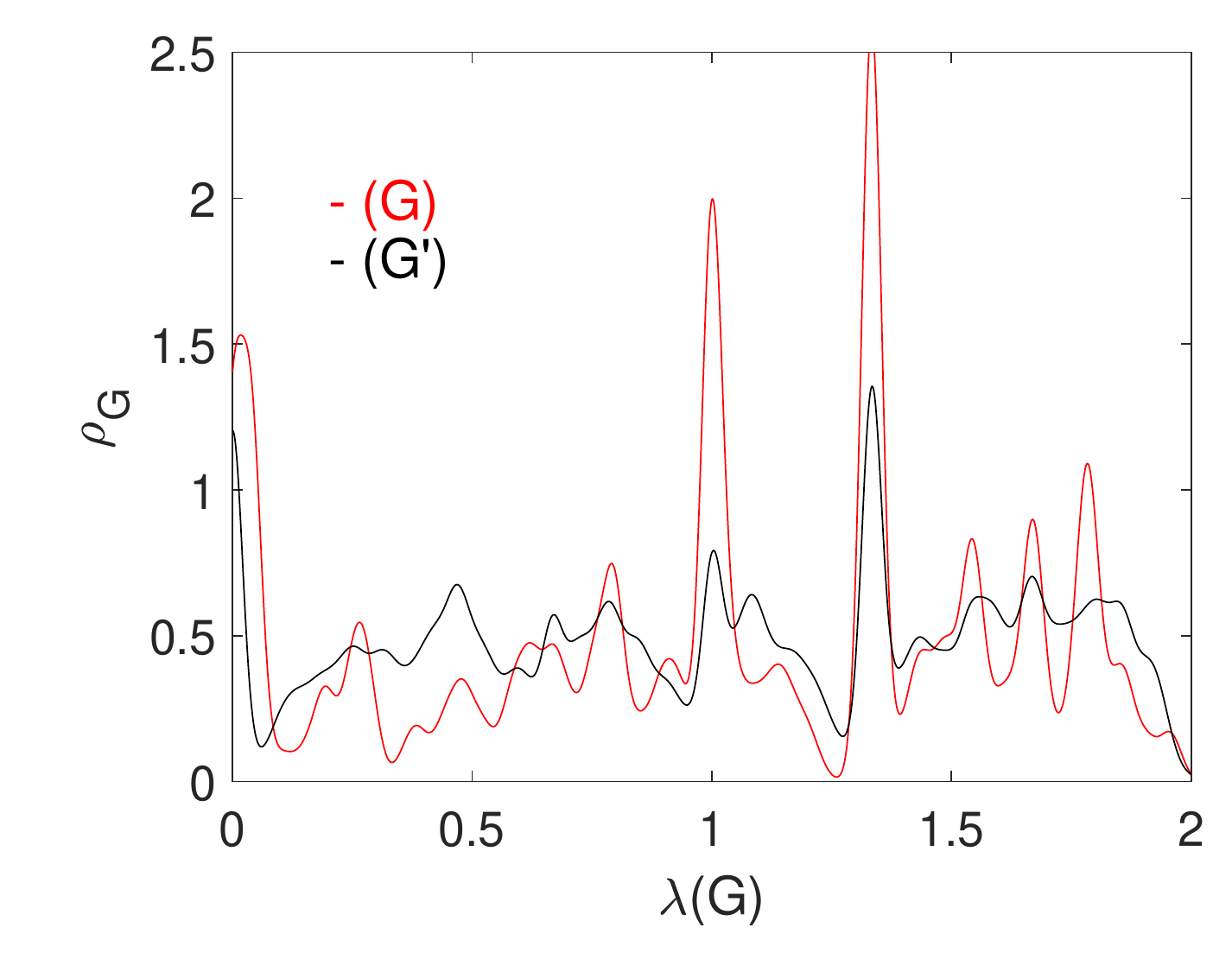} 

(\textbf{c}) 14 4 \hspace{4.7cm} (\textbf{f}) 16 3 \hspace{2.7cm}

\caption{\small{
The discrete eigenvalue distributions $f_{(G)}$  of the amplifiers constructors ($N_{eff}/N>1$) and  $f_{(G')}$ for the remaining graphs with ($N_{eff}/N<1$) and the smoothed spectral densities $\rho_{\mathcal{(G)}}$ and $\rho_{\mathcal{(G')}}$ according to Eq. \eqref{eq:density} for all quartic regular graphs with $N=14$, (a)-(c), and all cubic regular graphs with $N=16$, (d)-(f).
The red lines in (b) and (e) depict the shape of the Kersten-McKay distribution \eqref{eq:km}, which is rescaled in order to cover the same area as the bars of the histogram. The spectral density of the amplifier constructors is shown as red line in (c) and (f), while the black line is for the remaining graphs.
 See also Fig. \ref{fig:dense1} of the Supporting Information for results of the smoothed spectral densities $\rho_{\mathcal{(G)}}$ and $\rho_{\mathcal{(G')}}$  for the other $N$.}}
\label{fig:dense}
\end{figure}

\subsection*{Spectral analysis of amplifiers constructors}

In the previous section it has been argued that regular graphs  are a suitable input for a perturbation method to construct transient amplifiers of a dB updating process. In particular, it was shown that a small but significant subset of 
all pairwise non-isomorphic cubic and quartic  regular graphs up to a certain size ($N=22$ for cubic and $N=16$ for quartic) yields amplifier constructors. The results even suggest that the number of amplifier constructors is not limited on the graph size $N$ tested by the experimental settings of this paper. As the  population structure is expressed by the graph structure this naturally poses this question: Is there something in the structure of these graphs that makes them prone to construct  amplifiers? In the following we approach this question by methods of spectral graph theory~\cite{gu16,wilson08,wills20}, adding to the applications of spectral analysis of evolutionary graphs~\cite{rich17,rich19a,rich19b,rich20,allen19}.

The spectral analysis presented here is based on the $N$ eigenvalues $\lambda(\mathcal{G})$ of the normalized Laplacian $L_{\mathcal{G}}$, which gives us the spectrum  $0=\lambda_1\leq \lambda_2 \leq \ldots \lambda_N\leq 2$. The principal quantity for assessing structural properties of the graph is the spectral gap $\lambda_2$~\cite{hoff19,wilson08,wills20}. Fig. \ref{fig:lambda_2_1} gives the spectral gap $\lambda_2$ over $N_{eff}/N$ for the quartic graphs of size $N=14$ and the cubic graphs with $N=16$ as a scatter plot, for results on the remaining graphs, see Supporting Information, Fig. \ref{fig:lambda_2}. 
The value of $N_{eff}$ is the maximal value obtained by perturbing the regular graph by removing a single edge from the vertex $v_i$ with the largest remeeting time $\tau_i$.  
The amplifier constructors producing $N_{eff}/N>1$
are given red dots, while the remaining graphs are indicated by black dots. The main characteristics is that all amplifier constructors have small values of $\lambda_2$. Small values of $\lambda_2$ imply large mixing times, bottlenecks, clusters and low conductance~\cite{ban08,ban09,hoff19,wills20}. Moreover, a low spectral gap indicates path-like graphs which are rather easy to divide into disjointed subgraphs by removing edges or vertices. In other words, amplifiers constructors most likely possess cut and/or hinge vertices~\cite{chang97,ho96}.    There are, however, also a multitude of graphs with low $\lambda_2$ that do not produce transient amplifiers, thus a low spectral gap is necessary but not sufficient for the regular graphs under study to be amplifier constructors. It can be conjectured that this remains true for amplifiers constructors with $N>22$. Furthermore, it can be noted that the spectral gap versus $N_{eff}/N$ roughly appears as a half of a  parabola with the lowest values of $\lambda_2$ for   $N_{eff}/N>1$ and that the majority of values are concentrated for medium values of $\lambda_2$.

 \begin{figure}[htb]
\centering
\includegraphics[trim = 25mm 90mm 30mm 90mm,clip, width=6.5cm, height=5.5cm]{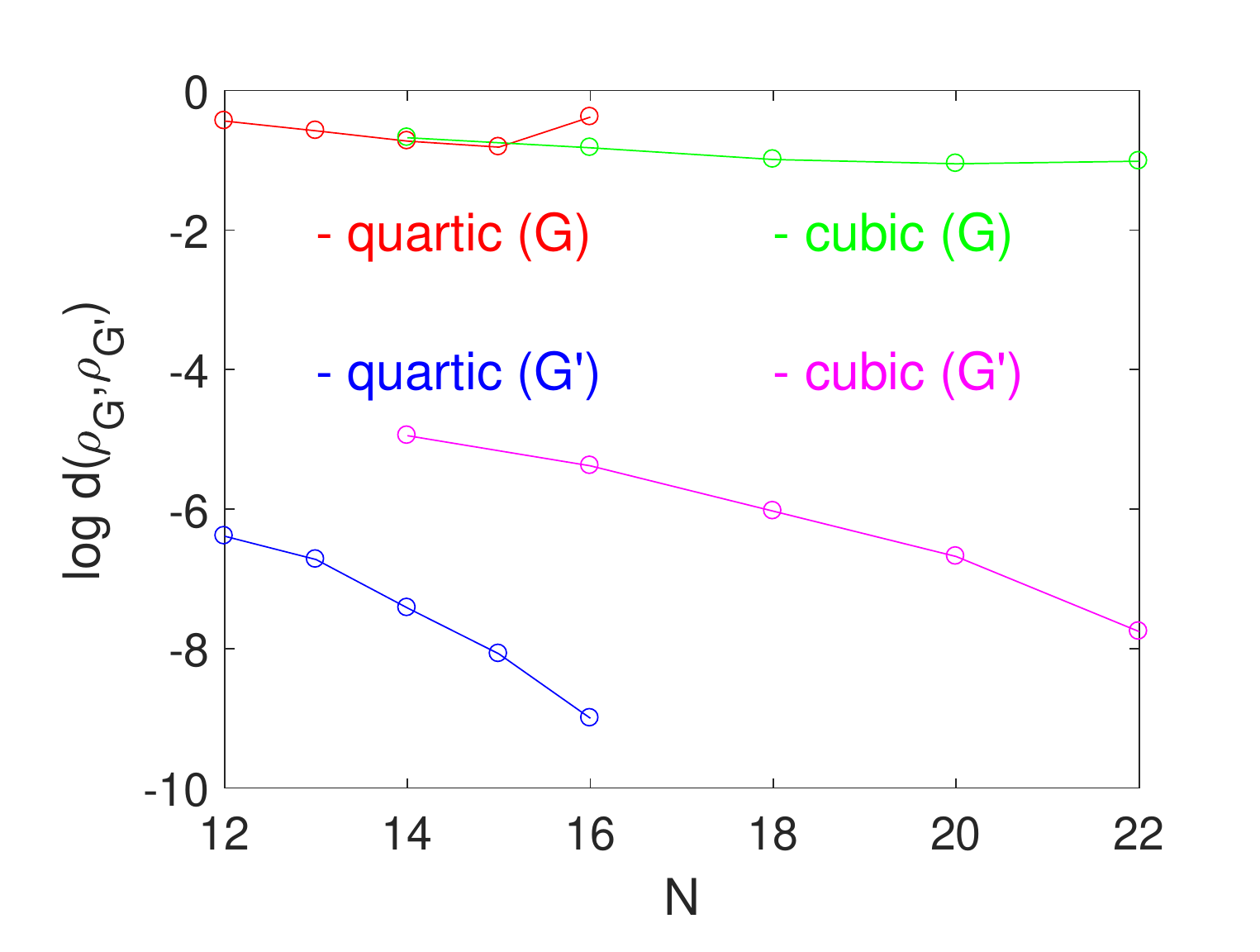}

\caption{\small{The spectral distance measure $d(\mathcal{G},\mathcal{G}')$ defined by Eq. \eqref{eq:distance} on logarithmic scale comparing regular graphs that construct amplifiers (cubic graphs as red line, quartic graphs as green line) with graphs that do not cubic graphs as blue line, quartic graphs as magenta line).}}
\label{fig:distance}
\end{figure}

As it seems unlikely that a single spectral measure (or any other scalar graph measure such as average path length or clustering coefficient) can uniquely identify an amplifier constructor, we next look at the whole spectrum.   Fig. \ref{fig:dense} shows the spectral distributions for the quartic graphs of size $N=14$ and the cubic graphs with $N=16$. The upper four panels give the discrete eigenvalue distributions $f_{(G)}$  of the amplifiers constructors ($N_{eff}/N>1$) and  $f_{(G')}$ for the remaining graphs with $N_{eff}/N<1$, while the lower panels show the smoothed spectral densities $\rho_{\mathcal{(G)}}$ and $\rho_{\mathcal{(G')}}$ according to Eq. \eqref{eq:density}. The smoothed spectral density $\rho_{\mathcal{(G)}}$ for the remaining graphs is given in the Supporting Information, Fig. \ref{fig:dense1}.  In all these figures the spectral density of the amplifier constructors $\rho_{\mathcal{(G)}}$ is shown as red line, while the black line is for the remaining graphs ($\rho_{\mathcal{(G')}}$).

We can now compare the results for amplifiers constructors with other regular graphs. 
Most noticeable it is that the spectra of amplifier constructors (upper panels of Fig. \ref{fig:dense})  differ substantially from the remaining  graphs (middle panels of Fig. \ref{fig:dense}) and also from random regular graphs in general~\cite{far01,oren09,bauer19}. 
It is known that for the limit case of $N \rightarrow \infty$, the spectral density of random regular graphs can be described by the Kesten-McKay distribution~\cite{mckay81,oren09,bauer19}, which for the normalized Laplacian and regular graphs is     \begin{equation}\rho_{KM}(x)= \frac{\sqrt{4(k-1)-k^2(1-x)^2}}{2 \pi k (2-x)x} \label{eq:km}\end{equation} with $1-2\sqrt{k-1}/k \leq x \leq 1+2\sqrt{k-1}/k$.  The distribution  $f_{(G')}$ and the  spectral density $\rho_{\mathcal{(G')}}$  obtained for the regular graphs that are no amplifier contractors have at least some similarity to the Kesten-McKay distribution, see the shape of the Kesten-McKay distribution given in the Fig. \ref{fig:dense}(c) and (d).  The distribution  $f_{(G)}$ and the  spectral density $\rho_{\mathcal{(G)}}$  obtained for amplifiers constructors do not. According to the Kesten-McKay law \eqref{eq:km} the eigenvalue density is evenly and symmetrically distributed on the interval $[0,2]$. For low values of $k$ we find two maxima close to the upper and lower limits of the distribution.  For amplifier constructors the spectra do not conform to such a characteristics. We find two characteristic peaks, one at $\lambda_i=1$ and another at $1<\lambda_i<1.5$. This can be found for all cubic and quartic regular graphs tested, see also  Fig. \ref{fig:dense1}.  These peaks indicate a substantial multiplicity of the eigenvalues. It has been shown that multiplicity of the eigenvalues of the normalized Laplacian is connected to motif doubling and motif attachment, which can be seen as the process of building a graph from joining or repeating identical substructures~\cite{ban08,ban09,meha15}. In other words, amplifier constructors are graphs which with a high probability contain identical (or at least almost identical) subgraphs and thus exhibit a substantial degree of graph symmetry. The results for the graphs producing highest $N_{eff}$, see Fig. \ref{fig:18_20_22} and Fig.    \ref{fig:15_16}, certainly examplify this result.

As the spectral density of amplifier constructors differs from other regular graphs, we finally look at the spectral distance measure  $d(\mathcal{G},\mathcal{G}')$, see Eq. \eqref{eq:distance}, to capture and quantify the differences in the graph structure, see Fig. \ref{fig:distance}. Here the distance measure is given for all cubic graphs with $N=\{14,16,18,20,22\}$ and all quartic graphs with $N=\{12,13,14,15,16\}$. The distances are calculated by comparing the amplifier constructors $(G)$ of a given size $N$ and degree $k$ to all graph with the same size and degree. The same is done for the graphs $(G')$ which are no amplifier constructors. We see that amplifier constructors have  much higher spectral distances, which remains largely constant for $N$ getting larger. By contrast, graphs that are no amplifier constructors have a low spectral distance,  are very similar to the set of all regular graphs with given $N$ and $k$, and the distance gets small for $N$ increasing.  This can be interpreted as the increasing number of pair-wise non-isomorphic graphs for $N$ getting larger (see Tab. \ref{tab:graphs} for the values of $\mathcal{L}_k(N)$) means that the fraction of graphs that are structurally similar increases as well. It is also interesting to note that the spectral distances for quartic graphs are lower than for cubic graphs. A possible interpretation is that cubic regular graphs show a larger structural variety than quartic regular graphs.
 To conclude amplifiers constructors can be clearly distinguished from regular graphs that do not yield transient amplifiers by their spectral density profile. The distance measure $d(\mathcal{G},\mathcal{G}')$ can also be applied to measure the spectral distance between an amplifier constructor and the transient amplifier obtained  by the perturbation methods. Additional experiments (results not given in figures due to brevity) have shown that due to the smallness of the perturbation just removing one edge, the results give no further essential information about structural properties.

\section*{Discussion}

\subsection*{Constructing transient amplifiers}

Transient amplifiers of selection are structured networks that increase the fixation probability of beneficial mutations as compared to a well-mixed population. Thus, transient amplifiers provide a mechanism for shifting the balance between natural selection and genetic drift, and therefore have considerable significance in evolutionary dynamics~\cite{allen20,hinder15,pav18,tka20}. Until recently, it was assumed that transient amplifiers are rare or nonexistent for death-Birth (dB) updating. In a recent work~\cite{allen20} first examples of transient amplifiers for dB updating were presented together with a procedure executable with polynomial time complexity
to decide whether or not a given graph is an amplifier of weak selection. This procedure includes finding the vertex with the largest remeeting time and perturbing the graph by removing an edge from this vertex. There is a certain likelihood that the resulting perturbed graph is a transient amplifier.   
This paper extends the approach by using the perturbation method and checking all pairwise nonisomorphic cubic and quartic regular graph up to a certain size ($N\leq 22$ for cubic and $N\leq 16$ for quartic graphs).  The results show that a small but significant subset of 
all these regular graphs produce perturbed graphs that are transient amplifiers. The regular graphs that possess this property are called amplifier constructors. 

Thus, a first major finding of this study is that transient amplifiers for dB updating most likely are not as rare as previously assumed. Although the percentage of amplifier constructors is low (for instance $0.19\%$ for cubic graphs of size $N=22$ or  $0.04\%$ for cubic graphs of size $N=16$), the exponential growth with size $N$ of the number of pairwise nonisomorphic graphs ensures their number is not negligibly. The given percentages mean $\mathcal{A}_3(22)=13.889$ and $\mathcal{A}_4(16)=3.129$ amplifier constructors for the cubic and quartic graphs of size $N=22$ and $N=16$, respectably. As for the considered sizes $N$ the exact number of all connected pairwise nonisomorphic cubic and quartic graphs is known, and all instances have been tested, the study encompasses the whole structural range of these graphs. In other words,   if different graph structures constitute the space of different population structures, all possible structural variants expressible by regular graphs have been covered by this study.

The perturbation methods discussed and analyzed here only removes a single edge. Thus, the difference between the transient amplifiers produced by the perturbation method and a regular graph is small. 
This may also explain why the effective population size $N_{eff}$ of the transient amplifiers remains close to $N$ and as a consequence also the maximal fitness $r_{max}$ for which amplification occurs is relatively close to $r=1$. 
Possibly, a modification of the perturbation method that systematically  removes a larger number of edges would also yield larger $r_{max}$. Also, the transient amplifiers considered here are based on graphs that are not weighted. As unweighted graphs restrict the effect of substantial amplifications~\cite{pav18,tka19}, combining the perturbation method with designing weights may lead to larger values of $r_{max}$.  This may be a topic for future work. 

\subsection*{Implications of spectral analysis}

A spectral analysis  revealed that the amplifier constructors share certain structural properties. The low spectral gap indicates that these graphs are path-like and can be easily divided into subgraphs by removing a low number of edges and/or vertices. From a graph-theoretical point of view, amplifier constructors most likely have cut and/or hinge vertices~\cite{chang97,ho96}.   The examples of the transient amplifiers producing the largest $N_{eff}$, see Fig. \ref{fig:18_20_22} and Fig.    \ref{fig:15_16} in the Supporting Information, are illustrative examples for this property. For instance, the graphs producing the highest $N_{eff}$ for the cubic graphs,  see Fig. \ref{fig:18_20_22},  have a triangular hub as a center, while each of the three vertices forming the triangular hub is a cut vertex. 

The analysis of the spectral density of the amplifier constructors has further shown a multiplicity of the eigenvalues of the normalized Laplacian. This spectral property  is connected to motif doubling and motif attachment, which can be seen as the process of building a graph from joining or repeating identical substructures~\cite{ban08,ban09,meha15}. This allows the conclusion that with a high probability amplifier constructors  contain identical (or at least almost identical) subgraphs and thus exhibit a substantial degree of graph symmetry. Also this can be clearly seen in the graphs producing highest $N_{eff}$, see Fig. \ref{fig:18_20_22} and Fig.    \ref{fig:15_16}. It might be interesting to note that similar results have been reported for other biological networks, for instance  metabolistic networks, transcription networks,  food-webs and phylogenetic trees ~\cite{ban07,lew16,milo02}.
This suggests the speculation that motif doubling and motif attachment is a general property of the mechanisms in biological networks, which also manifests itself in the interplay between the graph structure and the evolutionary dynamics of mutants invading the graph. In this respect, the 
graph structures identified by Allen et al.~\cite{allen20} and denoted as fans, separated hubs and stars of islands, are also fitting these structural patterns. In each of these three structures, a hub consisting of one (the fan), two (the separated hub) or three (the star of islands) vertices is surrounded by blades which are triangles of vertices. The number of these blades may vary. Clearly, such structures can be build by  motif doubling and motif attachment (the motif being the triangular blade) and  with some likelihood a spectral analysis would yield similar results as the analysis given here. Similar consideration also apply to other graphs structures identified as (transient) amplifiers such as (super-) stars and comets~\cite{jam15,pav17}

The graphs identified in this paper as amplifier constructors have, generally speaking, no expander properties. This is in contrast to the findings of a recent work which deals with evolutionary games on isothermal graphs, analyses relationships to the spectral gap of the graphs and discusses an intriguing link to expanders~\cite{allen19}. 
Although the spectral gap of the adjacency matrix $A$ was studied rather than the spectral gap of the normalized Laplacian $L_{\mathcal{G}}$, which is considered here, the findings relate to each other as for regular graphs  $L_{\mathcal{G}}=I-1/k \cdot A$, and thus the spectral gaps scale linearly to each other. The reverse is not necessarily true as all regular graphs are isothermal but not the other way around. For these evolutionary games it was shown that the prevalence of cooperative behavior is connected to the effective degree for random isothermal graphs. Moreover, it was demonstrated that there are bounds on the effective degree in terms of the spectral gap for expander graphs. In other words, expander properties might be particularly relevant for understanding cooperative behaviour on graphs. The results given in this paper do not point at expanders, which suggests some speculations. Isothermal graphs neither amplify nor suppress the effect of selection, as do regular graphs. As demonstrated here perturbed 
regular graphs (which can be seen as almost regular graphs as the transient amplifiers only have a single edge removed from a regular graph) do yield transient amplifiers. It might be that those ``weakly'' disturbed and thus ``almost regular'' graphs require different structural properties as random isothermal graphs which usually have a lesser degree of regularity.   Also these relationships should be addressed by future work.

To summarize the transient amplifiers constructed from cubic and quartic regular graphs share certain structural properties.  They are rather path-like graphs with low conductance which are relatively easy to divide into subgraphs by removing edges and/or vertices. Frequently, the subgraphs are identical (or at least very similar) and can be viewed as building blocks which are connected by cut and/or hinge vertices. This suggest the question of why and how  these structural properties promote the spread of beneficial mutations.  A possible explanation for this  question comes from viewing these structural properties  from the point of evolutionary dynamics on the graph. The dynamics of random walks on graphs with these structural properties implies large mixing times, bottlenecks and the emergence of clusters. 
On the other hand, it is known that 
clusters of  mutants (or cooperators in the case of evolutionary  games) promote survival and facilitate 
 spatial invasion,
  as shown for lattice  
grids~\cite{hau01,hau04,lang08,page00}, circle graphs~\cite{xiao19} and selected regular graphs~\cite{rich19b}. Thus, it appears plausible that for this reason the structural properties identified for amplifier constructors also promote the spread of beneficial mutations.  Moreover, searching for these structural properties could also guide the design process of transient amplifiers.  This may mean either to look for graphs with prescribed spectral characteristics, or to direct algorithms generating random graphs towards the relevant structural patterns.  
Additional work is needed to further clarify these relationships.

\section*{Acknowledgments} I wish to thank Markus Meringer  for making available the \texttt{genreg} software~\cite{mer99} used for generating the regular graphs according to Tab.~\ref{tab:graphs} and for helpful~discussions. Further thanks goes to Benjamin Allen for sharing the algorithm to calculate the coalescence times.

\section*{Supporting Information} The results of this paper are calculated and visualized with MATLAB. The adjacency matrices of the set of all amplifier constructors  as well as code to produce the results are available at \\ \small{\url{https://github.com/HendrikRichterLeipzig/TransientAmplifiersRegularGraphs}.}
Additional graphs and a table with the numbers $\mathcal{L}_k(N)$ of simple connected pairwise nonisomorphic $k$-regular graphs on  $N$ vertices are given in the following

 \begin{table}[htb]
\centering
\caption{The numbers $\mathcal{L}_k(N)$ of simple connected pairwise nonisomorphic $k$-regular graphs on  $N$ vertices see e.g.,~\cite{mer99}, which corresponds to the number of regular interaction networks with $N$ individuals and $k$  interacting neighbors for $11 \leq N \leq 22$.   }
\label{tab:graphs}
\center
\begin{tabular}{ccccc}

\hline
 \boldmath{$\: {}_N \: \backslash \: {}^k$} & \textbf{3} & \textbf{4} & \textbf{5} &\textbf{6} \\ \hline 11 &
0 & 265 & 0 & 266 \\
 12 & 85 & 1.544 & 7.848 & 7.849  \\
 13 & 0 & 10.778  & 0 & 367.860   \\
 14 &509 & 88.168 & 3.459.383 & 21.609.300  \\
 15 & 0 & 805.491 \\
 16 & 4.060& 8.037.418   \\
 18 & 41.301 \\
 20 & 510.489 \\
 22 & 7.319.447\\

\hline

\end{tabular}
\end{table}

 \begin{figure}[htb]
\centering
\includegraphics[trim = 35mm 90mm 40mm 20mm,clip, width=6.5cm, height=9.5cm]{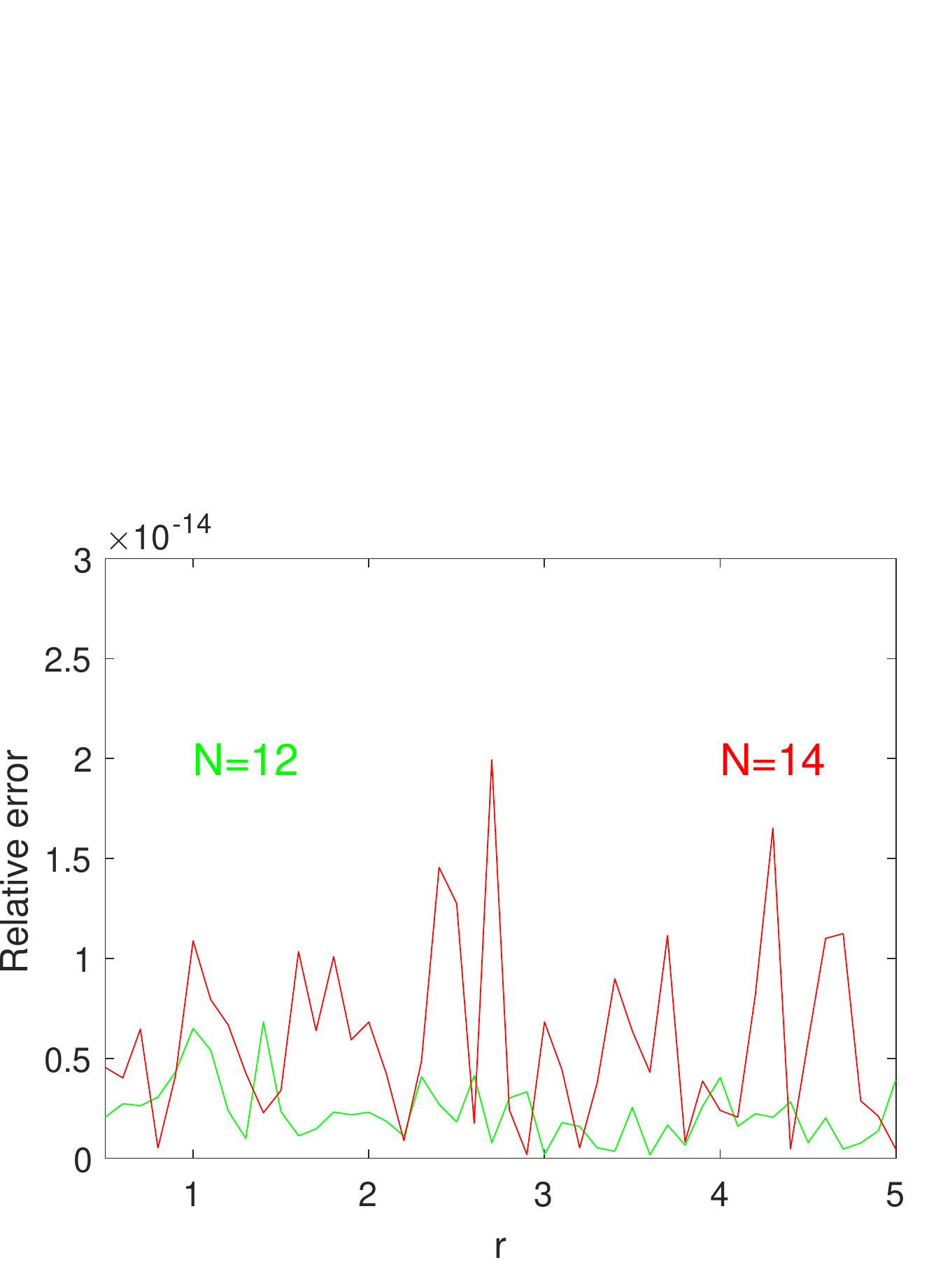} 
\includegraphics[trim = 35mm 90mm 40mm 20mm,clip, width=6.5cm, height=9.5cm]{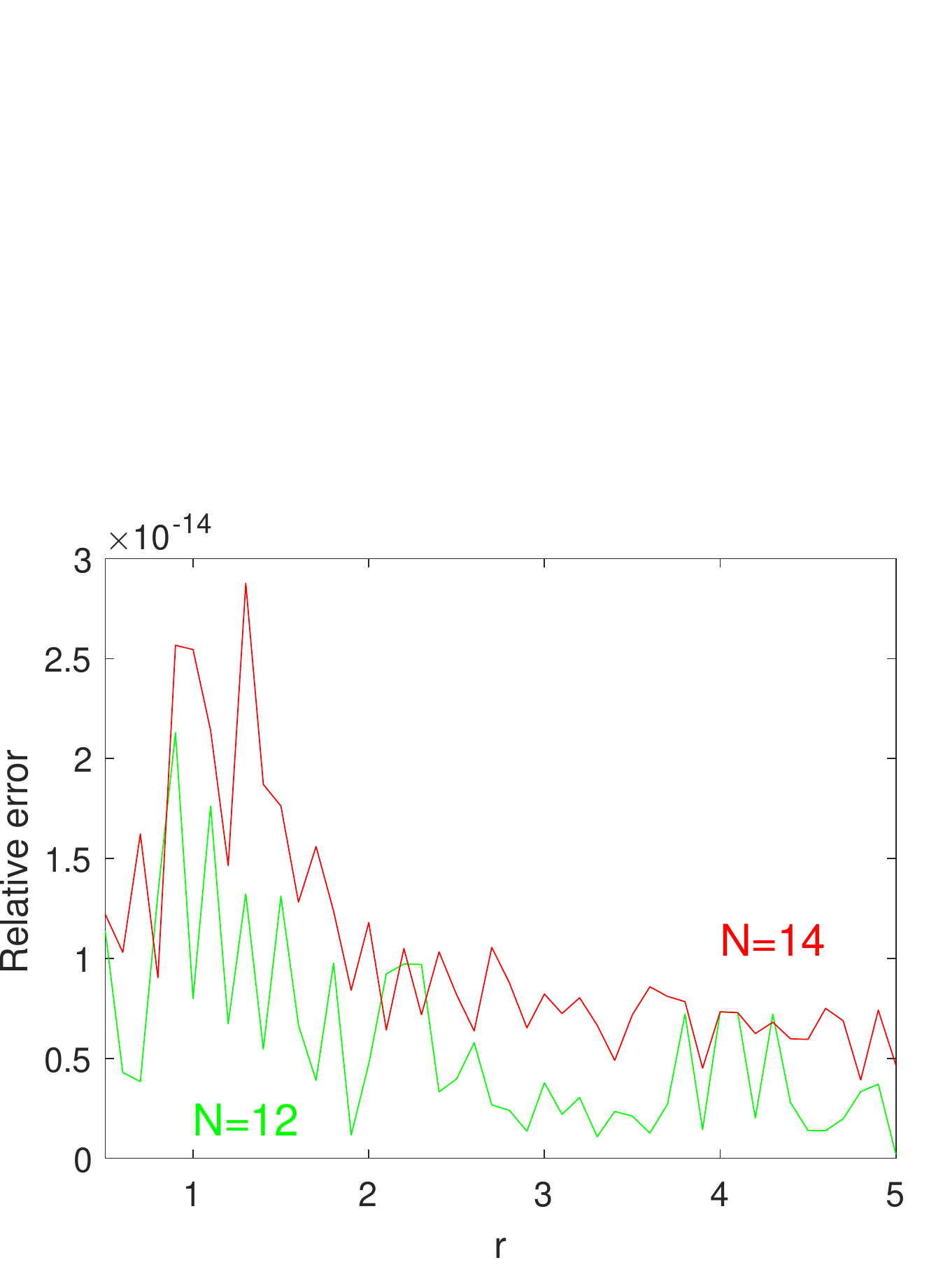} 

 (\textbf{a}) Complete graph \hspace{2.7cm} (\textbf{b}) Cycle graph  \hspace{2.7cm}

\caption{\small{The relative error between the numerical computation of the fixation probability using a Markov state transition matrix and the analytical result over mutant fitness $r$ for $N=12$ and $N=14$. (a) For the complete graph with the analytical results according to Eq. \eqref{eq:complete}. (b) For the cycle graph with analytical results according to Eq. \eqref{eq:ring}.  }}
\label{fig:error}
\end{figure}

 \begin{figure}[htb]
\centering
\includegraphics[trim = 25mm 90mm 40mm 100mm,clip, width=6.5cm, height=5cm]{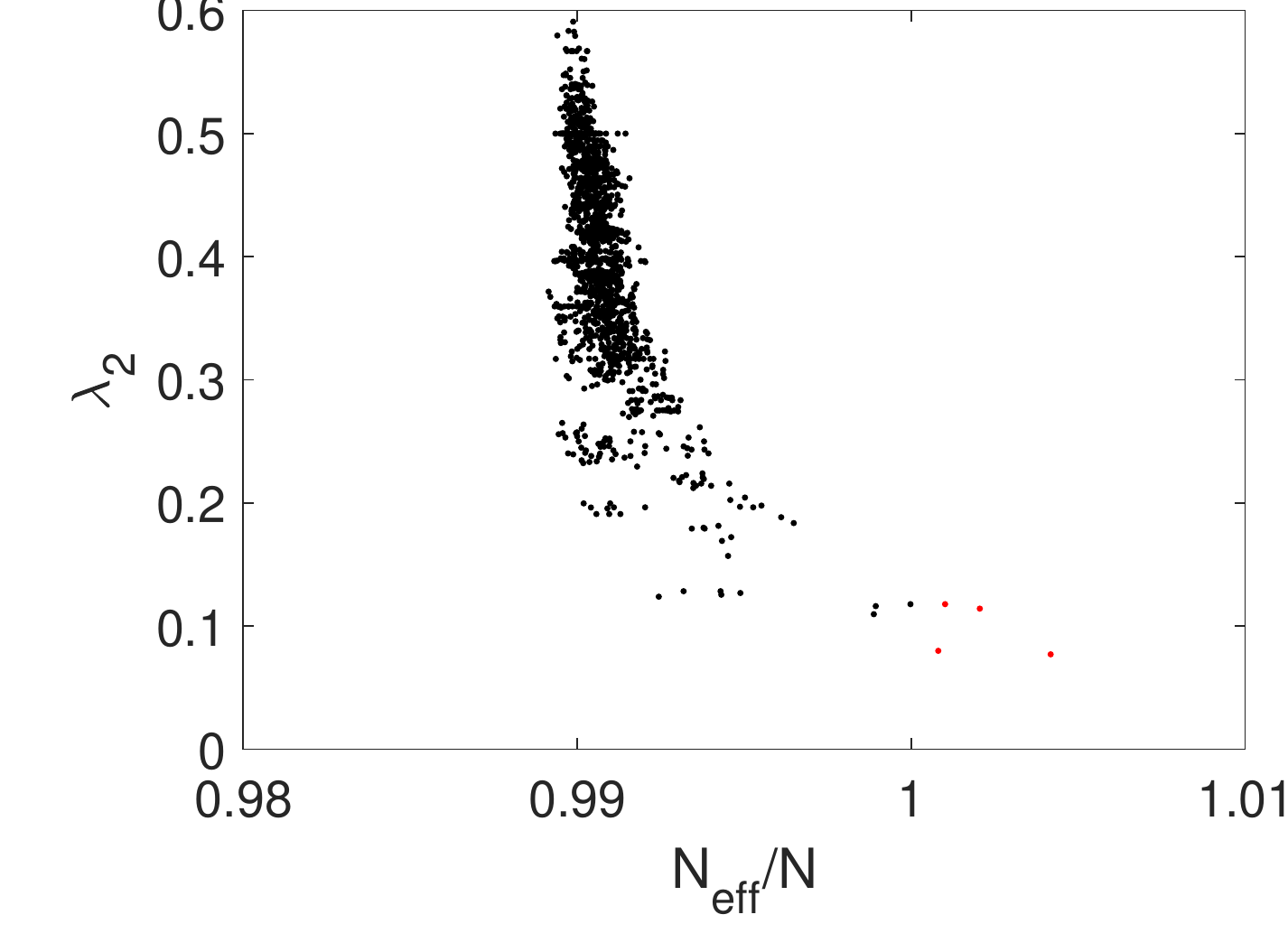} 
\includegraphics[trim = 25mm 90mm 40mm 100mm,clip, width=6.5cm, height=5cm]{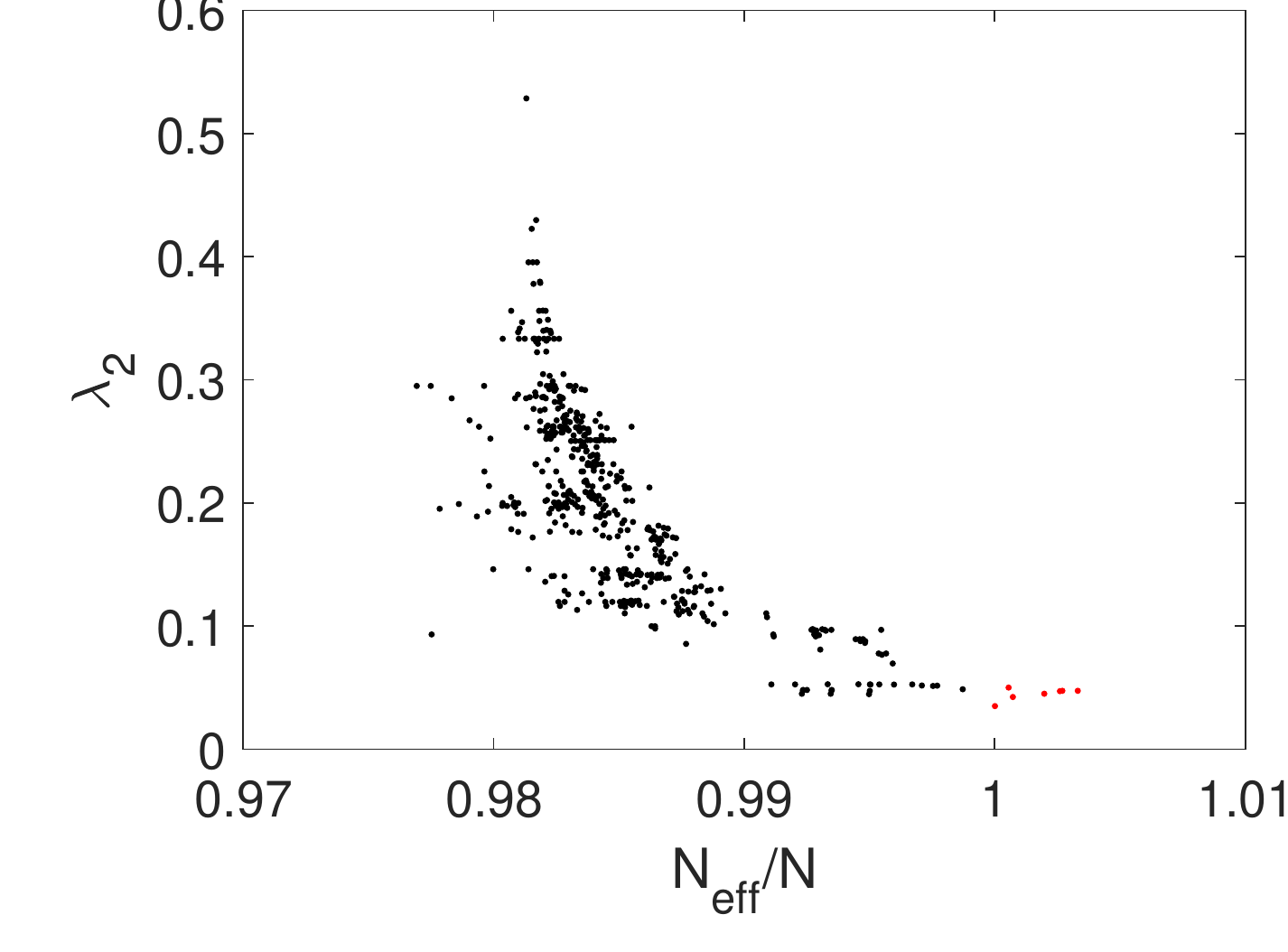} 

(\textbf{a}) 12 4 \hspace{2.7cm} (\textbf{e}) 14 3

\includegraphics[trim = 25mm 90mm 40mm 100mm,clip, width=6.5cm, height=5cm]{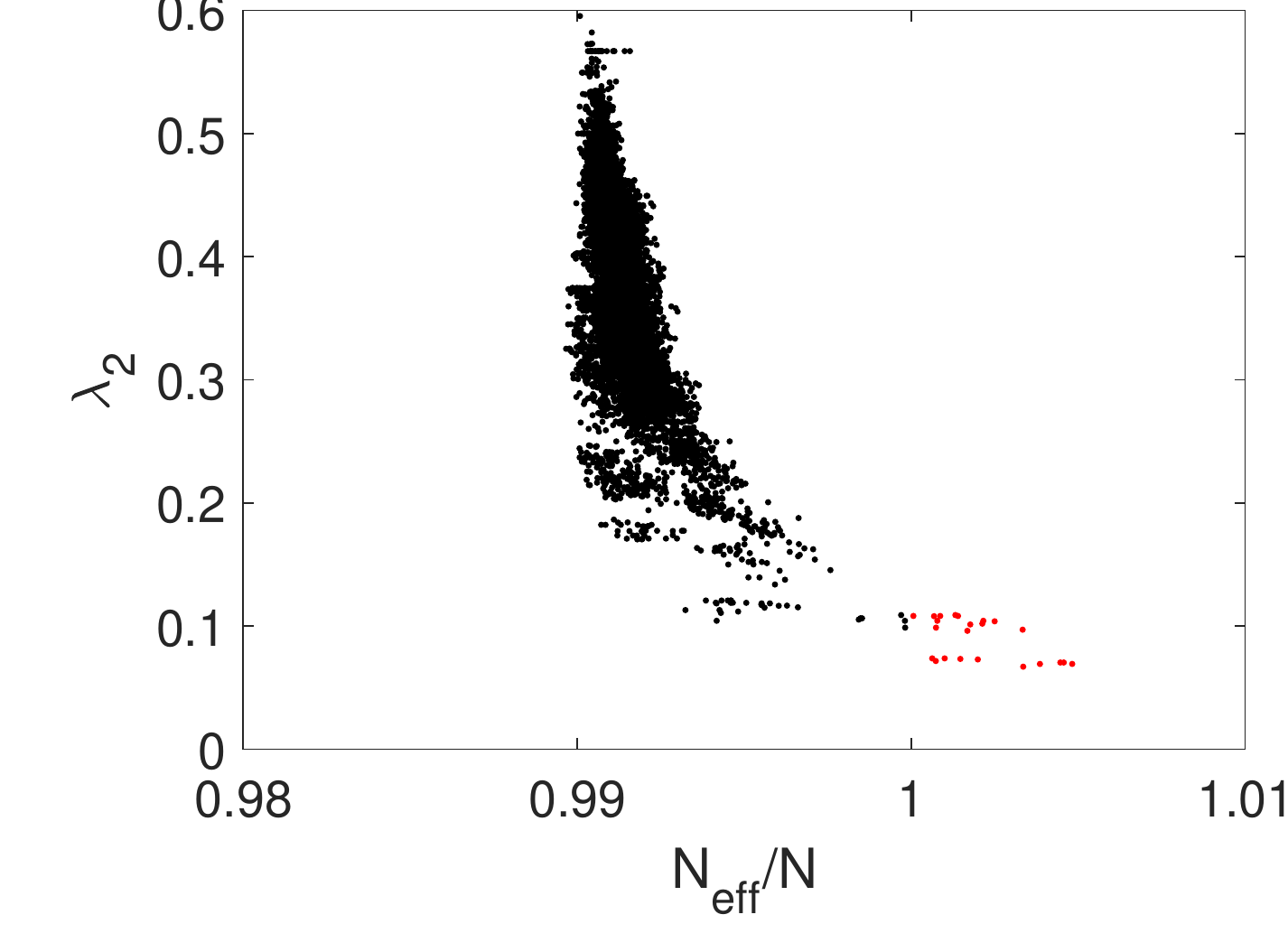} 
\includegraphics[trim = 25mm 90mm 40mm 100mm,clip, width=6.5cm, height=5cm]{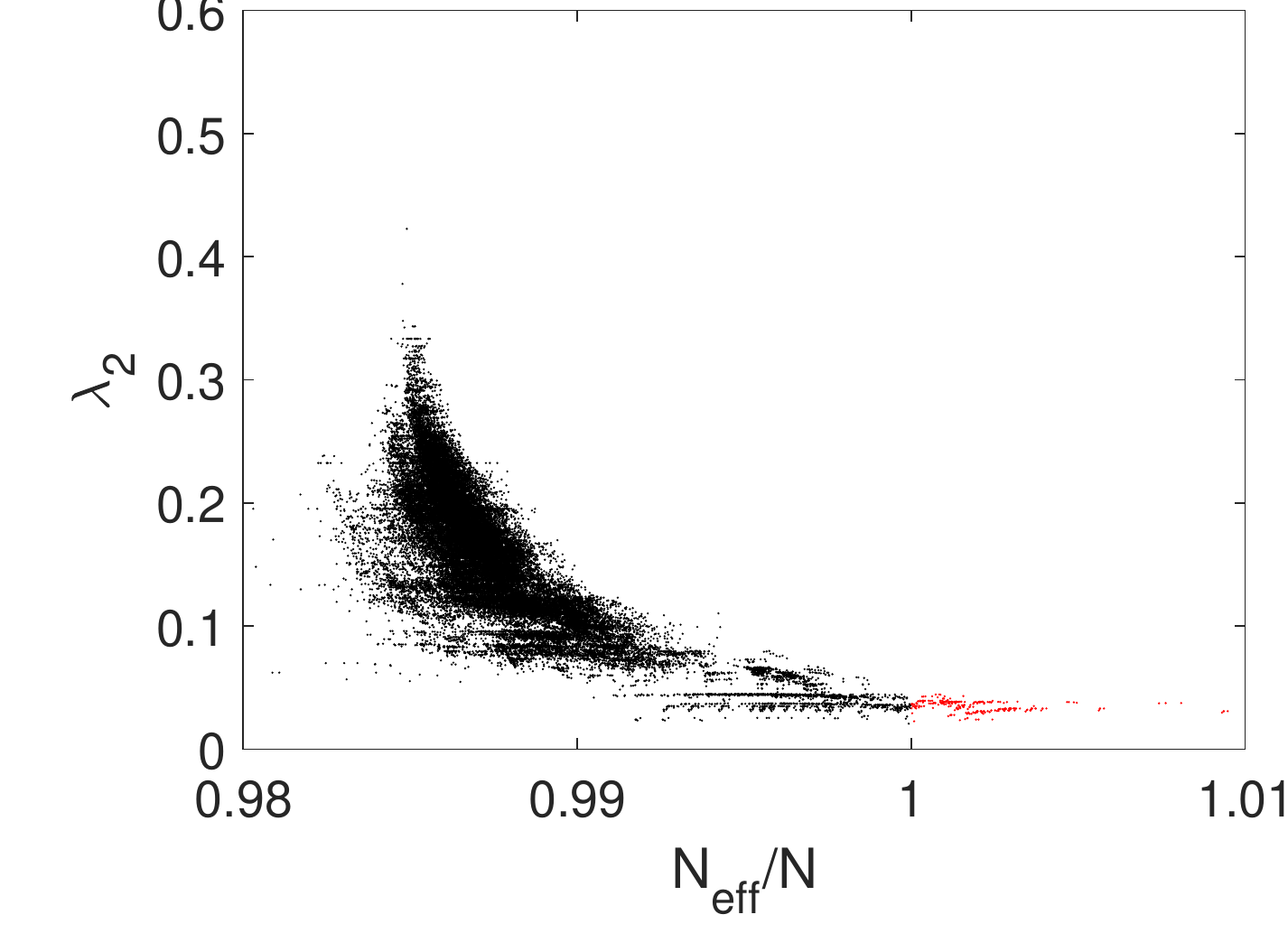} 

(\textbf{b})  13 4\hspace{2.7cm} (\textbf{f}) 18 3

  \includegraphics[trim = 25mm 90mm 40mm 100mm,clip, width=6.5cm, height=5cm]{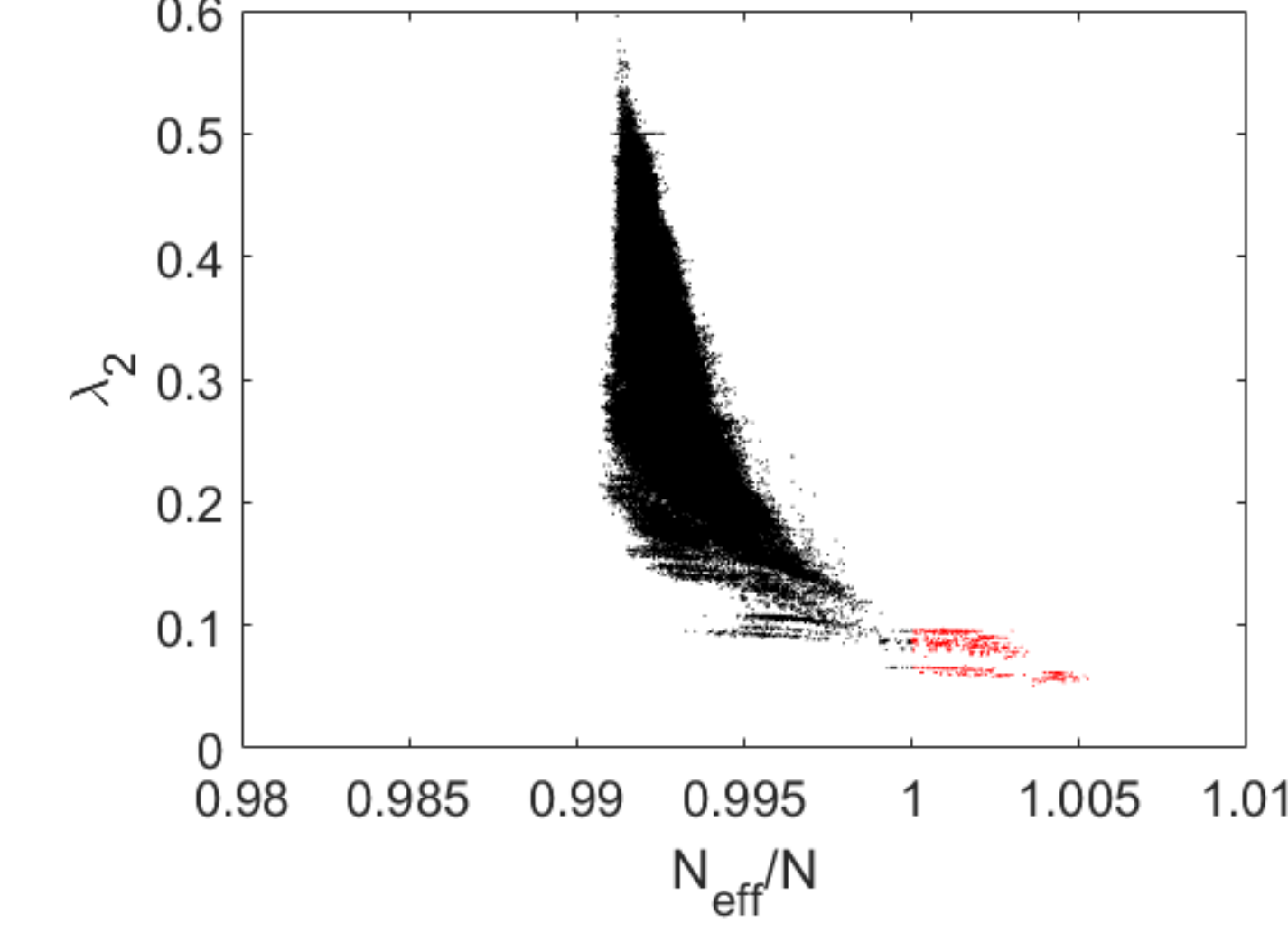} 
\includegraphics[trim = 25mm 90mm 40mm 100mm,clip, width=6.5cm, height=5cm]{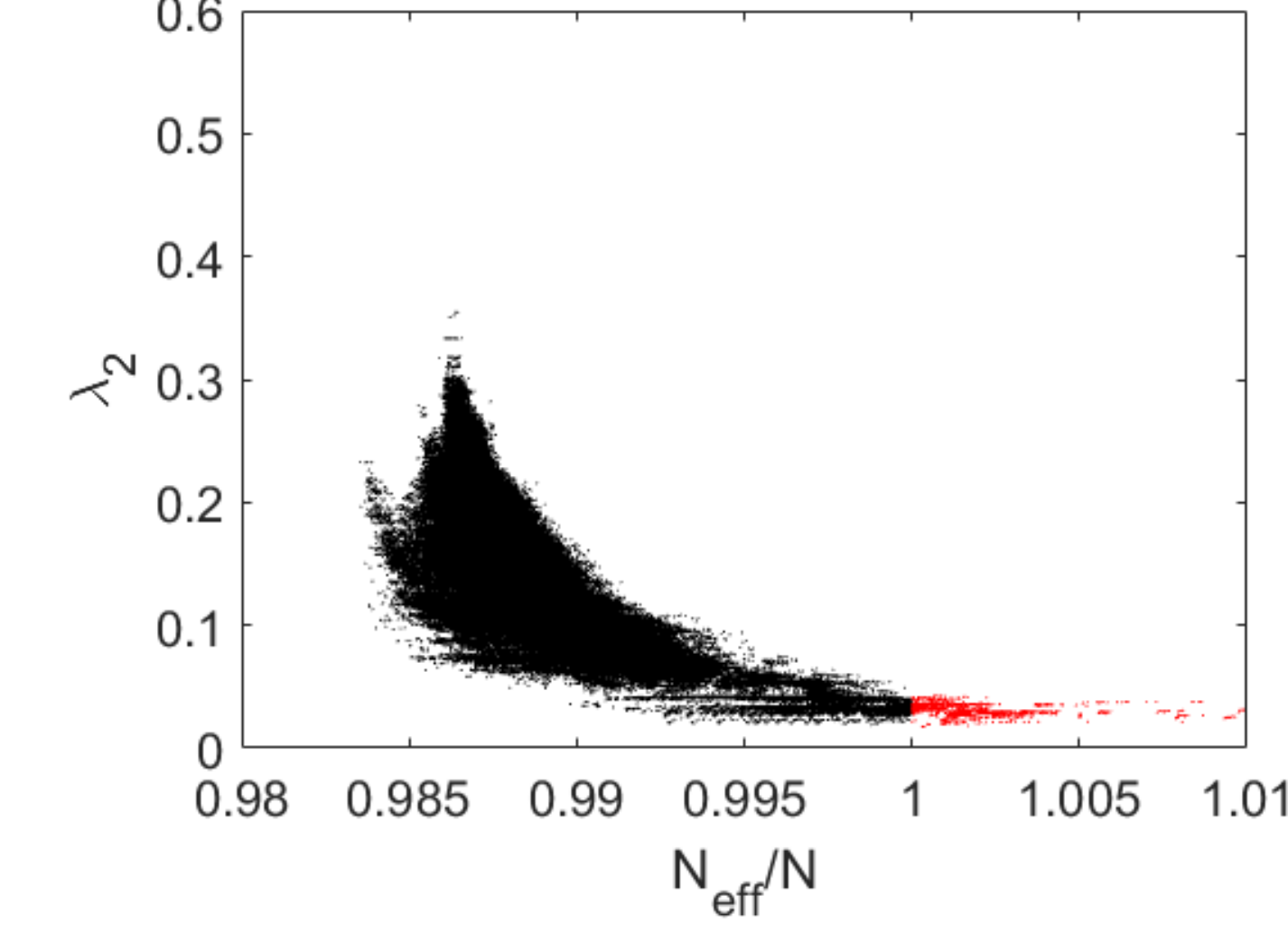} 

(\textbf{c}) 15 4 \hspace{2.7cm} (\textbf{g})  20 3 
 
   \includegraphics[trim = 25mm 90mm 40mm 100mm,clip, width=6.5cm, height=5cm]{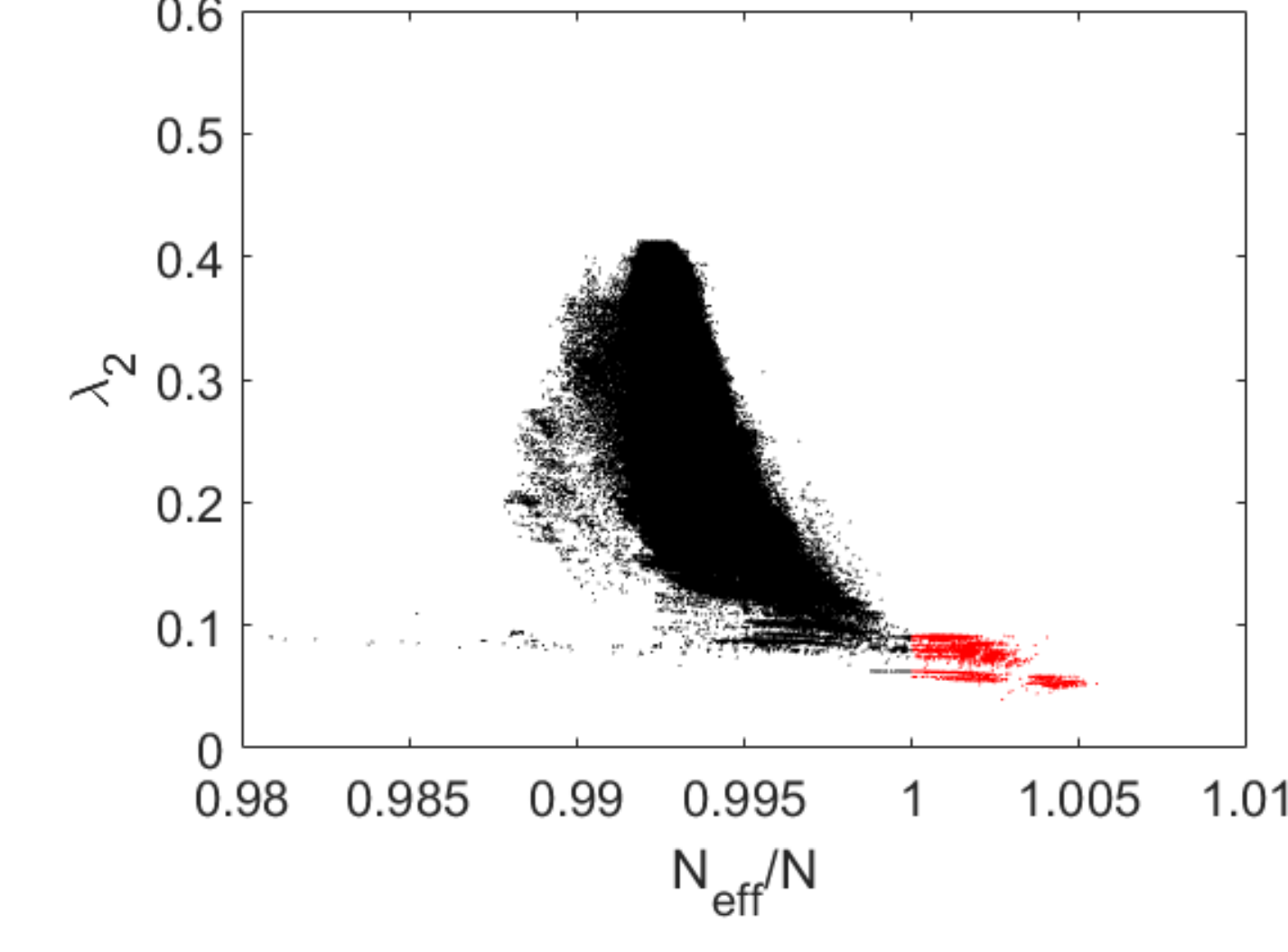} 
\includegraphics[trim = 25mm 90mm 40mm 100mm,clip, width=6.5cm, height=5cm]{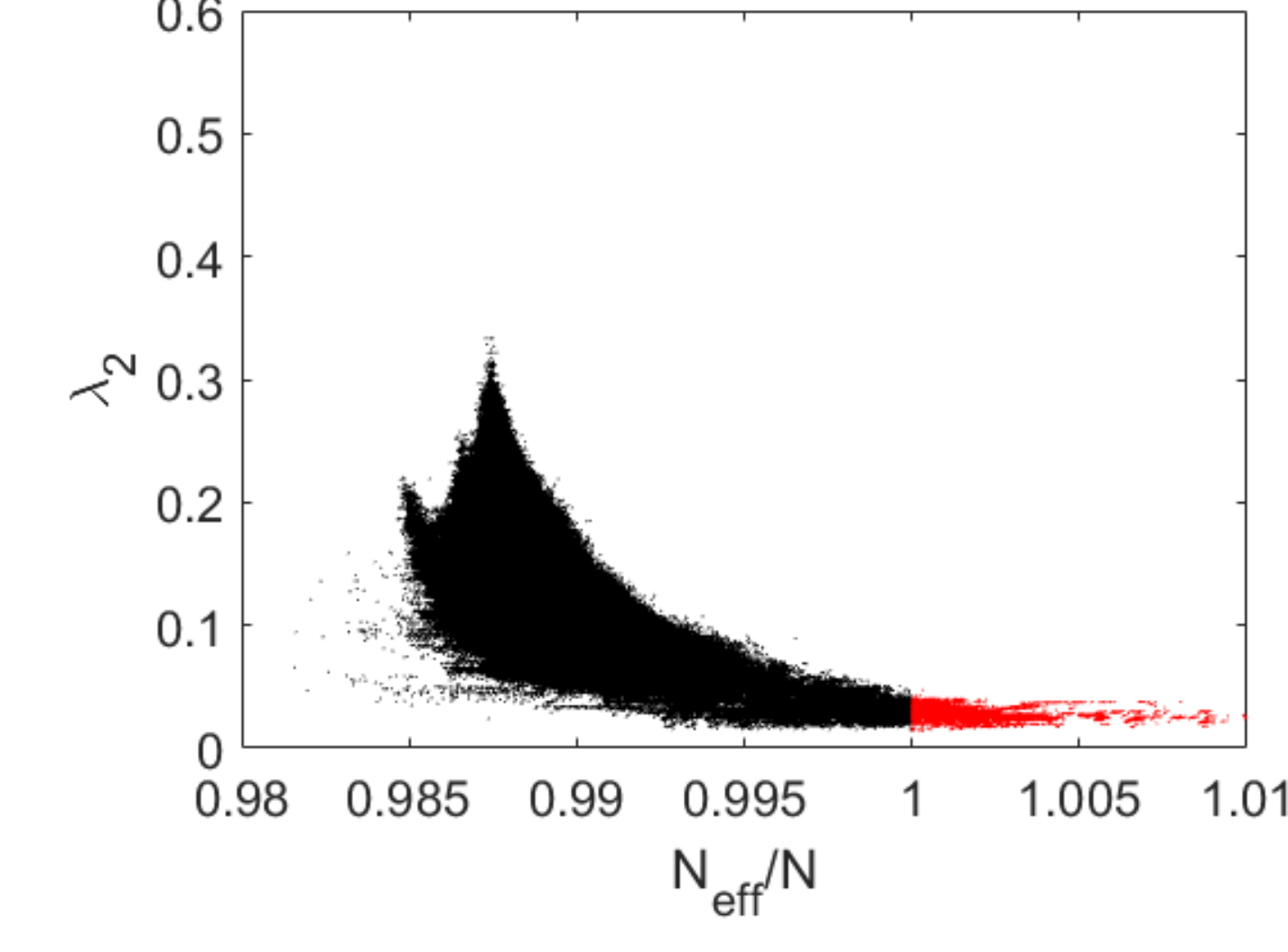} 

(\textbf{d}) 16 4 \hspace{2.7cm} (\textbf{h})  22 3 
\caption{\small{The spectral gap $\lambda_2$ versus the ratio $N_{eff}/N$  as a scatter plot for: (a)-(d) the quartic graphs of size $N=\{12,13,15,16\}$ and (e)-(h) the cubic graphs of size $N=\{14,18,20,22\}$. Supplement to Fig. \ref{fig:lambda_2_1} }}
\label{fig:lambda_2}
\end{figure}

 \begin{figure}[htb]
\centering
\includegraphics[trim = 53mm 100mm 52mm 100mm,clip, width=6cm, height=4.5cm]{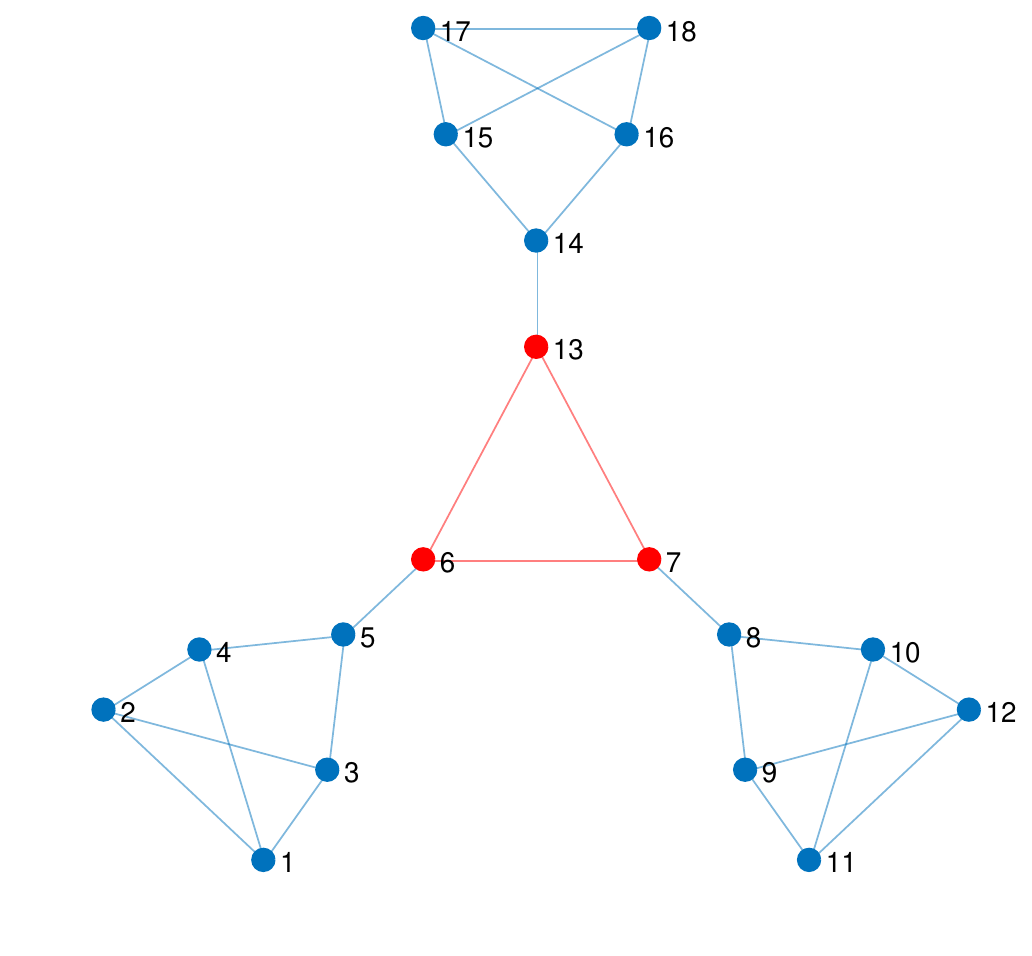}
\includegraphics[trim = 53mm 95mm 50mm 100mm,clip, width=6cm, height=4.5cm]{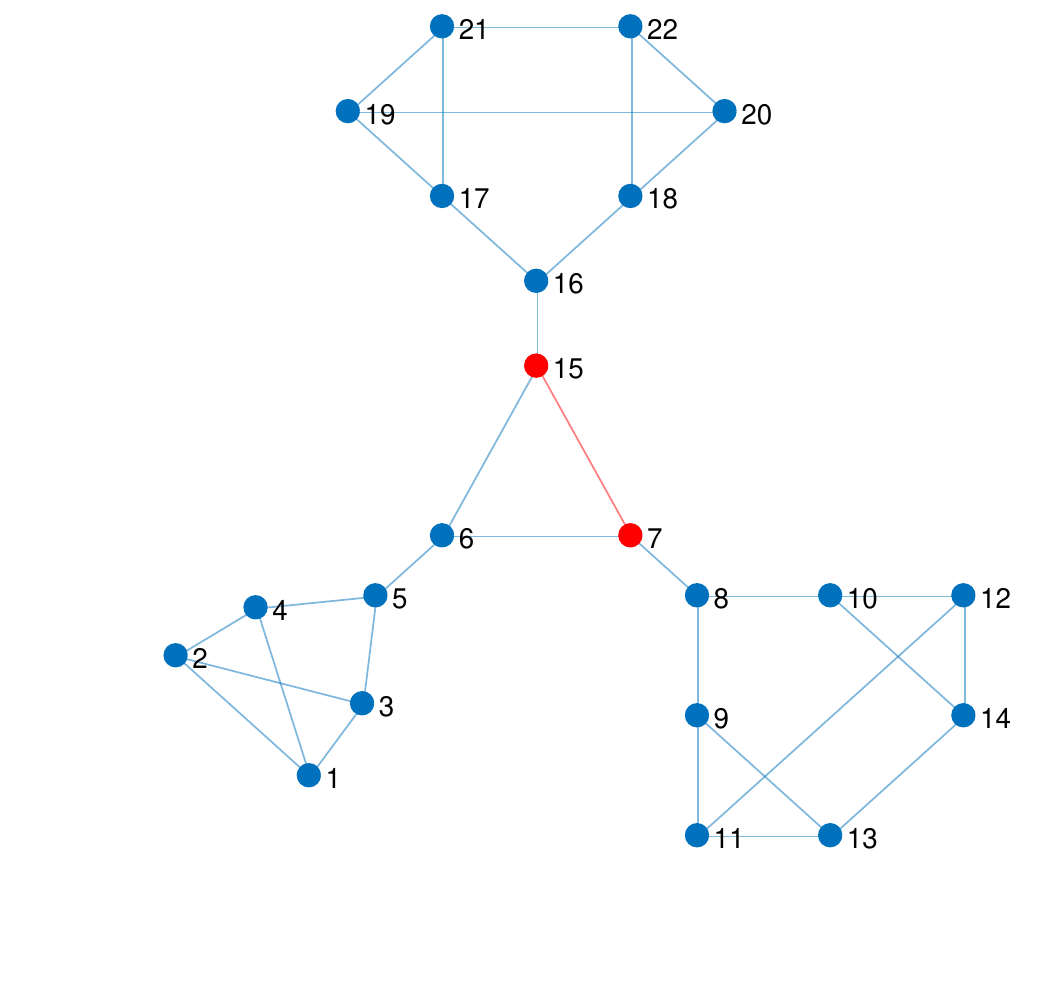}

(\textbf{a})  \hspace{2.7cm} (\textbf{c}) 

\includegraphics[trim = 53mm 95mm 50mm 90mm,clip, width=6cm, height=4.5cm]{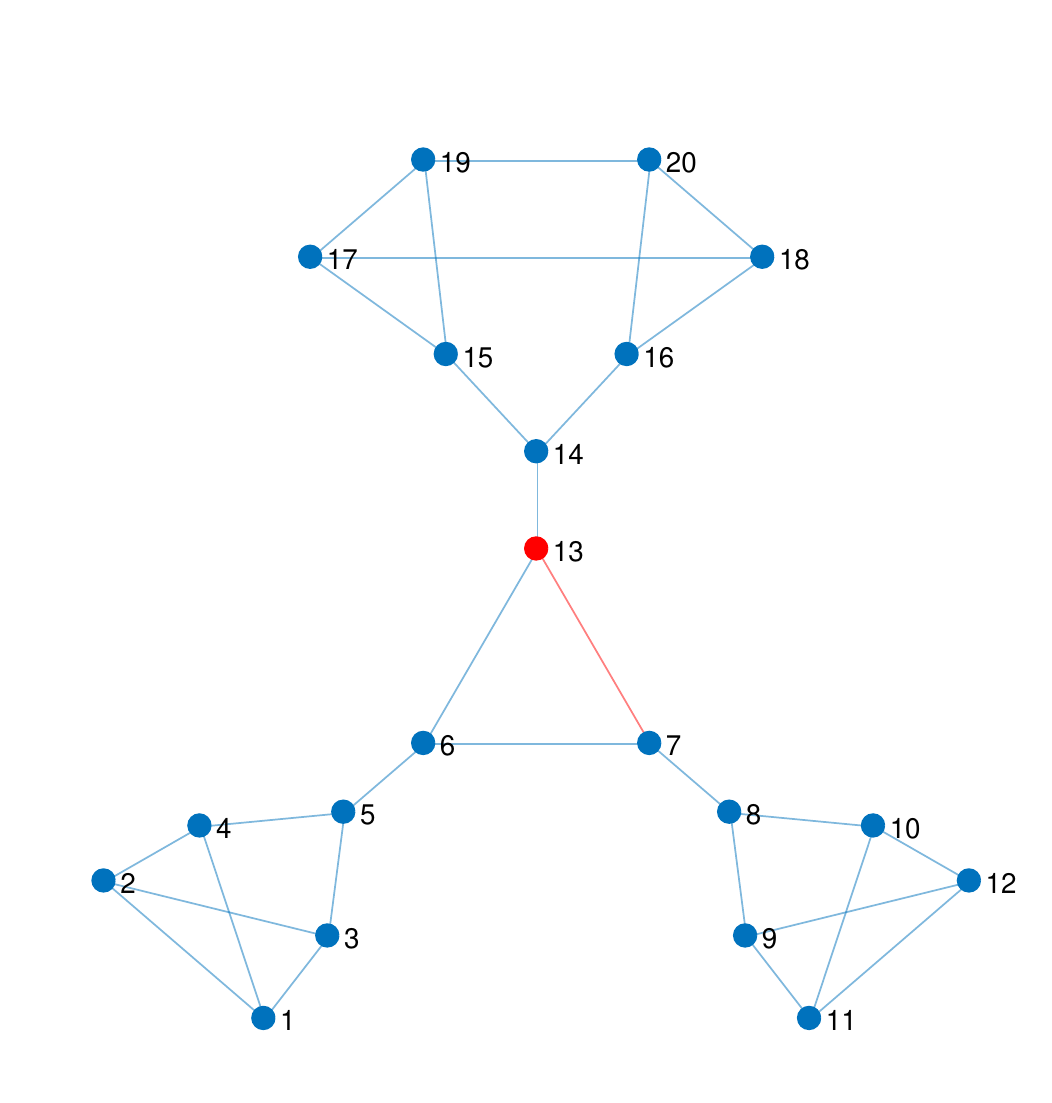}
\includegraphics[trim = 33mm 95mm 30mm 90mm,clip, width=6cm, height=4.5cm]{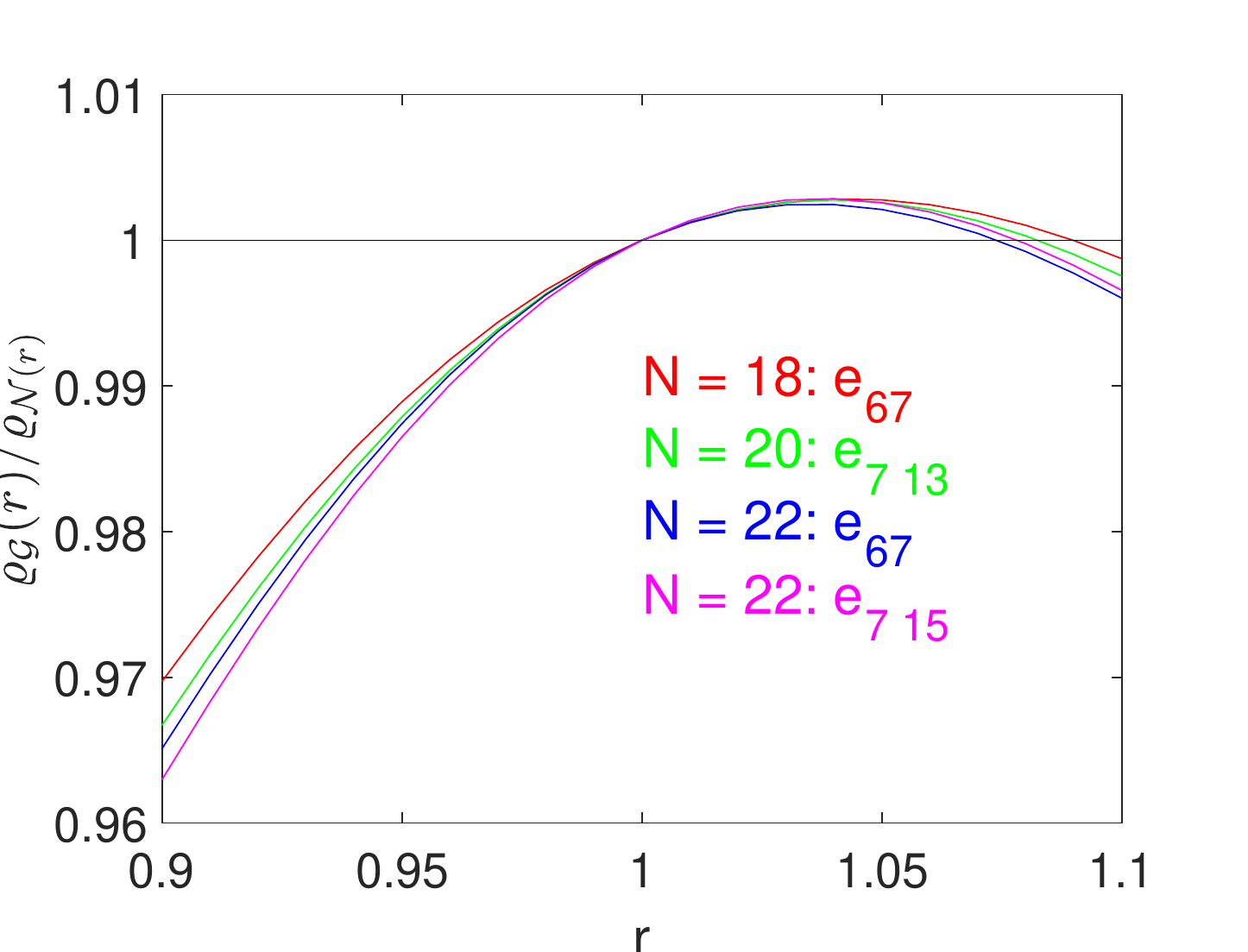}

(\textbf{b})  \hspace{2.7cm} (\textbf{d})  
 
\caption{\small{Cubic amplifiers constructor graphs producing the largest $N_{eff}$ among all $\mathcal{A}_3(N)$ amplifiers constructors according to Tab. \ref{tab:graphs_amp}. (a) $N=18$: Largest remeeting times for all three vertices in the triangle center of the graph,   $\tau_6=\tau_7=\tau_{13}=31.9937$. By removing either the edge $e_{67}$ or $e_{6\:13}$ or $e_{7\:13}$ we get $N_{eff}=18.2721$. (b), $N=20$: The largest remeeting time is $\tau_{13}=36.9143$ and removing the edge $e_{6\:13}$ or $e_{7\:13}$ yields $N_{eff}=20.2884$. (c) $N=22$: The largest remeeting times are $\tau_7=\tau_{15}=40.8256$ and removing the edge $e_{7\:15}$ yields $N_{eff}=22.3149$, while removing the edge $e_{67}$ or $e_{6\:15}$ gives $N_{eff}=22.2873$. (d)
The quantity $\varrho_{\mathcal{G}}(r)/\varrho_{\mathcal{N}}(r)$  over $r$. Values of $\varrho_{\mathcal{G}}(r)/\varrho_{\mathcal{N}}(r)>1$ for $r>1$ indicate a transient amplifier.  }  }
\label{fig:18_20_22}
\end{figure}

 \begin{figure}[htb]

\includegraphics[trim = 53mm 110mm 52mm 110mm,clip, width=6cm, height=4.5cm]{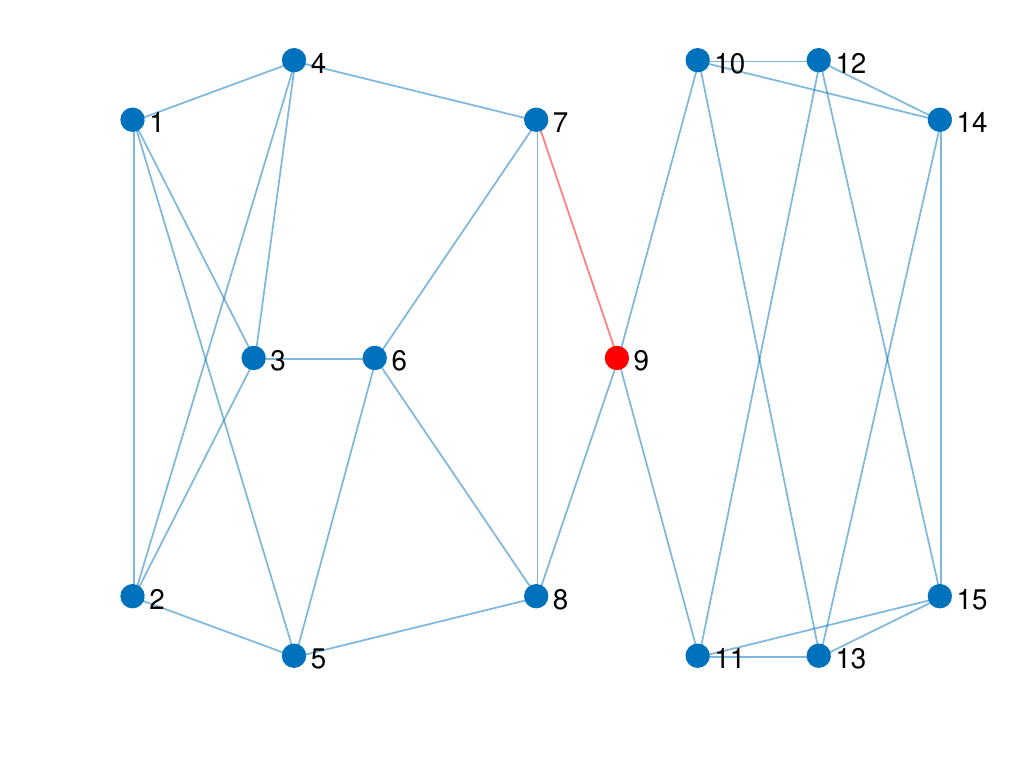}
\includegraphics[trim =33mm 95mm 30mm 90mm,clip, width=6cm, height=4.5cm]{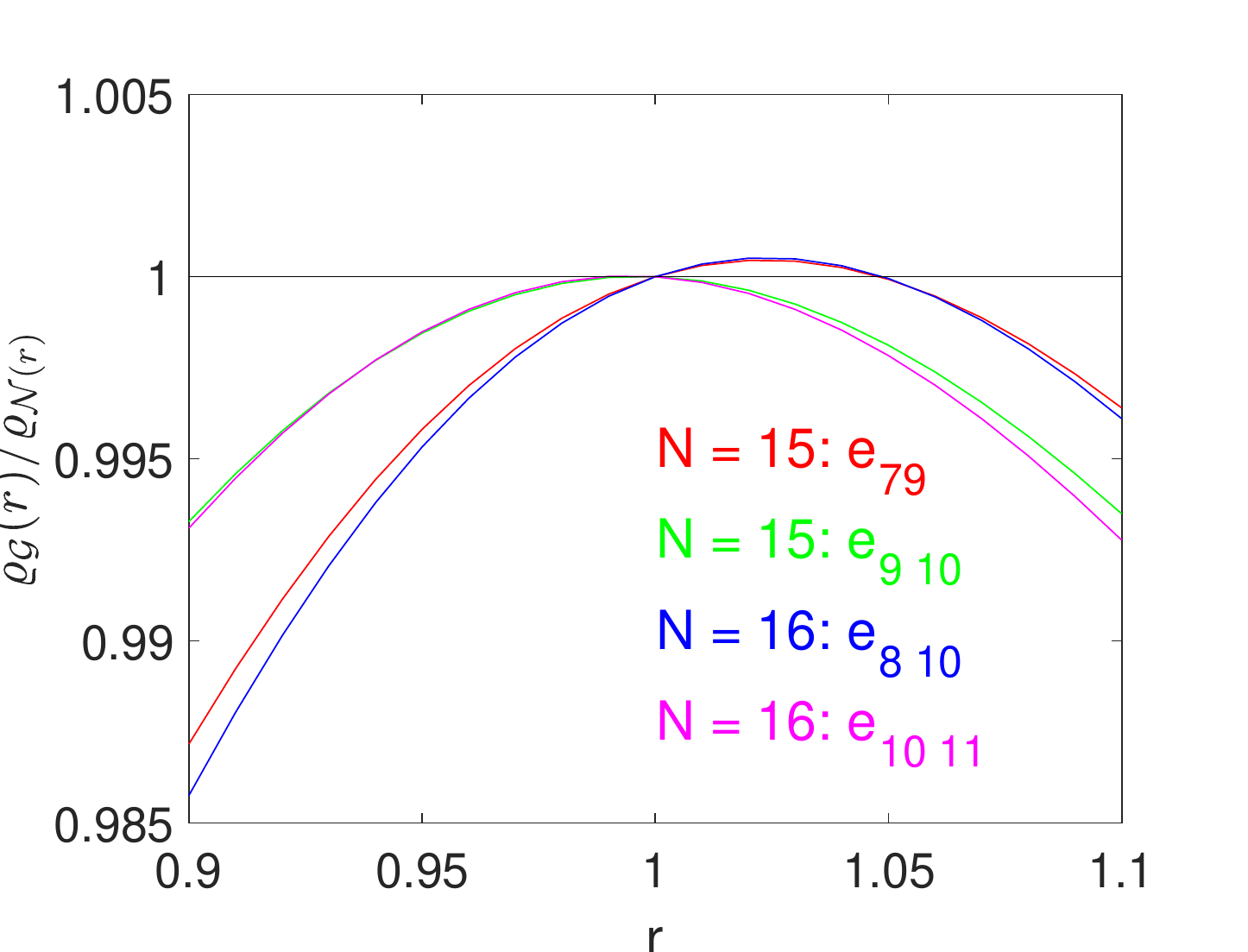}

(\textbf{a})  \hspace{6.7cm} (\textbf{c}) 

\includegraphics[trim =  53mm 110mm 50mm 110mm,clip, width=6cm, height=4.5cm]{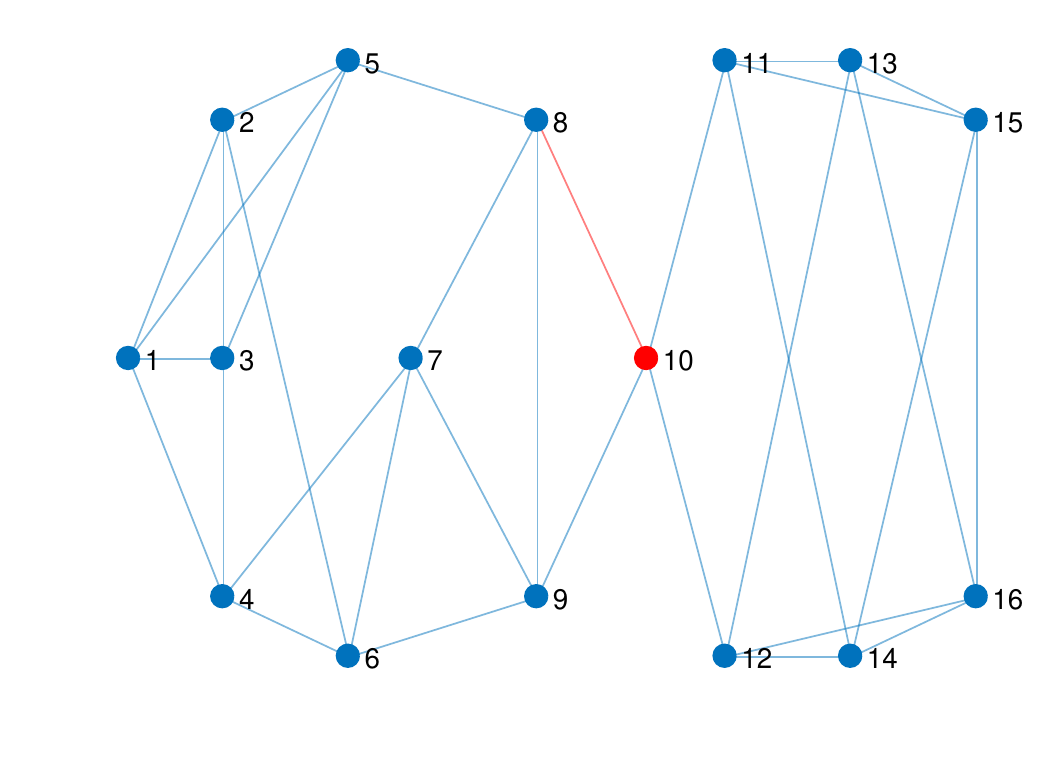}

(\textbf{b})  

\caption{\small{Quartic amplifiers constructor graphs producing the largest $N_{eff}$ among all $\mathcal{A}_4(N)$ amplifiers constructors according to Tab. \ref{tab:graphs_amp}. (a) $N=15$: Largest remeeting times for vertex $v_9$ with   $\tau_9=24.3560$. By removing either the edge $e_{79}$  we get  $N_{eff}=15.0787$ and a transient amplifier, while removing $e_{9 \: 10}$ or $e_{9\:11}$ gives $N_{eff}=14.9991$ and a fixation probability $\varrho_{\mathcal{G}}(r)$ that always smaller than $\varrho_{\mathcal{N}}(r)$ of the complete graph except for $r=1$, where it equals. (b), $N=16$: The largest remeeting time is $\tau_{10}=25.8981$ and removing the edge $e_{8\:10}$  yields $N_{eff}=16.0883$, while removing $e_{10\:11}$ or $e_{10\:12}$ only yields  $N_{eff}=15.9833$. (c)
The quantity $\varrho_{\mathcal{G}}(r)/\varrho_{\mathcal{N}}(r)$  over $r$. Values of $\varrho_{\mathcal{G}}(r)/\varrho_{\mathcal{N}}(r)>1$ for $r>1$ indicate a transient amplifier.  }  }
\label{fig:15_16}
\end{figure}

  \begin{figure}[htb]
\centering

 \includegraphics[trim = 25mm 90mm 40mm 100mm,clip, width=6.5cm, height=5cm]{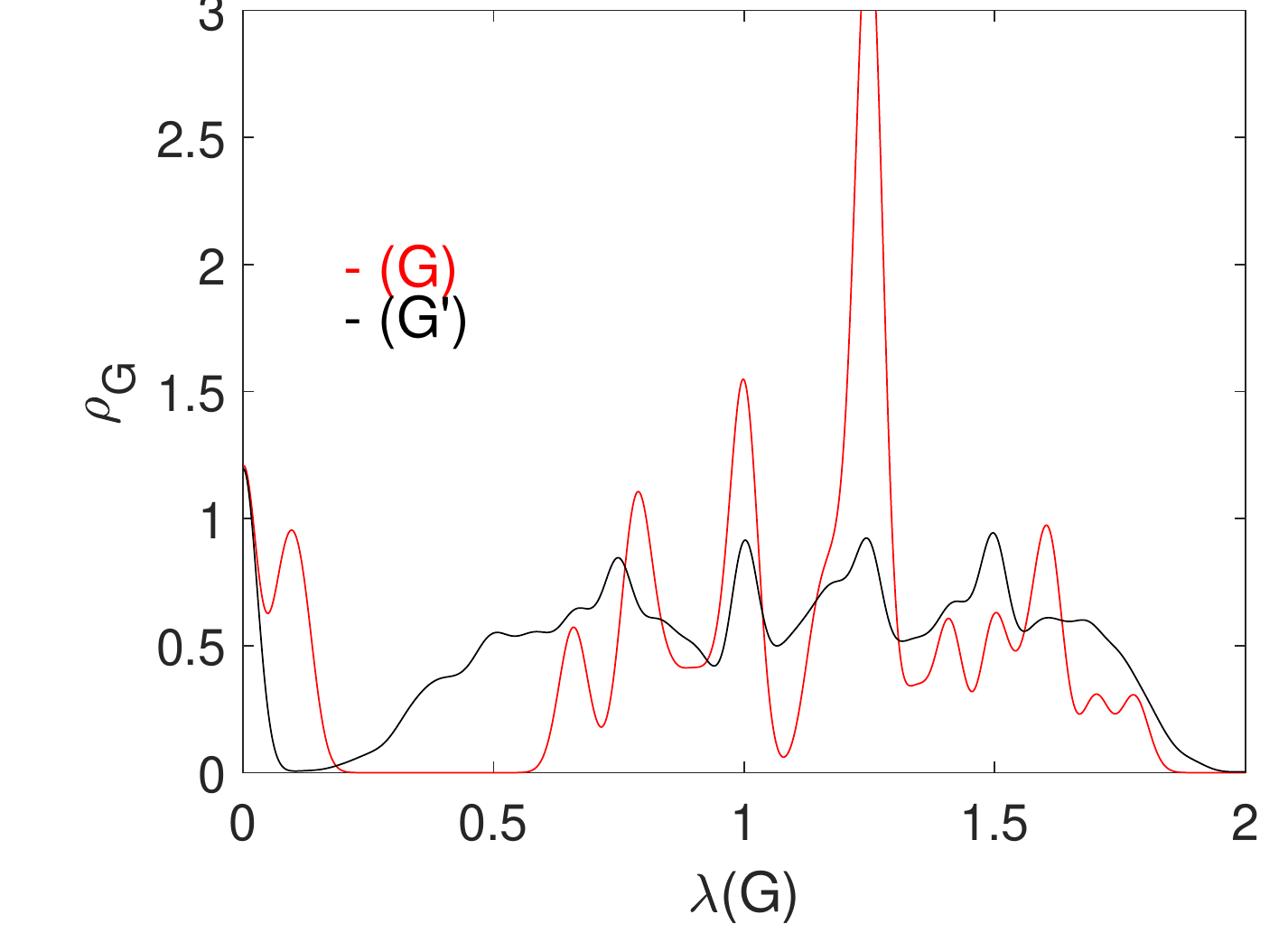} 
\includegraphics[trim = 25mm 90mm 40mm 100mm,clip, width=6.5cm, height=5cm]{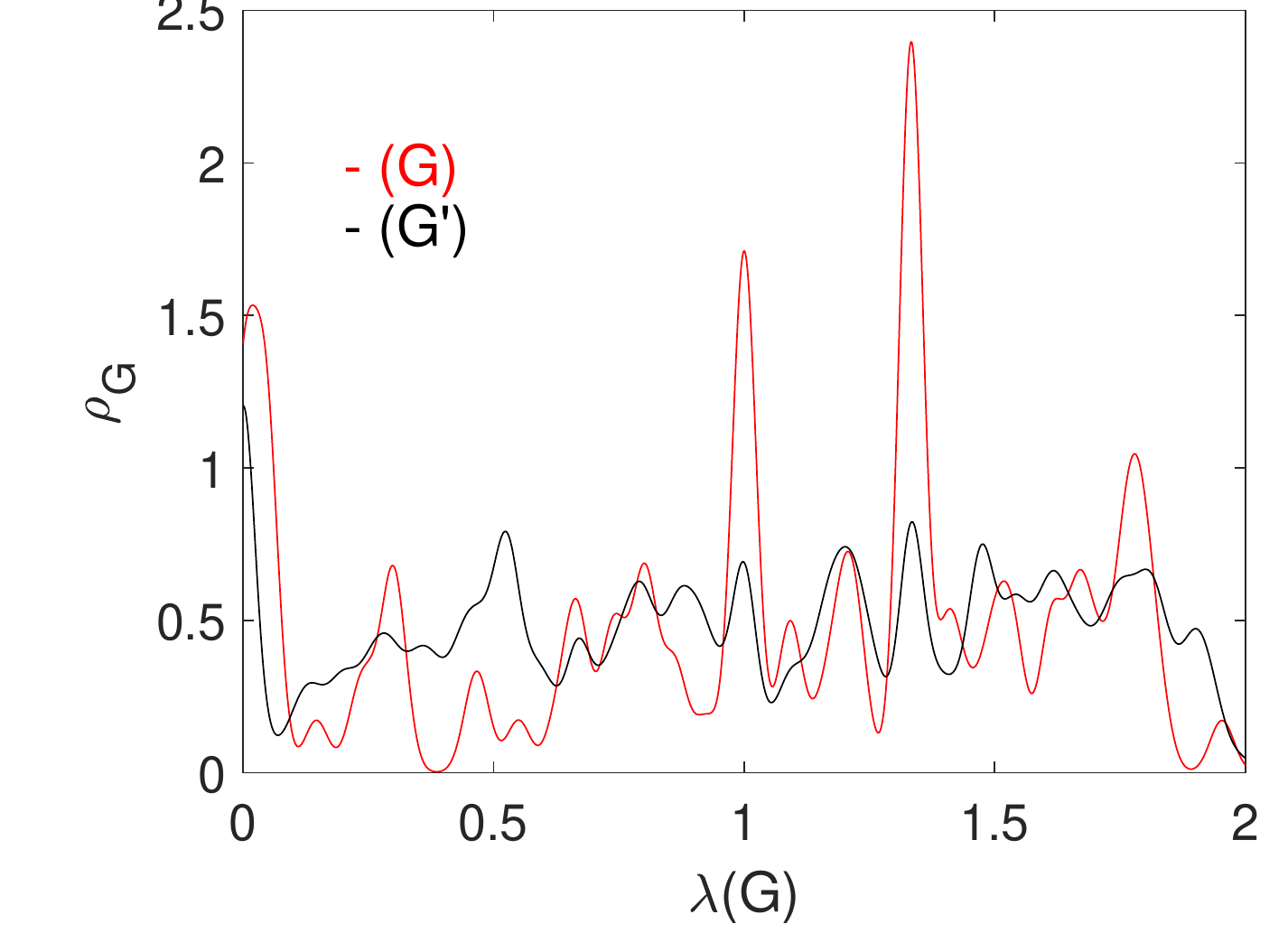} 

(\textbf{a}) 12 4   \hspace{2.7cm} (\textbf{e})   14 3

 \includegraphics[trim = 25mm 90mm 40mm 100mm,clip, width=6.5cm, height=5cm]{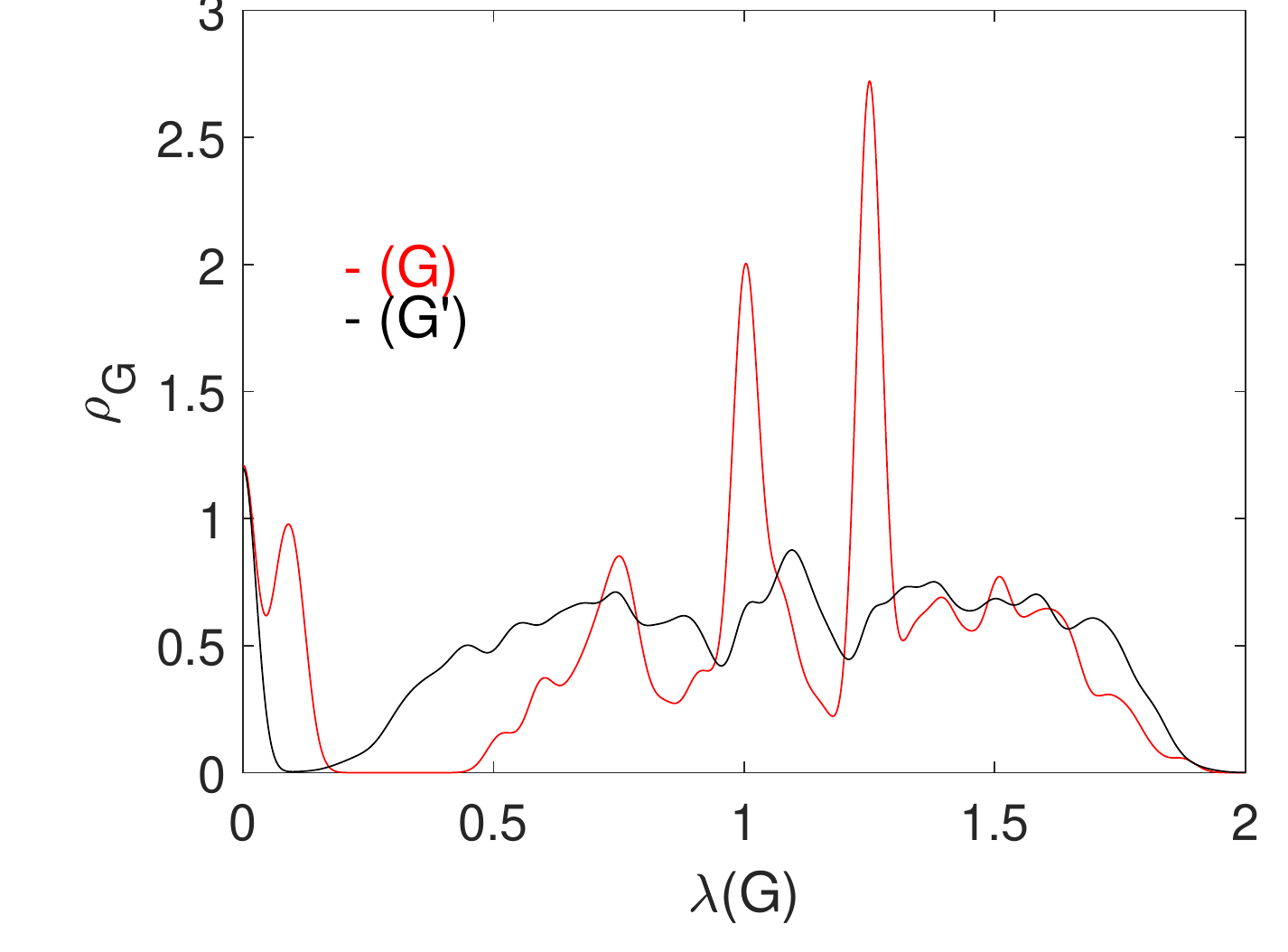} 
\includegraphics[trim = 25mm 90mm 40mm 100mm,clip, width=6.5cm, height=5cm]{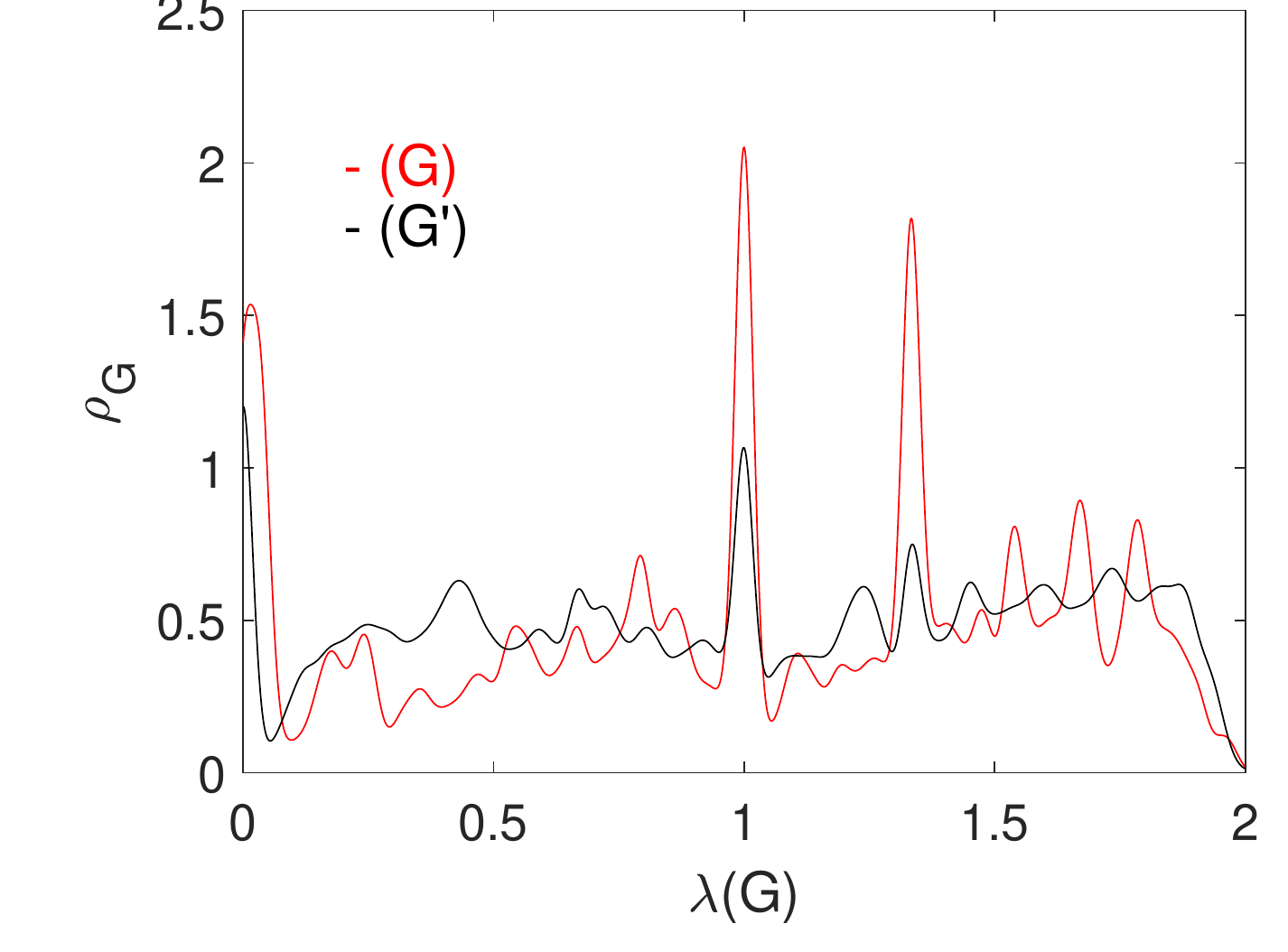} 

(\textbf{b}) 13 4   \hspace{2.7cm} (\textbf{f})  18 3 

 \includegraphics[trim = 25mm 90mm 40mm 100mm,clip, width=6.5cm, height=5cm]{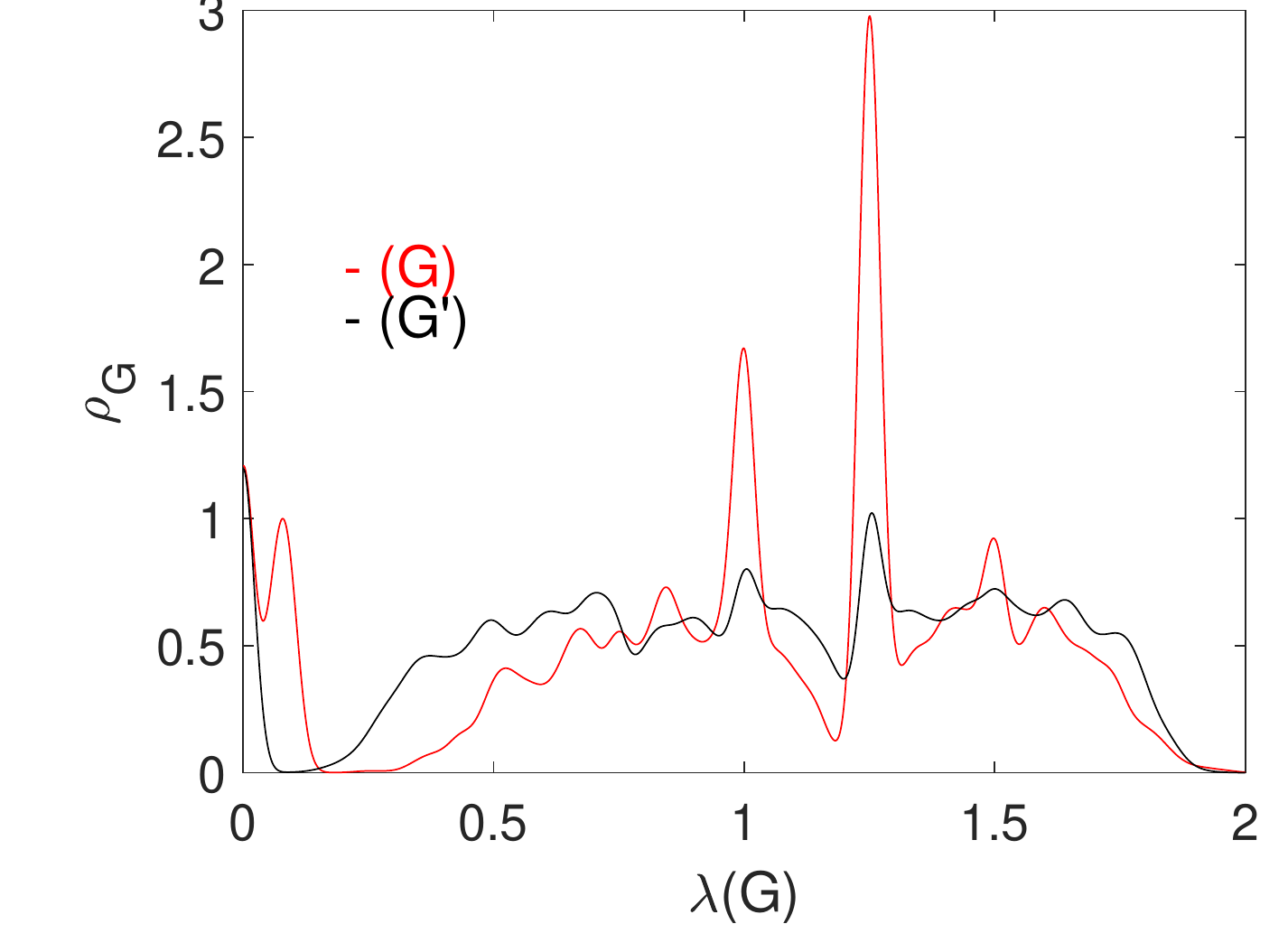} 
\includegraphics[trim = 25mm 90mm 40mm 100mm,clip, width=6.5cm, height=5cm]{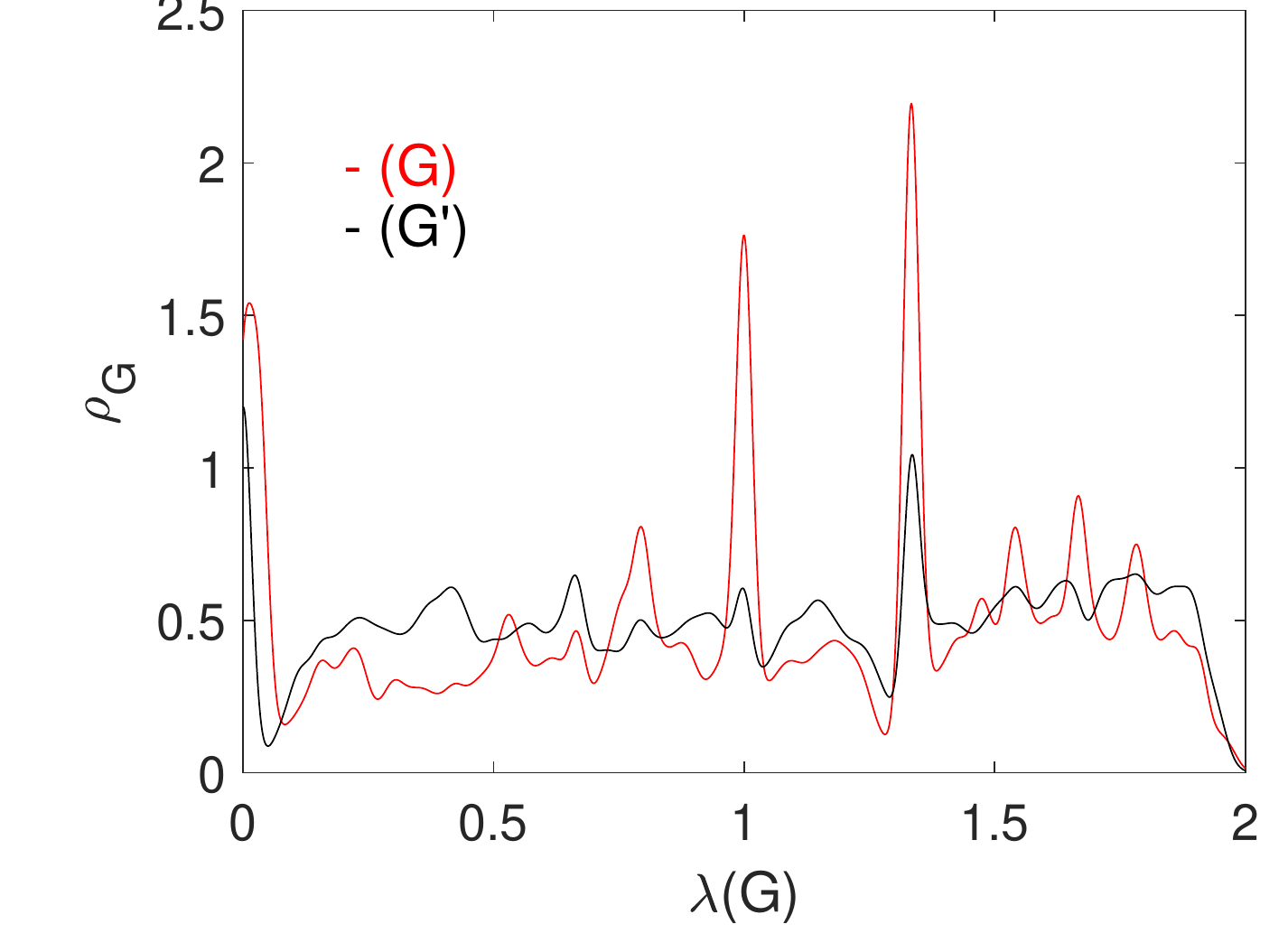} 

(\textbf{c}) 15 4  \hspace{2.7cm} (\textbf{g})   20 3 
 
 \includegraphics[trim = 25mm 90mm 40mm 100mm,clip, width=6.5cm, height=5cm]{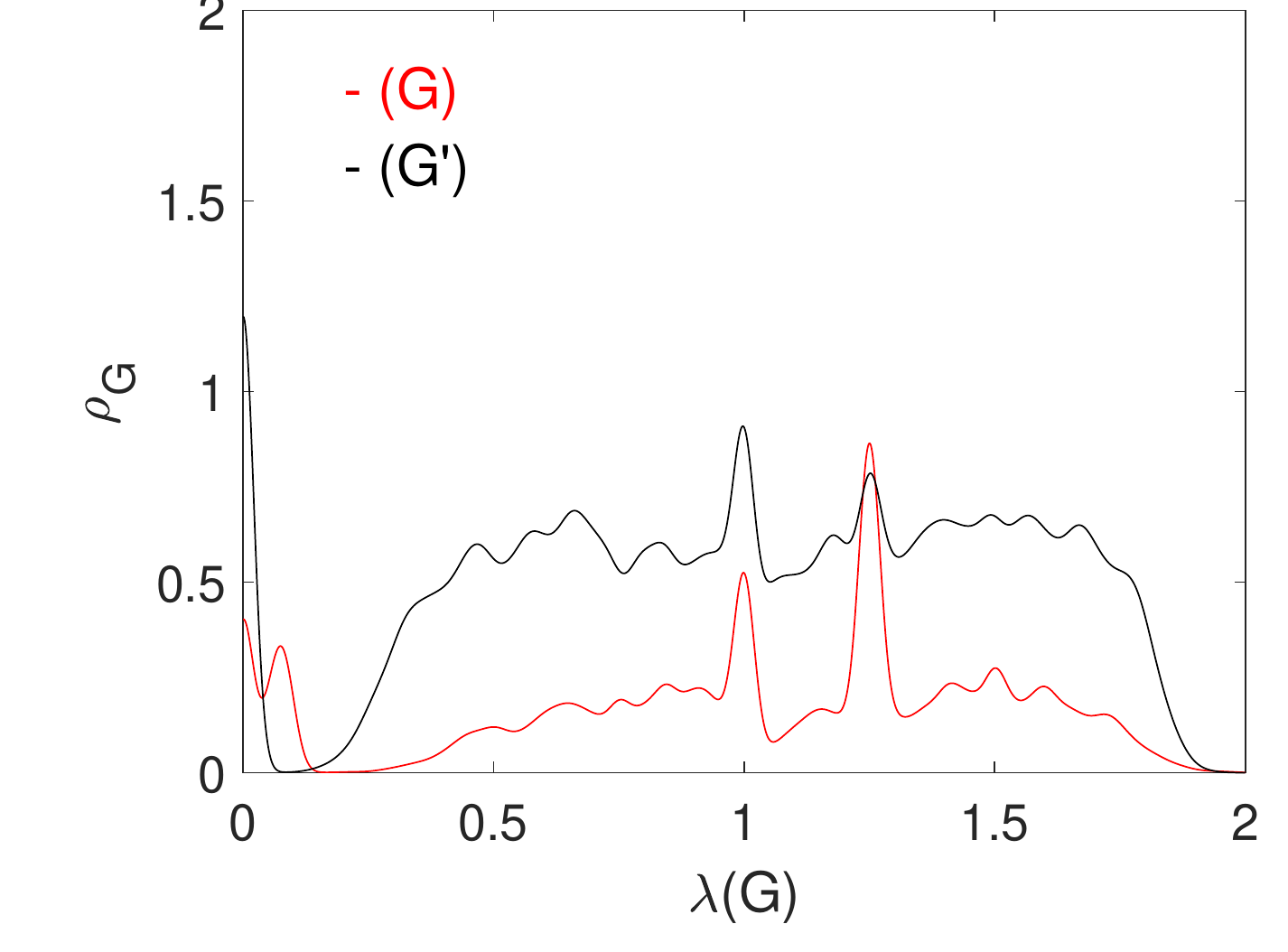} 
\includegraphics[trim = 25mm 90mm 40mm 100mm,clip, width=6.5cm, height=5cm]{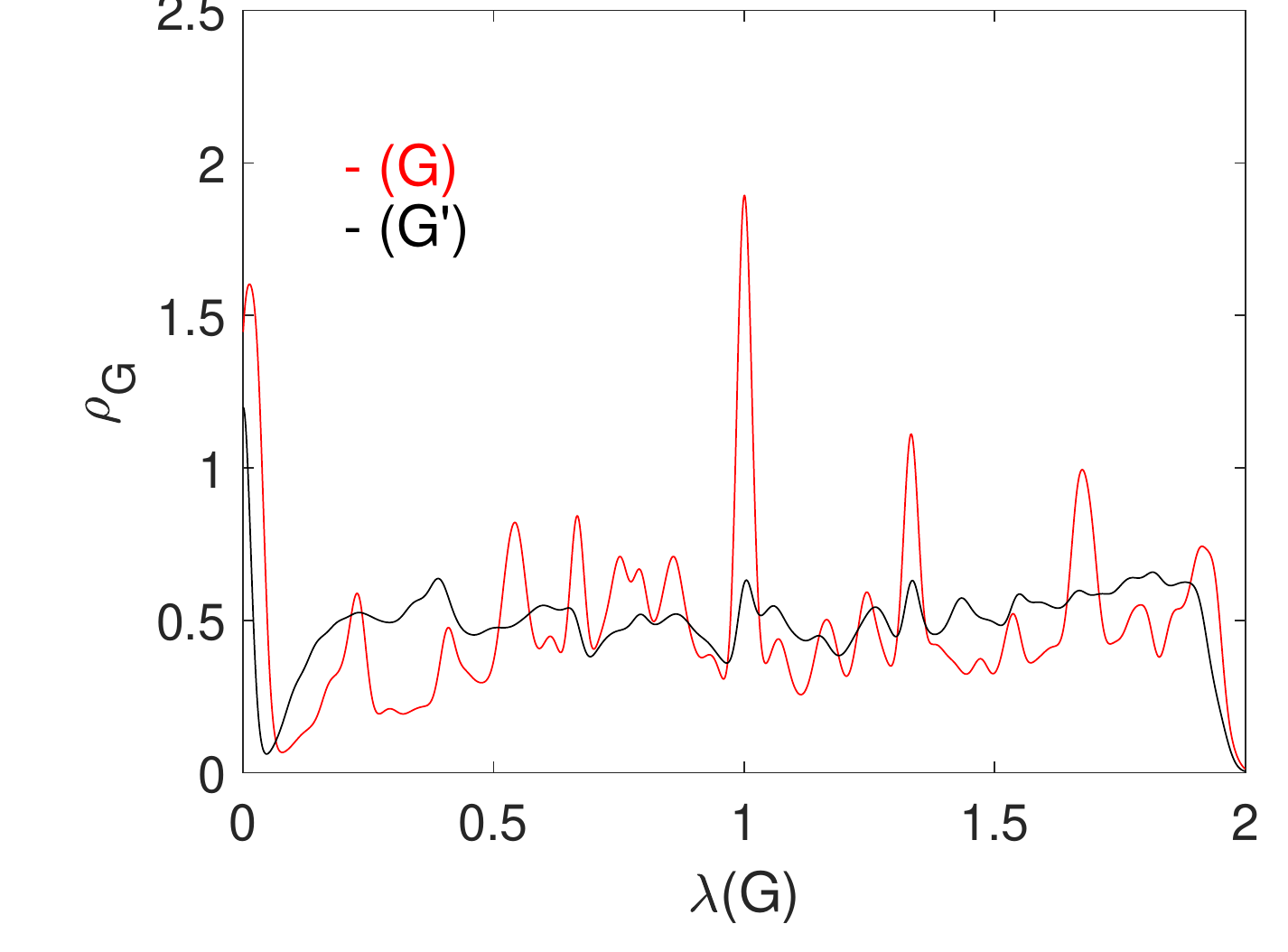} 

(\textbf{d}) 16 4   \hspace{2.7cm} (\textbf{h}) 22 3

\caption{\small{The smoothed spectral density $\rho_G$ for amplifier constructors (red line) and the remaining regular graphs (black line) for: (a)-(d) the quartic graphs of size $N=\{12,13,15,16\}$ and (e)-(h) the cubic graphs of size $N=\{14,18,20,22\}$. Supplement to Fig. \ref{fig:dense}}}
\label{fig:dense1}
\end{figure}

% Either type in your references using
% \begin{thebibliography}{}
% \bibitem{}
% Text
% \end{thebibliography}
%
% or
%
% Compile your BiBTeX database using our plos2015.bst
% style file and paste the contents of your .bbl file
% here. See http://journals.plos.org/plosone/s/latex for 
% step-by-step instructions.
% 

\end{document}